\documentclass[reqno,11pt]{amsart}
\usepackage{graphicx}
\usepackage{amscd}
\usepackage{slashed}
\usepackage{amssymb}
\usepackage{esint}
\usepackage{enumitem}
\usepackage[usenames,dvipsnames]{pstricks}
\usepackage{pst-grad} 
\usepackage{pst-plot} 
\usepackage[mathscr]{eucal}

\numberwithin{equation}{section}
\allowdisplaybreaks[1]

\title[Causal Fermion Systems -- An Overview]{Causal Fermion Systems -- An Overview}

\author[F.\ Finster]{Felix Finster \\ \\ May 2015}
\address{Fakult\"at f\"ur Mathematik \\ Universit\"at Regensburg \\ D-93040 Regensburg \\ Germany}
\email{finster@ur.de}

\newtheorem{Def}{Definition}[section]
\newtheorem{Thm}[Def]{Theorem}
\newtheorem{Prp}[Def]{Proposition}
\newtheorem{Lemma}[Def]{Lemma}
\newtheorem{Remark}[Def]{Remark}

\newtheorem{Example}[Def]{Example}

\newcommand{\Thanks}{\vspace*{.5em} \noindent \thanks}
\newcommand{\beq}{\begin{equation}}
\newcommand{\eeq}{\end{equation}}
\newcommand{\Proof}{\begin{proof}}
\newcommand{\QED}{\end{proof} \noindent}
\newcommand{\QEDrem}{\ \hfill $\Diamond$}

\newcommand{\la}{\langle}
\newcommand{\ra}{\rangle}
\newcommand{\bra}{\mathopen{<}}
\newcommand{\ket}{\mathclose{>}}
\newcommand{\Sl}{\mbox{$\prec \!\!$ \nolinebreak}}
\newcommand{\Sr}{\mbox{\nolinebreak $\succ$}}

\newcommand{\C}{\mathbb{C}}
\newcommand{\R}{\mathbb{R}}
\newcommand{\1}{\mbox{\rm 1 \hspace{-1.05 em} 1}}

\newcommand{\N}{\mathbb{N}}

\newcommand{\Pdd}{\mbox{$\partial$ \hspace{-1.2 em} $/$}}

\newcommand{\Aslsh}{\mbox{ $\!\!A$ \hspace{-1.2 em} $/$}}

\newcommand{\np}{n_{\mathrm{p}}}
\newcommand{\na}{n_{\mathrm{a}}}

\newcommand{\K}{{\mathscr{K}}}
\newcommand{\D}{\mathscr{D}}

\newcommand{\Cl}{{\mathscr{C}}\ell}
\renewcommand{\H}{\mathscr{H}}
\newcommand{\F}{{\mathscr{F}}}
\newcommand{\Lin}{\text{\rm{L}}}
\newcommand{\itemD}{\item[{\raisebox{0.125em}{\tiny $\blacktriangleright$}}]}

\newcommand{\as}{{\mathfrak{a}}}
\newcommand{\bs}{{\mathfrak{b}}}
\DeclareMathOperator{\re}{Re}
\DeclareMathOperator{\im}{Im}
\DeclareMathOperator{\Tr}{Tr}
\DeclareMathOperator{\tr}{tr}

\DeclareMathOperator{\Symm}{Symm}

\renewcommand{\O}{{\mathscr{O}}}
\renewcommand{\L}{{\mathcal{L}}}

\newcommand{\Sact}{{\mathcal{S}}}
\newcommand{\T}{{\mathcal{T}}}
\newcommand\B{{\mathscr{B}}}
\newcommand{\U}{\text{\rm{U}}}
\newcommand{\SU}{\text{\rm{SU}}}
\renewcommand{\u}{\text{\rm{u}}}

\newcommand{\UMNS}{U_\text{\tiny{MNS}}}

\DeclareMathOperator{\norm}{| \hspace*{-0.1em}| \hspace*{-0.1em}|}
\DeclareMathOperator{\supp}{supp}

\newcommand{\scrM}{\myscr M}
\newcommand{\scrN}{\myscr N}

\newcommand{\scrU}{{\mathscr{U}}}
\newcommand{\scrA}{{\mathscr{A}}}
\newcommand{\Sig}{\mathscr{S}}

\DeclareFontFamily{OT1}{rsfso}{}
\DeclareFontShape{OT1}{rsfso}{m}{n}{ <-7> rsfso5 <7-10> rsfso7 <10-> rsfso10}{}
\DeclareMathAlphabet{\myscr}{OT1}{rsfso}{m}{n}

\begin{document}
\maketitle
\begin{abstract}
The theory of causal fermion systems is an approach to describe fundamental physics.
We here introduce the mathematical framework and give an overview of the objectives
and current results.
\end{abstract}
\tableofcontents

Causal fermion systems were introduced in~\cite{rrev} as a reformulation and generalization
of the setting used in the fermionic projector approach~\cite{PFP}. 
The theory of causal fermion systems is an approach to describe fundamental physics. It gives quantum mechanics, general relativity and quantum field theory as limiting cases and is therefore a
candidate for a unified physical theory.
In this article,
we introduce the mathematical framework
and give an overview of the different limiting cases.
The presentation is self-contained and includes references to
the corresponding research papers.
The aim is not only to convey the underlying physical picture,
but also to lay the mathematical foundations in a conceptually convincing way.
This includes technical issues like specifying the topologies on the different spaces of
functions and operators, giving a mathematical definition of an ultraviolet regularization,
or specifying the maps which identify the objects of the
causal fermion system with corresponding objects in Minkowski space.
Also, we use a basis-independent notation whenever possible.
The reader interested in a non-technical introduction is referred to~\cite{dice2014}.

\section{The Abstract Framework} \label{secframe}

\subsection{Basic Definitions} \label{secbasicdef}
For conceptual clarity, we begin with the general definitions.
\begin{Def} \label{defparticle} (causal fermion system) {\em{
Given a separable complex Hilbert space~$\H$ with scalar product~$\la .|. \ra_\H$
and a parameter~$n \in \N$ (the {\em{``spin dimension''}}), we let~$\F \subset \Lin(\H)$ be the set of all
self-adjoint operators on~$\H$ of finite rank, which (counting multiplicities) have
at most~$n$ positive and at most~$n$ negative eigenvalues. On~$\F$ we are given
a positive measure~$\rho$ (defined on a $\sigma$-algebra of subsets of~$\F$), the so-called
{\em{universal measure}}. We refer to~$(\H, \F, \rho)$ as a {\em{causal fermion system}}.
}}
\end{Def} \noindent
We remark that the separability of the Hilbert space (i.e.\ the assumption that~$\H$ admits an at most
countable Hilbert space basis) is not essential and could be left out.
We included the separability assumption because it seems to cover all cases of physical interest
and is useful if one wants to work with basis representations.

A causal fermion system describes a space-time together
with all structures and objects therein (like the causal and metric structures, spinors
and interacting quantum fields).
In order to single out the physically admissible
causal fermion systems, one must formulate physical equations. This is accomplished with the help
of an action principle which we now introduce. For any~$x, y \in \F$, the product~$x y$ is an operator
of rank at most~$2n$. We denote its non-trivial eigenvalues (counting algebraic multiplicities)
by~$\lambda^{xy}_1, \ldots, \lambda^{xy}_{2n} \in \C$.
We introduce the {\em{spectral weight}}~$| \,.\, |$ of an operator as the sum of the absolute values
of its eigenvalues. In particular, the spectral weight of the operator
products~$xy$ and~$(xy)^2$ is defined by
\[ |xy| = \sum_{i=1}^{2n} \big| \lambda^{xy}_i \big|
\qquad \text{and} \qquad \big| (xy)^2 \big| = \sum_{i=1}^{2n} \big| \lambda^{xy}_i \big|^2 \:. \]
We introduce the Lagrangian and the action by
\begin{align}
\text{\em{Lagrangian:}} && \L(x,y) &= \big| (xy)^2 \big| - \frac{1}{2n}\: |xy|^2 \label{Lagrange} \\
\text{\em{action:}} && \Sact(\rho) &= \iint_{\F \times \F} \L(x,y)\: d\rho(x)\, d\rho(y) \:. \label{Sdef}
\end{align}
The {\em{causal action principle}} is to minimize~$\Sact$ by varying the universal measure
under the following constraints:
\begin{align}
\text{\em{volume constraint:}} && \rho(\F) = \text{const} \quad\;\; & \label{volconstraint} \\
\text{\em{trace constraint:}} && \int_\F \tr(x)\: d\rho(x) = \text{const}& \label{trconstraint} \\
\text{\em{boundedness constraint:}} && \T := \iint_{\F \times \F} |xy|^2\: d\rho(x)\, d\rho(y) &\leq C \:, \label{Tdef}
\end{align}
where~$C$ is a given parameter (and~$\tr$ denotes the trace of a linear operator on~$\H$).

In order to make the causal action principle mathematically well-defined, one needs
to specify the class of measures in which to vary~$\rho$. To this end,
on~$\F$ we consider the topology induced by the operator norm
\beq \label{supnorm}
\|A\| := \sup \big\{ \|A u \|_\H \text{ with } \| u \|_\H = 1 \big\} \:.
\eeq
In this topology, the Lagrangian as well as the integrands in~\eqref{trconstraint}
and~\eqref{Tdef} are continuous.
The $\sigma$-algebra generated by the open sets of~$\F$ consists of the so-called Borel sets.
A {\em{regular Borel measure}} is a measure on the Borel sets with the property that
it is continuous under approximations by compact sets from inside and by open sets from outside
(for basics see for example~\cite[\S52]{halmosmt}).
The right prescription is to vary~$\rho$ within the class of regular Borel measures of~$\F$.
In the so-called {\em{finite-dimensional setting}} when~$\H$ is finite-dimensional and the total volume~$\rho(\F)$
is finite, the existence of minimizers is proven in~\cite{discrete, continuum}, and the properties of
minimizing measures are analyzed in~\cite{support, lagrange}.

The causal action principle also makes mathematical sense in the so-called
{\em{infinite-dimensional setting}} when~$\H$ is infinite-dimensional and the total volume~$\rho(\F)$
is infinite. In this case, the volume constraint~\eqref{volconstraint}
is implemented by demanding that all variations~$(\rho(\tau))_{\tau \in (-\varepsilon, \varepsilon)}$
should for all~$\tau, \tau' \in (-\varepsilon, \varepsilon)$ satisfy the conditions
\beq \label{totvol}
\big| \rho(\tau) - \rho(\tau') \big|(\F) < \infty \qquad \text{and} \qquad
\big( \rho(\tau) - \rho(\tau') \big) (\F) = 0
\eeq
(where~$|.|$ denotes the total variation of a measure; see~\cite[\S28]{halmosmt}).
The existence theory in the infinite-dimensional setting has not yet been developed.
But it is known that the Euler-Lagrange equations corresponding to the
causal action principle still have a mathematical meaning (as will be explained in~\S\ref{secvary}
below).
This makes it possible to analyze the causal action principle without restrictions on the
dimension of~$\H$ nor on the total volume.
One way of getting along without an existence theory in the infinite-dimensional
setting is to take the point of view that on a fundamental physical level,
the Hilbert space~$\H$ is finite-dimensional, whereas the
infinite-dimensional setting merely is a mathematical idealization needed in
order to describe systems involving an infinite number of quantum particles.


We finally explain the significance of the constraints. Generally speaking,
the constraints~\eqref{volconstraint}--\eqref{Tdef} are needed to avoid trivial minimizers and in order for the variational principle to be well-posed.
More specifically, if we dropped the constraint of fixed total volume~\eqref{volconstraint}, the measure~$\rho=0$
would be trivial minimizer. Without the boundedness constraint~\eqref{Tdef},
the loss of compactness discussed in~\cite[Section~2.2]{continuum} implies that no minimizers exist.
If, on the other hand, we dropped the trace constraint~\eqref{trconstraint},
a trivial minimizer could be constructed as follows. We let~$x$ be the operator with the matrix representation
\[ x = \text{diag} \big( \underbrace{1, \ldots, 1}_{\text{$n$ times}}, 
\underbrace{-1, \ldots, -1}_{\text{$n$ times}}, 0, 0, \ldots \big) \]
and choose~$\rho$ as a multiple of the Dirac measure supported at~$x$.
Then~$\T > 0$ but~$\Sact=0$.

\subsection{Space-Time and Causal Structure} \label{seccausal}
A causal fermion system~$(\H, \F, \rho)$ encodes a large amount of information.
In order to recover this information, one can for example form
products of linear operators in~$\F$, compute the eigenvalues of such operator products
and integrate expressions involving these eigenvalues with respect to the universal measure.
However, it is not obvious what all this information means. In order to clarify the situation, we now
introduce additional mathematical objects.
These objects are {\em{inherent}} in the sense that we only use information already encoded
in the causal fermion system.

We first define {\em{space-time}}, denoted by~$M$, as the support of the universal measure,
\[ M := \text{supp}\, \rho \subset \F \:. \]
On~$M$ we consider the topology induced by~$\F$ (generated by the $\sup$-norm~\eqref{supnorm}
on~$\Lin(\H)$). Moreover, the universal measure~$\rho|_M$ restricted to~$M$ can be regarded as a volume
measure on space-time. This makes space-time into a {\em{topological measure space}}.
Furthermore, one has the following notion of causality:

\begin{Def} (causal structure) \label{def2}
{\em{ For any~$x, y \in \F$, the product~$x y$ is an operator
of rank at most~$2n$. We denote its non-trivial eigenvalues (counting algebraic multiplicities)
by~$\lambda^{xy}_1, \ldots, \lambda^{xy}_{2n}$. The points~$x$ and~$y$ are
called {\em{spacelike}} separated if all the~$\lambda^{xy}_j$ have the same absolute value.
They are said to be {\em{timelike}} separated if the~$\lambda^{xy}_j$ are all real and do not all 
have the same absolute value.
In all other cases (i.e.\ if the~$\lambda^{xy}_j$ are not all real and do not all 
have the same absolute value),
the points~$x$ and~$y$ are said to be {\em{lightlike}} separated. }}
\end{Def} \noindent
Restricting the causal structure of~$\F$ to~$M$, we get causal relations in space-time.
To avoid confusion, we remark that in earlier papers (see~\cite{lqg}, \cite{rrev})
a slightly different definition of the causal structure was used.
But the modified definition used here seems preferable.

The Lagrangian~\eqref{Lagrange} is compatible with the above notion of causality in the
following sense. Suppose that two points~$x, y \in \F$ are spacelike separated.
Then the eigenvalues~$\lambda^{xy}_i$ all have the same absolute value.
Rewriting~\eqref{Lagrange} as
\[ \L = \sum_{i=1}^{2n} |\lambda^{xy}_i|^2 - \frac{1}{2n} \sum_{i,j=1}^{2n}
 |\lambda^{xy}_i|\: |\lambda^{xy}_j|
 = \frac{1}{4n} \sum_{i,j=1}^{2n} \Big( \big|\lambda^{xy}_i \big| - \big|\lambda^{xy}_j \big| \Big)^2 \:, \]
one concludes that the Lagrangian vanishes. Thus pairs of points with spacelike
separation do not enter the action. This can be seen in analogy to the usual notion of causality where
points with spacelike separation cannot influence each other\footnote{For clarity, we point
out that our notion of causality does allow for nonlocal correlations and entanglement
between regions with space-like separation. This will become clear in~\S\ref{secwave}
and Section~\ref{secQFT}.}.
This analogy is the reason for the notion ``causal'' in ``causal fermion system''
and ``causal action principle.''

The above notion of causality is {\em{symmetric}} in~$x$ and~$y$, as we now explain.
Since the trace is invariant under cyclic permutations, we know that
\beq \label{trsymm}
\tr \big( (xy)^p \big) = \tr \big( x \,(yx)^{p-1}\, y \big) = \tr \big( (yx)^{p-1} \,yx \big)
= \tr \big( (yx)^p \big)
\eeq
(where~$\tr$ again denotes the trace of a linear operator on~$\H$).
Since all our operators have finite rank, there is a finite-dimensional subspace~$I$
of~$\H$ such that~$xy$ maps~$I$ to itself and vanishes on the orthogonal complement of~$I$.
Then the non-trivial eigenvalues of the operator product~$xy$ are given
as the zeros of the characteristic polynomial of the restriction~$xy|_I : I \rightarrow I$.
The coefficients of this characteristic polynomial (like the trace, the determinant, etc.)
are symmetric polynomials in the eigenvalues and can therefore be expressed in terms of traces of
powers of~$xy$. 
As a consequence, the identity~\eqref{trsymm} implies that the operators~$xy$
and~$yx$ have the same characteristic polynomial and are thus
isospectral. This shows that the causal notions are indeed symmetric in the
sense that~$x$ and~$y$ are spacelike separated if and only if~$y$ and~$x$ are
 (and similarly for timelike and lightlike separation). One also sees that the Lagrangian~$\L(x,y)$ is symmetric
in its two arguments.

A causal fermion system also distinguishes a {\em{direction of time}}.
To this end, we let~$\pi_x$ be the orthogonal projection in~$\H$ on the subspace~$x(\H) \subset \H$
and introduce the functional
\beq \label{Cform}
{\mathscr{C}} \::\: M \times M \rightarrow \R\:,\qquad
{\mathscr{C}}(x, y) := i \Tr \big( y\,x \,\pi_y\, \pi_x - x\,y\,\pi_x \,\pi_y \big)
\eeq
(this functional was first stated in~\cite[Section~7.5]{topology}, motivated by constructions
in~\cite[Section~3.5]{lqg}).
Obviously, this functional is anti-symmetric in its two arguments.
This makes it possible to introduce the notions
\beq \label{tdir}
\left\{ \begin{array}{cl} \text{$y$ lies in the {\em{future}} of~$x$} &\quad \text{if~${\mathscr{C}}(x, y)>0$} \\[0.2em]
\text{$y$ lies in the {\em{past}} of~$x$} &\quad \text{if~${\mathscr{C}}(x, y)<0$}\:. \end{array} \right.
\eeq
By distinguishing a direction of time, we get a structure similar to
a causal set (see for example~\cite{sorkin}). But in contrast to a causal set, our notion of
``lies in the future of'' is not necessarily transitive.
This corresponds to our physical conception that the transitivity of the causal relations
could be violated both on the cosmological scale (there might be closed timelike curves)
and on the microscopic scale (there seems no compelling reason why the causal
relations should be transitive down to the Planck scale).
This is the reason why we consider other structures (namely the universal measure and the causal
action principle) as being more fundamental. In our setting, causality merely is a
derived structure encoded in the causal fermion system.

\subsection{The Kernel of the Fermionic Projector} \label{secker}
The causal action principle depends crucially on the eigenvalues of the operator
product~$xy$ with~$x,y \in \F$. For computing these eigenvalues, it is convenient not to
consider this operator product on the (possibly infinite-dimensional) Hilbert space~$\H$, 
but instead to restrict attention to a finite-dimensional subspace of~$\H$, chosen such that
the operator product vanishes on the orthogonal complement of this subspace.
This construction leads us to the spin spaces and to the kernel of the fermionic projector,
which we now introduce.
For every~$x \in \F$ we define the {\em{spin space}}~$S_x$ by~$S_x = x(\H)$; it is a subspace of~$\H$ of dimension at most~$2n$. For any~$x, y \in M$ we define the
{\em{kernel of the fermionic operator}}~$P(x,y)$ by
\beq \label{Pxydef}
P(x,y) = \pi_x \,y|_{S_y} \::\: S_y \rightarrow S_x
\eeq
(where~$\pi_x$ is again the orthogonal projection on the subspace~$x(\H) \subset \H$).
Taking the trace of~\eqref{Pxydef} in the case~$x=y$, one
finds that~$\tr(x) = \Tr_{S_x}(P_\tau(x,x))$, making it possible to express
the integrand of the trace constraint~\eqref{trconstraint} in terms of the kernel of the fermionic operator.
In order to also express the eigenvalues of the operator~$xy$, we define the
{\em{closed chain}}~$A_{xy}$ as the product
\beq \label{Axydef}
A_{xy} = P(x,y)\, P(y,x) \::\: S_x \rightarrow S_x\:.
\eeq
Computing powers of the closed chain, one obtains
\[ A_{xy} = (\pi_x y)(\pi_y x)|_{S_x} = \pi_x\, yx|_{S_x} \:,\qquad
(A_{xy})^p = \pi_x\, (yx)^p|_{S_x} \:. \]
Taking the trace, one sees in particular that~$\Tr_{S_x}(A_{xy}^p) = \tr \big((yx)^p \big)$.
Repeating the arguments after~\eqref{trsymm}, one concludes that
the eigenvalues of the closed chain coincide with the non-trivial
eigenvalues~$\lambda^{xy}_1, \ldots, \lambda^{xy}_{2n}$ of the operator~$xy$ in
Definition~\ref{def2}. Therefore, the kernel of the fermionic operator encodes the causal structure of~$M$.
The main advantage of working with the kernel of the fermionic operator is that the closed chain~\eqref{Axydef}
is a linear operator on a vector space of dimension at most~$2n$, making it possible
to compute the~$\lambda^{xy}_1, \ldots, \lambda^{xy}_{2n}$ as the eigenvalues of 
a finite matrix. 

Next, it is very convenient to arrange that the kernel of the fermionic operator is symmetric in the sense that
\beq \label{Pxysymm}
P(x,y)^* = P(y,x) \:.
\eeq
To this end, one chooses on the spin space~$S_x$ the {\em{spin scalar product}} $\Sl .|. \Sr_x$ by
\beq \label{ssp}
\Sl u | v \Sr_x = -\la u | x u \ra_\H \qquad \text{(for all $u,v \in S_x$)}\:.
\eeq
Due to the factor~$x$ on the right, this definition really makes the kernel of the fermionic operator symmetric,
as is verified by the computation
\begin{align*}
\Sl u \,|\, P(x,y) \,v \Sr_x &= - \la u \,|\, x\, P(x,y) \,v \ra_\H = - \la u \,|\, x y \,v \ra_\H \\
&= -\la \pi_y \,x\, u \,|\, y \,v \ra_\H = \Sl P(y,x)\, u \,|\,  v \Sr_y
\end{align*}
(where~$u \in S_x$ and~$v \in S_y$).
The spin space~$(S_x, \Sl .|. \Sr_x)$ is an {\em{indefinite}} inner product of signature~$(p,q)$ with~$p,q \leq n$.
In this way, indefinite inner product spaces arise naturally when analyzing the
mathematical structure of the causal action principle.

The kernel of the fermionic operator as defined by~\eqref{Pxydef} is also referred to as
the kernel of the {\em{fermionic projector}}, provided that suitable normalization conditions are
satisfied. Different normalization conditions have been proposed and analyzed
(see the discussion in~\cite[Section~2.2]{norm}). More recently, it was observed in~\cite{noether} that
one of these normalization conditions is automatically satisfied if the universal measure
is a minimizer of the causal action principle (see~\S\ref{secnoether} below).
With this in mind, we no longer need to be so careful about the normalization.
For notational simplicity, we always refer to~$P(x,y)$ as the kernel of the fermionic projector.

\subsection{Wave Functions and Spinors} \label{secwave}
For clarity, we sometimes denote the spin space~$S_x$ at a space-time point~$x \in M$
by~$S_xM$. A {\em{wave function}}~$\psi$ is defined as a function
which to every~$x \in M$ associates a vector of the corresponding spin space,
\beq \label{psirep}
\psi \::\: M \rightarrow \H \qquad \text{with} \qquad \psi(x) \in S_xM \quad \text{for all~$x \in M$}\:. 
\eeq
We now want to define what we mean by {\em{continuity}} of a wave function.
For the notion of continuity, we need to compare the wave function at different space-time points,
being vectors~$\psi(x) \in S_xM$ and~$\psi(y) \in S_y$M in different spin spaces.
Using that both spin spaces~$S_xM$ and~$S_yM$ are subspaces of the same
Hilbert space~$\H$, an obvious idea is to simply work with the Hilbert space norm
$\|\psi(x) - \psi(y)\|_\H$. However, in view of the factor~$x$ in the spin scalar product~\eqref{ssp},
it is preferable to insert a corresponding power of the operator~$x$.
Namely, the natural norm on the spin space~$(S_x, \Sl .|. \Sr_x)$ is given by
\[ \big| \psi(x) \big|_x^2 := \big\la \psi(x) \,\big|\, |x|\, \psi(x) \big\ra_\H = \Big\| \sqrt{|x|} \,\psi(x) \Big\|_\H^2 \]
(where~$|x|$ is the absolute value of the symmetric operator~$x$ on~$\H$, and~$\sqrt{|x|}$
the square root thereof). This leads us to defining that
the wave function~$\psi$ is {\em{continuous}} at~$x$ if
for every~$\varepsilon>0$ there is~$\delta>0$ such that
\[ \big\| \sqrt{|y|} \,\psi(y) -  \sqrt{|x|}\, \psi(x) \big\|_\H < \varepsilon
\qquad \text{for all~$y \in M$ with~$\|y-x\| \leq \delta$} \:. \]
Likewise, $\psi$ is said to be continuous on~$M$ if it continuous at every~$x \in M$.
We denote the set of continuous wave functions by~$C^0(M, SM)$.
Clearly, the space of continuous wave functions is a complex vector space with pointwise operations, i.e.\
$(\alpha \psi + \beta \phi)(x) := \alpha \psi(x) + \beta \phi(x)$ with~$\alpha, \beta \in \C$.

It is an important observation that every vector~$u \in \H$ of the Hilbert space gives rise to a unique
wave function. To obtain this wave function, denoted by~$\psi^u$, we simply project the vector~$u$
to the corresponding spin spaces,
\beq \label{psiudef}
\psi^u \::\: M \rightarrow \H\:,\qquad \psi^u(x) = \pi_x u \in S_xM \:.
\eeq
We refer to~$\psi^u$ as the {\em{physical wave function}} of~$u \in \H$.
The estimate%
\footnote{For completeness, we derive the inequality~($\star$):
Since the operator~$\sqrt{|y|} - \sqrt{|x|}$ is symmetric and has finite rank, there is a normalized
vector~$u \in \H$ such that
\beq \label{sqrtpos}
\Big(\sqrt{|y|} - \sqrt{|x|} \Big) u = \pm \Big\|\sqrt{|y|} - \sqrt{|x|} \Big\|\,u \:.
\eeq
Possibly by exchanging the roles of~$x$ and~$y$ we can arrange the plus sign. Then
\[ \Big\|\sqrt{|y|} - \sqrt{|x|} \Big\|
= \big\la u \,\big|\, \Big(\sqrt{|y|} - \sqrt{|x|} \Big) u \big\ra
\leq \big\la u \,\big|\, \Big(\sqrt{|y|} + \sqrt{|x|} \Big) u \big\ra \:, \]
where in the last step we used that the operator~$\sqrt{|x|}$ is positive.
Multiplying by~$\big\|\sqrt{|y|} - \sqrt{|x|} \big\|$ and using~\eqref{sqrtpos} with the plus sign,
we obtain
\begin{align*}
\Big\|\sqrt{|y|} - \sqrt{|x|} \Big\|^2
&\leq \frac{1}{2} \bigg(  \big\la u \,\big|\, \Big(\sqrt{|y|} + \sqrt{|x|} \Big) \Big(\sqrt{|y|} - \sqrt{|x|} \Big) u \big\ra
+ \big\la \Big(\sqrt{|y|} - \sqrt{|x|} \Big) u \,\big|\,\Big(\sqrt{|y|} + \sqrt{|x|} \Big) u \big\ra \bigg) \\
&= \frac{1}{2}
\:\big\la u \,\big|\, \left\{ \Big(\sqrt{|y|} + \sqrt{|x|} \Big), \Big(\sqrt{|y|} - \sqrt{|x|} \Big) \right\} u \big\ra
= \big\la u \,\big|\, \big(|y| -|x| \big) \,u \big\ra \leq \big\| |y| -|x| \big\| \:.
\end{align*}
We thus obtain the inequality~$\big\| \sqrt{|y|} - \sqrt{|x|} \big\|^2 \leq \big\| |y|- |x| \big\|$.
Applying this inequality with~$x$ replaced by~$x^2$ and~$y$ replaced by~$y^2$,
it also follows that~$\big\| |y|- |x| \big\|^2 \leq \big\| y^2- x^2 \big\|
\leq \big\| y-x \big\| \,\big\| y+x \big\|$. Combining these inequalities gives~($\star$).}
\begin{align*}
\Big\| &\sqrt{|y|} \,\psi^u(y) -  \sqrt{|x|}\, \psi^u(x) \Big\|_\H = \Big\| \sqrt{|y|} \,u -  \sqrt{|x|}\,u \Big\|_\H \\
&\leq \Big\| \sqrt{|y|} - \sqrt{|x|} \Big\| \, \|u\|_\H \overset{(\star)}{\leq}
\|y-x\|^\frac{1}{4} \:\|y+x\|^\frac{1}{4} \: \|u\|_\H
\end{align*}
shows that~$\psi^u$ is indeed continuous.
The physical picture is that the physical wave functions~$\psi^u$ are those wave functions
which are realized in the physical system.
Using a common physical notion, one could say that the vectors in~$\H$ correspond
to the ``occupied states'' of the system, and that an occupied state~$u \in \H$ is represented
in space-time by the corresponding physical wave function~$\psi^u$.
The shortcoming of this notion is that an ``occupied state'' is defined only
for free quantum fields, whereas the
physical wave functions are defined also in the interacting theory.
For this reason, we prefer not use the notion of ``occupied states.''

For a convenient notation, we also introduce the {\em{wave evaluation operator}}~$\Psi$
as an operator which to every Hilbert space vector associates the corresponding physical wave function,
\beq \label{weo}
\Psi \::\: \H \rightarrow C^0(M, SM)\:, \qquad u \mapsto \psi^u \:.
\eeq
Evaluating at a fixed space-time point gives the mapping
\[ \Psi(x) \::\: \H \rightarrow S_xM\:, \qquad u \mapsto \psi^u(x) \:. \]
The kernel of the fermionic projector can be expressed in terms of the wave evaluation operator:
\begin{Lemma} \label{lemmaPxyrep} For any~$x, y \in M$,
\begin{align}
x &= - \Psi(x)^* \,\Psi(x) \label{Fid} \\
P(x,y) &= -\Psi(x)\, \Psi(y)^*\:. \label{Pid}
\end{align}
\end{Lemma}
\Proof For any~$v \in S_xM$ and~$u \in \H$,
\[ \Sl v \,|\, \Psi(x)\, u \Sr_x = \Sl v \,|\, \pi_x\, u \Sr_x
\overset{\eqref{ssp}}{=} -\la v \,|\, x\, u \ra_\H = \la (-x)\, v \,|\, u \ra_\H \]
and thus
\[ \Psi(x)^* = -x|_{S_xM} \::\: S_xM \rightarrow \H \:. \]
Hence
\[ \Psi(x)^* \,\Psi(x) \,u = \Psi(x)^* \,\psi^u_x = -x \,\psi^u_x \overset{\eqref{psiudef}}{=} -x \,\pi_x u = -x u \:, \]
proving~\eqref{Fid}. Similarly, the relation~\eqref{Pid} follows from the
computation
\[ \Psi(x)\, \Psi(y)^* = -\pi_x\, y|_{S_y} = -P(x,y) \:. \]
This completes the proof.
\QED

The structure of the wave functions~\eqref{psirep} taking values in the spin spaces
is reminiscent of sections of a vector bundle. The only difference is that our setting
is more general in that the base space~$M$ does not need to be a manifold, and the
fibres~$S_xM$ do not need to depend smoothly on the base point~$x$.
However, comparing to the setting of spinors in Minkowski space or on a Lorentzian
manifold, one important structure is missing: we have no Dirac matrices
and no notion of Clifford multiplication. The following definition is a step towards
introducing these additional structures.

\begin{Def} (Clifford subspace) \label{defcliffsubspace} {\em{
We denote the space of symmetric linear operators on~$(S_x, \Sl .|. \Sr_x)$
by~$\Symm(S_x) \subset \Lin(S_x)$.
A subspace~$K \subset \Symm(S_x)$ is called
a {\em{Clifford subspace}} of signature~$(r,s)$ at the point~$x$ (with~$r,s \in \N_0$)
if the following conditions hold:
\begin{itemize}
\item[(i)] For any~$u, v \in K$, the anti-commutator~$\{ u,v \} \equiv u v + v u$ is a multiple
of the identity on~$S_x$.
\item[(ii)] The bilinear form~$\la .,. \ra$ on~$K$ defined by
\beq \label{anticomm}
\frac{1}{2} \left\{ u,v \right\} = \la u,v \ra \, \1 \qquad {\text{for all~$u,v \in K$}}
\eeq
is non-degenerate and has signature~$(r,s)$.
\end{itemize} }}
\end{Def}
In view of the anti-commutation relations~\eqref{anticomm}, a Clifford subspace can be
regarded as a generalization of the space spanned by the usual Dirac matrices.
However, the above definition has two shortcomings:
First, there are many different Clifford subspaces, so that there is no unique
notion of Clifford multiplication. Second, we are missing the structure of tangent vectors
as well as a mapping which would associate a tangent vector to an element of
the Clifford subspace.

These shortcomings can be overcome by using either geometric or measure-theoretic
methods. In the geometric approach, one gets along with the non-uniqueness of the Clifford subspaces
by working with suitable equivalence classes. Using geometric information
encoded in the causal fermion system, one can then construct mappings between
the equivalence classes at different space-time points. This method will be
outlined in~\S\ref{secgeom}. In the measure-theoretic approach, on the other hand, one
uses the local form of the universal measure with the aim of 
constructing a unique Clifford subspace at every space-time point.
This will be outlined in~\S\ref{sectopology}.
Before entering these geometric and measure-theoretic constructions, we
introduce additional structures on the space of wave functions.

\subsection{The Fermionic Projector on the Krein Space} \label{secKrein}
The space of wave functions can be endowed with an inner product and a topology.
The inner product is defined by
\beq \label{Sprod}
\bra \psi | \phi \ket = \int_M \Sl \psi(x) | \phi(x) \Sr_x \: d\rho(x) \:.
\eeq
In order to ensure that the last integral converges, we also introduce the scalar product~$\la\!\la .|. \ra\!\ra$
by
\beq \label{ndef}
\la\!\la \psi | \phi \ra\!\ra = \int_M \la \psi(x) |\, |x|\, \phi(x) \ra_\H \:d\rho(x)
\eeq
(where~$|x|$ is again the absolute value of the symmetric operator~$x$ on~$\H$).
The {\em{one-particle space}}~$(\K, \bra .|. \ket)$ is defined as the space of wave functions for which
the corresponding norm~$\norm . \norm$ is finite, with the topology induced by this norm, and endowed with
the inner product~$\bra .|. \ket$. Such an indefinite inner product space with a topology induced
by an additional scalar product is referred to as a {\em{Krein space}} (see for example~\cite{bognar, langer}).

When working with the one-particle Krein space, one must keep in mind that
the physical wave function~$\psi^u$ of a vector~$u \in \H$ does not need to be a vector in~$\K$
because the corresponding integral in~\eqref{Sprod} may diverge. Similarly, the
scalar product~$\la\!\la \psi^u | \psi^u \ra\!\ra$ may be infinite.
One could impose conditions on the causal fermion system which ensure that the integrals
in~\eqref{Sprod} and~\eqref{ndef} are finite for all physical wave functions.
Then the mapping~$u \mapsto \psi^u$ would give rise to an embedding~$\H \hookrightarrow \K$
of the Hilbert space~$\H$ into the one-particle Krein space. However, such conditions seem too
restrictive and are not really needed. Therefore, here we shall not impose any conditions on the causal
fermion systems but simply keep in mind that the physical wave functions are in general no
Krein vectors.

Despite this shortcoming, the Krein space is useful because the kernel of the fermionic projector
gives rise to an operator on~$\K$. Namely, choosing a suitable dense domain of
definition\footnote{For example, one may choose~$\D(P)$ as the set of all vectors~$\psi \in \K$
satisfying the conditions
\[ \phi := \int_M x\, \psi(x)\, d\rho(x) \:\in \: \H \qquad \text{and} \qquad \norm \phi \norm < \infty\:. \]
}~$\D(P)$, we can regard~$P(x,y)$ as the integral kernel of a corresponding operator~$P$,
\beq \label{Pdef}
P \::\: \D(P) \subset \K \rightarrow \K \:,\qquad (P \psi)(x) =
\int_M P(x,y)\, \psi(y)\, d\rho(y)\:,
\eeq
referred to as the {\em{fermionic projector}}. The fermionic projector has the following two useful
properties:
\begin{itemize}
\itemD $P$ is {\em{symmetric}} in the sense that~$\bra P \psi | \phi \ket = \bra \psi | P \phi \ket$
for all~$\psi, \phi \in \D(P)$: \\[0.2em] \label{ABdef}
The symmetry of the kernel of the fermionic projector~\eqref{Pxysymm} implies that
\[ \Sl P(x,y) \psi(y)\,|\, \psi(x) \Sr_x = \Sl \psi(y)\,|\, P(y,x) \psi(x) \Sr_y \:. \]
Integrating over~$x$ and~$y$ and
applying~\eqref{Pdef} and~\eqref{Sprod} gives the result.
\itemD $(-P)$ is {\em{positive}} in the sense that~$\bra \psi | (-P) \psi \ket \geq 0$~for
all $\psi \in \D(P)$: \\[0.2em]
This follows immediately from the calculation
\begin{align*}
\bra \psi | (-P) \psi \ket &= - \iint_{M \times M} \Sl \psi(x) \,|\, P(x,y)\, \psi(y) \Sr_x\:
d\rho(x)\, d\rho(y) \\
&= \iint_{M \times M} \la \psi(x) \,|\, x \, \pi_x \,y\, \psi(y) \ra_\H \:
d\rho(x)\, d\rho(y) = \la \phi | \phi \ra_\H \geq 0 \:,
\end{align*}
where we again used~\eqref{Sprod} and~\eqref{Pxydef} and set
\[ \phi = \int_M x\, \psi(x)\: d\rho(x)\:. \]
\end{itemize}

\subsection{Geometric Structures} \label{secgeom}
A causal fermion system also encodes geometric information on space-time.
More specifically, in the paper~\cite{lqg} notions of connection and curvature are introduced
and analyzed. We now outline a few constructions from this paper.
Recall that the kernel of the fermionic projector~\eqref{Pxydef} is a mapping from
one spin space to another, thereby inducing relations between different space-time points.
The idea is to use these relations for the construction of a spin connection~$D_{x,y}$, being a unitary
mapping between the corresponding spin spaces,
\[ D_{x,y} \::\: S_y \rightarrow S_x \]
(we consistently use the notation that
the subscript~$_{xy}$ denotes an object at the point~$x$, whereas
the additional comma $_{x,y}$ denotes an operator which maps an object at~$y$
to an object at~$x$).
The simplest method for constructing the spin connection would be to
form a polar decomposition, $P(x,y) = A_{xy}^{-\frac{1}{2}}\,U$, and to
introduce the spin connection as the unitary part, $D_{x,y} = U$.
However, this method is too naive, because we want the spin connection to be compatible
with a corresponding metric connection~$\nabla_{x,y}$ which should map Clifford subspaces
at~$x$ and~$y$ (see Definition~\ref{defcliffsubspace} above) isometrically to each other.
A complication is that, as discussed at the end of~\S\ref{secwave},
the Clifford subspaces at~$x$ and~$y$ are not unique.
The method to bypass these problems is to work with several Clifford subspaces
and to use so-called splice maps, as we now briefly explain.

First, it is useful to restrict the freedom in choosing the Clifford subspaces
with the following construction.
Recall that for any~$x \in M$, the operator~$(-x)$ on~$\H$ has at most $n$~positive
and at most~$n$ negative eigenvalues. We denote its positive and negative spectral subspaces
by~$S_x^+$ and~$S_x^-$, respectively. In view of~\eqref{ssp}, these subspaces are also orthogonal
with respect to the spin scalar product,
\[ S_x = S_x^+ \oplus S_x^- \:. \]
We introduce the {\em{Euclidean sign operator}}~$s_x$ as a symmetric operator on~$S_x$
whose eigenspaces corresponding to the eigenvalues~$\pm 1$ are the spaces~$S_x^+$
and~$S_x^-$, respectively.
Since~$s_x^2=\1$, the span of the Euclidean sign operator is a one-dimensional Clifford subspace of
signature~$(1,0$). The idea is to extend~$s_x$ to obtain higher-dimensional Clifford subspaces.
We thus define a {\em{Clifford extension}} as a Clifford subspace which contains~$s_x$.
By restricting attention to Clifford extensions, we have reduced the freedom in choosing
Clifford subspaces. However, there is still not a unique Clifford extension, even for fixed
dimension and signature. But one can define the {\em{tangent space}}~$T_x$
as an equivalence class of Clifford extensions; for details see~\cite[Section~3.1]{lqg}.
The bilinear form~$\la .,. \ra$ in~\eqref{anticomm} induces a Lorentzian metric on the
tangent space.

Next, for our constructions to work, we need to
assume that the points~$x$ and~$y$ are both regular and are properly timelike
separated, defined as follows:
\begin{Def} \label{defregular} {\em{
A space-time point~$x \in M$ is said to be {\em{regular}} if~$x$ has the maximal possible rank,
i.e.~$\dim x(\H) = 2n$. Otherwise, the space-time point is called {\em{singular}}.}}
\end{Def} \noindent
In most situations of physical interest (like Dirac see configurations to be
discussed in Sections~\ref{secmink} and~\ref{seclimit} below),
all space-time points are regular. Singular points, on the other hand,
should be regarded as exceptional points or ``singularities'' of space-time.

\begin{Def} \label{defproptl} {\em{
The space-time points~$x,y \in M$ are {\em{properly timelike}}
separated if the closed chain~$A_{xy}$, \eqref{Axydef}, has a strictly positive spectrum and if all
eigenspaces are definite subspaces of~$(S_x, \Sl .|. \Sr_x)$. }}
\end{Def} \noindent
By a definite subspace of~$S_x$ we mean a subspace on which
the inner product~$\Sl .|. \Sr_x$ is either positive or negative definite.

The two following observations explain why the last definition makes sense:
\begin{itemize}[leftmargin=2em]
\itemD Properly timelike separation implies timelike separation (see Definition~\ref{def2}): \\[0.2em]
Before entering the proof, we give a simple counter example which shows why the
assumption of definite eigenspaces in Definition~\ref{defproptl} is necessary for the implication
to hold. Namely, if the point~$x$ is regular and~$A_{xy}$ is the identity, then
the eigenvalues~$\lambda_1,\ldots, \lambda_{2n}$ are all strictly positive, but they are all equal.

If~$I \subset S_x$ is a definite invariant subspace of~$A_{xy}$,
then the restriction~$A_{xy}|_I$ is a symmetric operator on the Hilbert space~$(I, \pm \Sl .|. \Sr_{I \times I})$,
which is diagonalizable with real eigenvalues. Moreover, the orthogonal complement~$I^\perp$ of~$I \subset S_x$
is again invariant. If~$I^\perp$ is non-trivial, the restriction~$A_{xy}|_{I^\perp}$ has at least one eigenspace. Therefore, the assumption in Definition~\ref{defproptl} that all eigenspaces are definite makes it possible to proceed
inductively to conclude that the operator~$A_{xy}$ is diagonalizable and has real eigenvalues.

If~$x$ and~$y$ are properly timelike separated, then its eigenvalues are
by definition all real and positive. Thus it remains to show that they are not all the same.
If conversely they were all the same, i.e.~$\lambda_1 = \cdots = \lambda_{2n} = \lambda>0$,
then~$S_x$ would necessarily have the maximal dimension~$2n$.
Moreover, the fact that~$A_{xy}$ is diagonalizable implies that~$A_{xy}$ would be a multiple
of the identity on~$S_x$. Therefore, the spin space~$(S_x, \Sl .|. \Sr)$ would have to be definite, in contradiction to
the fact that it has signature~$(n,n)$.
\itemD The notion is symmetric in~$x$ and~$y$: \\[0.2em]
Suppose that~$A_{xy} u = \lambda u$ with~$u \in S_x$ and~$\lambda \in \R \setminus \{0\}$.
Then the vector~$w := P(y,x) \,u \in S_y$ is an eigenvector of~$A_{yx}$ again to the
eigenvalue~$\lambda$,
\begin{align*}
A_{yx} \,w &= P(y,x) P(x,y) \,P(y,x) \, u \\
&= P(y,x) \,A_{xy} \,u = \lambda\, P(y,x)\, u = \lambda w \:.
\end{align*}
Moreover, the calculation
\begin{align*}
\lambda \,\Sl u | u \Sr &= \Sl u | A_{xy} u \Sr = \Sl u \,|\, P(x,y)\, P(y,x) \,u \Sr \\
&= \Sl P(y,x) u \,|\, P(y,x) u \Sr = \Sl w | w \Sr
\end{align*}
shows that~$w$ is a definite vector if and only if~$u$ is.
We conclude that~$A_{yx}$ has positive eigenvalues and definite eigenspaces
if and only if~$A_{xy}$ has these properties.
\end{itemize}

So far, the construction of the spin connection has been worked out only in the case
of spin dimension~$n=2$. Then for two regular and properly timelike separated points~$x,y \in M$,
the spin space~$S_x$ can be decomposed uniquely
into an orthogonal direct sum~$S_x = I^+ \oplus I^-$ of a two-dimensional
positive definite subspace~$I^+$ and a two-dimensional negative definite subspace~$I^-$ of~$A_{xy}$.
We define the {\em{directional sign operator}}~$v_{xy}$ of~$A_{xy}$ as the
unique operator with eigenvalues~$-1,1,0$ such that
the eigenspaces corresponding to the eigenvalues~$\pm 1$
are the subspaces~$I^\pm$.

Having the Euclidean sign operator~$s_x$ and the directional sign operator~$v_{xy}$
to our disposal, under generic assumptions one can distinguish two Clifford subspaces at
the point~$x$: a Clifford subspace~$K_{xy}$ containing~$v_{xy}$ and a Clifford extension~$K_x^{(y)}$
(for details see~\cite[Lemma~3.12]{lqg}). Similarly, at the point~$y$ we have a
distinguished Clifford subspace~$K_{yx}$ (which contains~$v_{yx}$)
and a distinguished Clifford extension~$K_y^{(x)}$.
For the construction of the {\em{spin connection}}~$D_{x,y} : S_y \rightarrow S_x$ one works
with the Clifford subspaces~$K_{xy}$ and~$K_{yx}$ and demands that these are mapped
to each other. More precisely, the spin connection is uniquely characterized by the following
properties (see~\cite[Theorem~3.20]{lqg}):
\begin{itemize}
\item[(i)] $D_{x,y}$ is of the form
\[ D_{x,y} = e^{i \varphi_{xy}\, v_{xy}}\: A_{xy}^{-\frac{1}{2}}\: P(x,y) \quad \text{with} \quad
\varphi_{xy} \in (-\frac{3\pi}{4}, -\frac{\pi}{2}) \cup (\frac{\pi}{2}, \frac{3\pi}{4}) \:. \]
\item[(ii)] The spin connection maps the Clifford subspaces~$K_{xy}$ and~$K_{yx}$ to each other, i.e.\
\[ D_{y,x} \,K_{xy}\, D_{x,y} = K_{yx} \:. \]
\end{itemize}
The spin connection has the properties
\[ D_{y,x} = (D_{x,y})^{-1} = (D_{x,y})^* \qquad \text{and} \qquad
A_{xy} = D_{x,y}\, A_{yx}\, D_{y,x} \:. \]
All the assumptions needed for the construction of the spin connection are combined
in the notion that~$x$ and~$y$ must be {\em{spin-connectable}}
(see~\cite[Definition~3.17]{lqg}). We remark that in the limiting case of a Lorentzian manifold,
the points~$x$ and~$y$ are spin-connectable if they are timelike separated
and sufficiently close to each other (see~\cite[Section~5]{lqg}).

By composing the spin connection along a discrete ``path'' of space-time points,
one obtains a ``parallel transport'' of spinors.
When doing so, it is important to keep track of the different Clifford subspaces and to carefully transform
them to each other. In order to illustrate in an example how this works,
suppose that we want to compose the spin connection~$D_{y,z}$
with~$D_{z,x}$. As mentioned above, the spin connection~$D_{z,x}$ at the point~$z$
is constructed using the Clifford subspace~$K_{zx}$. The spin connection~$D_{y,z}$, however,
takes at the same space-time point~$z$ the Clifford subspace~$K_{zy}$ as reference.
This entails that before applying~$D_{y,z}$ we must transform
from the Clifford subspace~$K_{zx}$ to the Clifford subspace~$K_{zy}$.
This is accomplished by the {\em{splice map}}~$U_z^{(y|x)}$,
being a uniquely defined unitary transformation of~$S_x$ with the property that
\[ K_{zy} = U_z^{(y|x)} \,K_{zx}\, \big( U_z^{(y|x)} \big)^* \:. \]
The splice map must be sandwiched between the spin connections in combinations like
\[ D_{y,z}\, U_z^{(y|x)} \,D_{z,x} \:. \]

In order to construct a corresponding metric connection~$\nabla_{x,y}$, one uses a similar procedure to
related the Clifford subspaces to corresponding Clifford extensions. More precisely, one
first unitarily transform the Clifford extension~$K_y^{(x)}$ to the Clifford subspace~$K_{yx}$.
Unitarily transforming with the spin connection~$D_{xy}$ gives the Clifford subspace~$K_{xy}$.
Finally, one unitarily transforms to the Clifford extension~$K_x^{(y)}$.
Since the Clifford extensions at the beginning and end are
representatives of the corresponding tangent spaces, we thus obtain an isometry
\[ \nabla_{x,y} \::\: T_y \rightarrow T_x \]
between the tangent spaces (for details see~\cite[Section~3.4]{lqg}).

In this setting, {\em{curvature}} is defined as usual as the holonomy of the connection.
Thus the curvature of the spin connection is given by
\[ \mathfrak{R}(x,y,z) = U_x^{(z|y)} \:D_{x,y}\: U_y^{(x|z)} \:D_{y,z}\: U_z^{(y|x)}
\:D_{z,x} \::\: S_x \rightarrow S_x \:, \]
and similarly for the metric connection.
In~\cite[Sections~4 and~5]{lqg} it is proven that the above notions in fact
reduce to the spinorial Levi-Civita connection and the Riemannian curvature on a globally hyperbolic Lorentzian manifold if the causal fermion system is constructed by regularizing solutions of the Dirac equation
(similar as will explained in the next section for the Minkowski vacuum) and the regularization
is suitably removed. These results show that the notions of connection and curvature
defined above indeed generalize the corresponding notions in Lorentzian spin geometry.

\subsection{Topological Structures} \label{sectopology}
From a mathematical perspective, causal fermion systems provide a framework for
non-smooth geometries or generalized ``quantum geometries.''
In this context, it is of interest how the topological notions on a differentiable manifold
or a spin manifold generalize to causal fermion systems.
Such topological questions are analyzed in~\cite{topology}, as we now briefly summarize.

By definition, space-time~$M$ is a topological space (see~\S\ref{seccausal}).
Attaching to every space-time point~$x \in M$ the corresponding spin space~$S_x$
gives the structure of a {\em{sheaf}}, making it possible to describe the topology by sheaf cohomology.
If one assumes in addition that all space-time points are regular (see Definition~\ref{defregular}), then
all spin spaces are isomorphic, giving rise to a {\em{topological vector bundle}}.

In order to get the connection to spinor bundles, one needs the additional structure of
Clifford multiplication. As explained in~\S\ref{secwave}, the notion of a Clifford subspace
(see Definition~\ref{defcliffsubspace}) makes it possible to define Clifford structures at
every space-time point, but the definition is not unique and does not give the
connection to tangent vectors of the base space.
In~\S\ref{secgeom} these shortcomings where bypassed by working with suitable equivalence
classes of Clifford subspaces.
From the topological point of view, the basic question is whether one can choose
a representative of this equivalence class at each space-time point in such a way that
the representative depends continuously on the base point.
This leads to the notion of a {\em{Clifford section}}~$\Cl$, being a continuous
mapping which to every space-time point~$x \in M$ associates a corresponding Clifford
subspace~$\Cl_x$ (for details see~\cite[Section~4.1]{topology}).
Choosing a Clifford section leads to the structure of a so-called {\em{topological spinor bundle}}.
An advantage of working with topological spinor bundles is
that no notion of differentiability is required.

If~$M$ has a differentiable structure, one would like to associate a tangent vector~$u \in T_xM$
to a corresponding element of the Clifford subspace~$\Cl_x$.
This leads to the notion of a {\em{spin structure}}~$\gamma$ on a topological spinor bundle,
being a continuous mapping which to every~$x \in M$ associates a mapping~$\gamma_x :
T_xM \rightarrow \Cl_x$.
The topological obstructions for the existence of a spin structure on a topological spinor bundle
generalize the spin condition on a spin manifold (for details see~\cite[Sections~4.2 and~4.5]{topology}).

A useful analytic tool for the construction of Clifford sections are so-called
{\em{tangent cone measures}} (see~\cite[Section~5]{topology}). These measures
make it possible to analyze the local structure of space-time in a neighborhood of a point~$x \in M$
(again without any differentiability assumptions).
The tangent cone measures can be used to distinguish a specific Clifford subspace~$\Cl_x$
and to relate~$\Cl_x$ to neighboring space-time points.

We close with two remarks. First, all the above constructions generalize to the
{\em{Riemannian setting}} if the definition of causal fermion systems is extended to
so-called {\em{topological fermion systems}} (see~\cite[Definition~2.1]{topology}).
We thus obtain a mathematical framework to describe {\em{spinors on singular spaces}}
(see~\cite[Sections~7 and~8]{topology} for many examples).
Second, one can introduce nontrivial topological notions even for discrete space-times
by constructing neighborhoods of~$M$ in~$\F$ (using the metric structure of~$\F$
induced by the norm on the Banach space~$\Lin(\H)$) and by studying the topology of these
neighborhoods.

\section{Correspondence to Minkowski Space} \label{secmink}
In order to put the abstract framework in a simple and concrete context, we
now explain how to describe Dirac spinors in Minkowski space as a causal fermion system.

\subsection{Concepts Behind the Construction of Causal Fermion Systems}
\label{secuvintro}
We let~$(\scrM, \la .,. \ra)$ be Minkowski space 
(with the signature convention~$(+ - - -)$) and~$d\mu$ the standard volume measure
(thus~$d\mu = d^4x$ in a reference frame~$x= (x^0, \ldots, x^3)$).
We denote the spinor space at a point~$x \in \scrM$ by~$S_x\scrM$, so that a Dirac wave function~$\psi$
takes values in
\[ \psi(x) \in S_x\scrM \simeq \C^4 \:. \]
The spinor space at~$x$ is endowed with an indefinite inner product of signature~$(2,2)$,
which as in physics textbooks we denote by~$\overline{\psi} \phi$
(where~$\overline{\psi} = \psi^\dagger \gamma^0$ is the usual adjoint spinor).
Clearly, in Minkowski space one has a trivial parallel transport of spinors, making it possible
to identify the spinor spaces at different space-time points. Thus the
space-time index~$S_x\scrM$ of the spinor space is added only for notational clarity.

On the solutions of the Dirac equation
\beq \label{Dirfree}
(i \gamma^j \partial_j - m) \psi = 0
\eeq
we consider the usual Lorentz invariant scalar product
\beq \label{sprodMin}
( \psi | \phi) := 2 \pi \int_{\R^3} (\overline{\psi} \gamma^0 \phi)(t, \vec{x})\: d^3x \:,
\eeq
making the solution space to a separable Hilbert space.
We choose~$\H$ as a closed subspace of this Hilbert space
with the induced scalar product~$\la .|. \ra_\H := (.|.)|_{\H \times \H}$.
Clearly, $\H$ is again a separable Hilbert space.
In order to describe the vacuum (i.e.\ the physical system where no particles and anti-particles are present),
one chooses~$\H$ as the subspace spanned by all the negative-energy solutions (the ``Dirac sea vacuum'').
To describe particles or anti-particles, one includes
positive-energy solutions or leaves out negative-energy solutions, respectively.
But any other closed subspace of the solution space may be chosen as well.
We remark for clarity that in this section, we only consider the vacuum
Dirac equation~\eqref{Dirfree}, so that
the Dirac particles do not interact (interacting systems will be discussed in Section~\ref{seclimit} below).

In order to get into the framework of causal fermion systems, to every space-time point~$x \in \scrM$
we want to associate a linear operator~$F(x) \in \F$. Once this has been accomplished, the resulting mapping
\beq \label{Fmap}
F \::\: \scrM \rightarrow \F \:.
\eeq
can be used to introduce a measure~$\rho$ on~$\F$. Namely, we say that a
subset~$\Omega \subset \F$
is measurable if and only if its pre-image~$F^{-1}(\Omega)$ is a measurable subset of~$\scrM$.
Moreover, we define the measure of~$\Omega$ as the space-time volume of the pre-image,
$\rho(\Omega) := \mu ( F^{-1}(\Omega) )$. This construction is commonly used in mathematical
analysis and is referred to as the {\em{push-forward measure}}, denoted by
\[ \rho= F_* \mu \:. \]
Then~$(\H, \F, \rho)$ will be a causal fermion system.

The basic idea for constructing~$F(x)$ is to represent the inner product on the spinors in terms of the Hilbert space scalar product, i.e.
\beq \label{Fdefnaive}
\la \psi | F(x) \phi \ra_\H =  -(\overline{\psi} \phi)(x)  \qquad \text{for all~$\psi, \phi \in \H$}\:.
\eeq
The operator~$F(x)$ gives information on the densities and correlations of the
Dirac wave functions at the space-time point~$x$. It is referred to as the {\em{local correlation operator}}
at~$x$. Relating the maximal number of positive and negative eigenvalues of~$F(x)$ to the
signature of the inner product~$(\overline{\psi} \phi)(x)$, one sees that~$F(x)$ indeed has at most two positive
and at most two negative eigenvalues. However, the equation~\eqref{Fdefnaive} suffers from the shortcoming
that the right side is in general ill-defined because
solutions~$\psi, \phi \in \H$ are in general not continuous and thus cannot be evaluated pointwise.
This is the reason why we need to introduce an {\em{ultraviolet regularization}} (UV regularization).
Before entering the analysis, we first outline our method and
explain the physical picture in a few remarks. The mathematical construction will be
given afterwards in~\S\ref{secuvreg}.

In order to put our constructions in the general physical context, we first
note that UV regularizations are frequently used in relativistic quantum field theory
as a technical tool to remove divergences.
A common view is that the appearance of such divergences
indicates that the physical theory is incomplete and should be replaced for very small distances
by another, more fundamental theory.
The renormalization program is a method to get along with standard quantum field theory
by finding a way of dealing with the divergences.
The first step is the UV regularization,
which is usually a set of prescriptions which make divergent integrals finite.
The next step of the renormalization program is to show that the 
UV regularization can be taken out if other parameters of the theory (like masses
and coupling constants) are suitably rescaled.
Conceptually, in the renormalization program the UV regularization merely is a technical tool.
All predictions of theory should be independent of how the regularization is carried out.

In the context of causal fermion systems, however, the physical picture behind the UV regularization
is quite different. Namely, in our setting the {\em{regularized}} objects are to be considered as the
fundamental physical objects. Therefore, the regularization has a physical significance. It
should describe the microscopic structure of physical space-time.

Before explaining this physical picture in more detail, we need to introduce a microscopic length
scale~$\varepsilon>0$ on which the UV regularization should come into play.
Regularization lengths are often associated to the Planck length
$\ell_P \approx1.6 \cdot 10^{-35} \:{\mbox{m}}$. The analysis of the gravitational field
in~\cite{cfs}
suggests that~$\varepsilon$ should be chosen even much smaller than the Planck length
(see~\cite[Section~4.9 and \S5.4.3]{cfs}).
Even without entering a detailed discussion
of the length scales, it is clear that~$\varepsilon$ will be by many orders of magnitude smaller than most other
physical length scales of the system.
Therefore, it is a sensible method to analyze the causal action principle in the asymptotics
when~$\varepsilon$ is very small. In order to make such an asymptotics mathematically precise,
we necessarily need to consider the {\em{regularization length}}~$\varepsilon$ as a
{\em{variable parameter}} taking values in an
interval~$(0, \varepsilon_{\max})$. Only for such a variable parameter, one can
analyze the asymptotics as~$\varepsilon \searrow 0$.

For any~$\varepsilon \in (0, \varepsilon_{\max})$, similar to~\eqref{Fmap}
we shall construct a mapping~$F^\varepsilon : \scrM \rightarrow \F$
by suitably inserting an UV regularization in~\eqref{Fdefnaive}.
Then we construct the corresponding universal measure
as the push-forward by~$F^\varepsilon$, i.e.
\beq \label{pushforward}
\rho^\varepsilon := F^\varepsilon_* \mu \:.
\eeq
This shall give rise to a causal fermion system~$(\H, \F, \rho^\varepsilon)$.
We will also explain how to identify the objects in
Minkowski space with corresponding objects of the causal fermion system: \\

\begin{center}
\begin{tabular}{|c|c|}
\hline & \\[-0.8em]
$\quad$ {\bf{Minkowski space}} $\quad$ & {\bf{causal fermion system}} \\[0.2em]
\hline & \\[-0.8em]
space-time point~$x \in \scrM$ & space-time point~$x \in M^\varepsilon:=\supp \rho^\varepsilon$ \\[0.2em]
topology of~$\scrM$ & topology of~$M^\varepsilon$ \\[0.2em]
spinor space $S_x\scrM$ & spin space $S_xM^\varepsilon$ \\[0.2em]
causal structure of Minkowski space & causal structure of Definition~\ref{def2} \\[0.2em]
\hline
\end{tabular}
\end{center}
\hspace*{1em}

\noindent
With these identifications made, the structures of Minkowski space are no longer needed.
They are encoded in the causal fermion system, and we may describe the physical space-time
exclusively by the causal fermion system. We consider the objects with UV regularization
as described by the causal fermion system as the fundamental physical objects.

In the following remarks we elaborate on the physical picture behind the UV regularization
and explain why our setting is sufficiently general to describe the physical situation we have in mind.
\begin{Remark} {\bf{(method of variable regularization)}} \label{remmvr} {\em{
As just explained, the only reason for considering a family of causal fermion systems
is to give the asymptotics~$\varepsilon \searrow 0$ a precise mathematical meaning.
But from the physical point of view, a specific regularization for a specific value of~$\varepsilon$
should be distinguished by the fact that the corresponding causal fermion system~$(\H, \F, \rho^\varepsilon)$
describes our physical space-time. We again point out that this concept is different from standard quantum field
theory, where the regularization merely is a technical tool used in order to remove divergences.
In our setting, the regularization has a physical significance. The {\em{regularized}} objects are to be
considered as the {\em{fundamental}} physical objects, and the regularization is a method to describe
the microscopic structure of physical space-time.

This concept immediately raises the question how the ``physical regularization'' should look like.
Generally speaking, the regularized space-time should look like Minkowski space down to distances
of the scale~$\varepsilon$. For distances smaller than~$\varepsilon$, the structure of
space-time may be completely different. The simplest method of regularizing is to ``smear out''
or ``mollify'' all wave functions on the scale~$\varepsilon$
(this corresponds to Example~\ref{exmollify} below). But it is also conceivable that
space-time has a non-trivial microstructure on the scale~$\varepsilon$, which cannot be
guessed or extrapolated from the structures of Minkowski space.
Since experiments on the length scale~$\varepsilon$ seem out of reach, it is completely
unknown what the microscopic structure of space-time is.
Nevertheless, we can hope that we can get along without knowing this micro-structure,
because the detailed form of this micro-structure might have no influence
on the effective physical equations which are valid on the energy scales accessible to experiments.
More precisely, the picture is that the general structure of the effective physical equations should be independent
of the micro-structure of space-time. Values of mass ratios or coupling constants, however, may well
depend on the micro-structure (a typical example is the gravitational constant, which is
closely tied to the Planck length, which in turn is related to~$\varepsilon$
as explained in~\cite[Section~4.9]{cfs}).
In more general terms, the unknown micro-structure
of space-time should enter the effective physical equations only by a finite (hopefully
small) number of free parameters, which can then be taken as empirical free parameters
of the effective macroscopic theory.

Clearly, the above picture must be questioned and supported by mathematical results.
To this end, one needs to analyze in detail how the effective macroscopic theory
depends on the regularization. For this reason, it is not sufficient to consider a specific family of regularizations.
Instead, one must analyze a whole class of regularizations which is so large that it covers all relevant
regularization effects. This strategy is referred to as the
{\em{method of variable regularization}} (for a longer explanation see~\cite[\S4.1]{PFP}).
It is the reason why in Definition~\ref{defreg} below we shall only state properties of the regularization,
but we do not specify how precisely it should look like.
}} \QEDrem
\end{Remark}

\begin{Remark} {\bf{(sequences of finite-dimensional regularizations)}} \label{remdiscrete}
{\em{ The critical reader may wonder why we consider a family of regularizations~$(\H, \F, \rho^\varepsilon)$
parametrized by a continuous parameter~$(0, \varepsilon_{\max})$.
Would it not be more suitable to consider instead a sequence of
causal fermion systems~$(\H_\ell, \F_\ell, \rho_\ell)$ which asymptotically as~$\ell \rightarrow \infty$
describes Minkowski space?
A related question is why we constructed the measure~$\rho$ as the push-forward of the
Lebesgue measure~\eqref{pushforward}. Would it not be better to work with more general
measures such as to allow for the possibility of discrete micro-structures?
The answer to these questions is that it is no loss of generality and a simply a matter
of convenience to work with the family~$(\H, \F, \rho^\varepsilon)$ with~$\varepsilon \in (0, \varepsilon_{\max})$,
as we now explain.

We first point out that we do not demand our family~$(\H, \F, \rho^\varepsilon)$
to be in any sense ``continuous'' in the parameter~$\varepsilon$. Therefore, one can
also describe a sequence~$(\H, \F, \rho_\ell)$ simply by choosing the family~$\rho^\varepsilon$ to be
piecewise constant, for example
\[ \rho^\varepsilon = \rho_\ell \qquad \text{if} \qquad \frac{1}{\ell} \leq \varepsilon < \frac{1}{\ell+1}\:. \]
Similarly, it is no loss of generality to take~$\rho$ as the push-forward measure of the Lebesgue measure
because~$F^\varepsilon(x)$ need not depend continuously on~$x \in M$.
For example, one can arrange a discrete space-time like a space-time lattice by
choosing~$F^\varepsilon$ as a mapping which is piecewise constant on little cubes of
Minkowski space. Clearly, this mapping is not continuous, but it is continuous almost everywhere. Moreover, 
its image is a discrete set, corresponding to a discrete micro-structure of space-time.
For the method for representing a general measure~$\rho$ as the push-forward of
for example the Lebesgue measure we refer the interested reader
to the proof of~\cite[Lemma~1.4]{continuum}.

The remaining question is why we may keep the Hilbert space~$\H$ fixed.
In particular, we noted in~\S\ref{secbasicdef} that the existence of minimizers
of the causal action principle has been proven only if~$\H$ is finite-dimensional.
Therefore, should one not consider a filtration~$\H_1 \subset \H_2 \subset \cdots \subset \H$
of~$\H$ by finite-dimensional subspaces? Indeed, from the conceptual point of view,
this would be the correct way to proceed.
Nevertheless, the following consideration explains why we can just as well replace all
the Hilbert spaces~$\H_\ell$ by the larger space~$\H$:
For a given causal fermion system~$(\H_\ell, \F_\ell, \rho_\ell)$
with~$\H_\ell \subset \H$, by extending all operators by zero to the orthogonal complement
of~$\H_\ell$, one obtains the so-called {\em{extended causal fermion system}}~$(\H, \F, \rho_\ell)$.
The fact that the causal fermion system was extended can still be seen by forming the
so-called {\em{effective Hilbert space}} as
\[ \H^\text{eff} = \overline{ \text{span} \{ x(\H) \:|\: x \in \supp \rho \} }\:. \]
Namely, for an extended causal fermion system, the effective Hilbert space still is a subset
of the original Hilbert space, $\H^\text{eff} \subset \H_\ell$.
Moreover, the support of the extended causal fermion system is still contained
in~$\F_\ell \subset \Lin(\H_\ell)$. Therefore, we do not lose any
information by extending a causal fermion system. Conversely, when analyzing a causal fermion system,
it seems preferable to always make the Hilbert space as small as possible by taking~$\H^\text{eff}$
as the underlying Hilbert space.

The delicate point about extending causal fermion systems is that the causal action principle does
depend sensitively on the dimension of the underlying Hilbert space~$\H$.
More specifically, the infimum of the action is known to be strictly decreasing in the dimension of~$\H$
(see the estimates in~\cite[Lemma~5.1]{discrete}, which apply similarly in the 
more general setting of~\cite{continuum}).
Therefore, a minimizer~$\rho$ of the causal action principle will no longer be a minimizer
if the causal fermion system is extended.
However, the first order {\em{Euler-Lagrange equations}} (for details see~\S\ref{secvary} below)
are still satisfied for the extended causal fermion system.
Therefore, for convenience we fix the Hilbert space~$\H$ and consider
a family of causal fermion systems~$(\H, \F, \rho^\varepsilon)$ thereon.
In order for the causal action principle to be well-defined and for~$\rho^\varepsilon$ to be
a minimizer, one should replace~$\H$ by the corresponding effective Hilbert space~$\H^\text{eff}$,
which may depend on~$\varepsilon$ and should be arranged to be finite-dimensional.
For the analysis of the Euler-Lagrange equations, however, the restriction to~$\H^\text{eff}$ is unnecessary,
and it is preferable to work with the extended Hilbert space~$\H$.
}} \QEDrem
\end{Remark}

We finally remark that the hurried reader who wants to skip the following constructions
may read instead the introductory section~\cite[Section~1.1]{rrev}
where formal considerations without UV regularization are given.
Moreover, a more explicit analysis of four-dimensional Minkowski space with a particularly
convenient regularization is presented in~\cite[Section~4]{lqg}. For a somewhat simpler analysis of
two-dimensional Minkowski space we refer to~\cite[Section~7.2]{topology}.

\subsection{Introducing an Ultraviolet Regularization} \label{secuvreg}
We now enter the construction of the UV regularization.
We denote the continuous Dirac wave functions (i.e.\ the continuous sections of the spinor bundle,
not necessarily solutions of the Dirac equation)
by~$C^0(\scrM, S\scrM)$. Similarly, the smooth wave functions with compact support in
a subset~$K \subset \scrM$ are denoted by~$C^\infty_0(K, S\scrM)$. For the $C^k$-norms
we use the notation
\[ |\eta|_{C^k(K)} = \sum_{|\alpha| \leq k}\: \sup_{x \in K} |\partial^\alpha \eta(x)| \qquad
\text{for~$\eta \in C^\infty_0(K, S\scrM)$}\:, \]
where the~$\alpha$ are multi-indices. Here~$|.|$ is any pointwise norm on the spinor spaces (we again
identify all spinor spaces with the trivial parallel transport). Since any two such norms
can be estimated from above and below by a constant,
the $C^k$-norms corresponding to different choices of the norms~$|.|$ are also equivalent.
For example, one can choose~$|\psi|^2 := \overline{\psi} \gamma^0 \psi$ similar to the
integrand in the scalar product~\eqref{sprodMin}. But clearly, other choices are possible just as well.

The UV regularization is performed most conveniently with so-called regularization operators, which we now 
define.
\begin{Def} \label{defreg} Consider a family of linear operators~$({\mathfrak{R}}_\varepsilon)$
with~$0 < \varepsilon < \varepsilon_{\max}$
which map~$\H$ to the continuous wave functions,
\[ {\mathfrak{R}}_\varepsilon \::\: \H \rightarrow C^0(\scrM, S\scrM) \:. \]
The family is called a family of {\bf{regularization operators}} if the following conditions hold:
\begin{itemize}
\item[(i)] The image of every regularization operator is pointwise bounded,
meaning that for every~$\varepsilon \in (0, \varepsilon_{\max})$ and all~$x \in \scrM$
there is a constant~$c>0$ such that for all~$u \in \H$,
\beq \label{reges0}
\big| \big({\mathfrak{R}}_\varepsilon u \big)(x) \big| \leq c \:\|u\|_\H\ \:.
\eeq
\item[(ii)] The image of every regularization operator is equicontinuous almost everywhere
in the sense that
for every~$\varepsilon \in (0, \varepsilon_{\max})$, almost all~$x \in \scrM$ and every~$\delta>0$,
there is an open neighborhood~$U \subset \scrM$ of~$x$ such that for all~$u \in \H$ and all~$y \in U$,
\beq \label{reges1}
\big| \big({\mathfrak{R}}_\varepsilon u \big)(x) - \big({\mathfrak{R}}_\varepsilon u \big)(y) \big| \leq \delta \:\|u\|_\H\ \:.
\eeq
\item[(iii)] In the limit~$\varepsilon \searrow 0$, the family converges weakly to the identity,
meaning that for every compact subset~$K \subset \scrM$ and every~$\delta>0$
there is a constant~$\varepsilon_0>0$, such that
for all~$\varepsilon \in (0, \varepsilon_0)$, $u \in \H$ and~$\eta \in C^\infty_0(K, S\scrM)$,
\beq \label{reges2}
\Big| \int_\scrM \overline{\eta(x)} \big( {\mathfrak{R}}_\varepsilon(u) - u \big)(x)\: d^4x \Big| \leq \delta \:
\|u\|_\H\, |\eta|_{C^1(K)} \:.
\eeq
\end{itemize}
\end{Def} \noindent
We point out that we do not demand that the regularized wave function~${\mathfrak{R}}_\varepsilon \psi$
is again a solution of the Dirac equation. This could be imposed (as is done in~\cite[Section~4]{finite}),
but doing so seems too restrictive for the physical applications.
We also note that ``almost all'' in~(ii) refers to the standard volume measure~$d\mu$ on~$\scrM$.

For the mathematically interested reader we remark that the above properties~(i) and~(ii)
are very similar to the assumptions in the Arzel{\`a}-Ascoli theorem
(see for example~\cite[Section~VII.5]{dieudonne1} or~\cite[Theorem~7.25]{rudinprinciples}).
In fact, if we replaced ``almost all'' in~(ii) by ``all'', one could apply the Arzel{\`a}-Ascoli theorem
and restate the properties~(i) and~(ii) equivalently by
saying that taking the image~${\mathfrak{R}}_\varepsilon(B_1(0))$ of the unit ball in~$\H$
and restricting the resulting family of functions to any compact set~$K \subset \scrM$,
one obtains a relatively compact subset of~$C^0(K, S\scrM)$. It is remarkable that
the properties~(i) and~(ii) come up naturally as conditions for a sensible UV regularization,
although we shall never use compactness arguments in our proofs.
Weakening ``all'' by ``almost all'' in~(ii) makes it possible to describe discrete
space-times like space-time lattices, as was mentioned in Remark~\ref{remdiscrete} above.

Simple examples of regularization operators are obtained by mollifying the wave functions
on the scale~$\varepsilon$:
\begin{Example} {\bf{(regularization by mollification)}} \label{exmollify} {\em{Let~$h \in C^\infty_0(\scrM, \R)$ be a non-negative test function with
\[ \int_\scrM h(x)\: d^4x=1 \:. \]
We define the operators~${\mathfrak{R}}_\varepsilon$ for~$\varepsilon>0$ as the
convolution operators
\[ ({\mathfrak{R}}_\varepsilon u)(x) := \frac{1}{\varepsilon^4}
\int_\scrM h\Big(\frac{x-y}{\varepsilon}\Big)\: u(y)\: d^4y \:. \]

Let us prove that the family~$({\mathfrak{R}}_\varepsilon)_{0<\varepsilon<1}$ is a family of regularization operators. First, 
\[ \big| \big( {\mathfrak{R}}_\varepsilon u \big)(x) \big|
\leq \frac{|h|_{C^0}}{\varepsilon^4}\: \int_K |u(y)|\: d^4y \leq \frac{|h|_{C^0}}{\varepsilon^4}\:\sqrt{\mu(K)}\:
\Big( \int_K |u(y)|^2\: d^4y \Big)^\frac{1}{2}\:, \]
where in the last step we used the Schwarz inequality.
We now rewrite the obtained space-time integral of~$|u|^2$ with the help of Fubini's theorem
as a bounded time integral and a spatial integral. In view of~\eqref{sprodMin},
the spatial integral can be estimated by the Hilbert space norm. We thus obtain
\beq \label{intes}
\int_K |u(y)|^2 \: d^4y \leq C \int_K \big(\overline{u} \gamma^0 u \big)(y)\: d^4y \leq
C \int_{t_0}^{t_1} \|u\|_\H^2 = C \,(t_1-t_0)\: \|u\|_\H^2 \:,
\eeq
where~$t_0$ and~$t_1$ are chosen such that~$K$ is contained in the time strip~$t_0 < t < t_1$. We conclude that
\[ \big| \big( {\mathfrak{R}}_\varepsilon u \big) \big|
\leq \frac{|h|_{C^0}}{\varepsilon^4}\:\sqrt{\mu(K)\:C\, (t_1-t_0)}\: \|u\|_\H^2 \:, \]
proving~\eqref{reges0}.

In order to derive the inequality~\eqref{reges1}, we begin with the estimate
\[ \big| \big( {\mathfrak{R}}_\varepsilon u \big)(x) - \big( {\mathfrak{R}}_\varepsilon u \big)(y) \big|
\leq \frac{1}{\varepsilon^4}\: \sup_{z \in \scrM} \Big|
h\Big(\frac{x-z}{\varepsilon}\Big) - h\Big(\frac{y-z}{\varepsilon}\Big) \Big|
\int_K |u(y)|\: d^4y \:. \]
Again applying~\eqref{intes} and using that~$h$ is uniformly continuous, one obtains~\eqref{reges1}.

It remains to prove~\eqref{reges2}. We first write the integral on the left as
\beq \label{vorhol}
\int_\scrM \overline{\eta(x)} \big( {\mathfrak{R}}_\varepsilon(u) - u \big)(x)\: d^4x
= \int_\scrM \overline{ \big( \eta_\varepsilon(y) - \eta(y) \big) }\: u(y)\: d^4y \:,
\eeq
where we set
\[ \eta_\varepsilon(y) = \frac{1}{\varepsilon^4} \int_\scrM \eta(x) \:h\Big(\frac{x-y}{\varepsilon}\Big)\: d^4x\:. \]
Now we use the standard estimate for convolutions
\begin{align*}
| & \eta_\varepsilon(y) - \eta(y)| = \frac{1}{\varepsilon^4}
\bigg| \int_\scrM \big( \eta(x)-\eta(y) \big)\:h\Big(\frac{x-y}{\varepsilon}\Big)\: d^4x \bigg| \\
&= \bigg| \int_\scrM \Big( \eta(y+\varepsilon z)-\eta(y) \Big)\:h(z)\: d^4z \bigg|
\leq |\eta|_{C^1(K)} \int_\scrM |\varepsilon z|\: \:h(z)\: d^4z
\end{align*}
(where in the last step we used the mean value theorem).
This gives rise to the estimate
\[ |\eta_\varepsilon - \eta|_{C^0(K)} \leq c\, \varepsilon\, |\eta|_{C^1(K)} \:, \]
where~$c$ may depend on~$K$ and the choice of~$h$, but is independent of~$\eta$.
This makes it possible to estimate~\eqref{vorhol} by
\[ \Big| \int_\scrM \overline{\eta(x)} \big( {\mathfrak{R}}_\varepsilon(u) - u \big)(x)\: d^4x \Big| \\
\leq \varepsilon\, |\eta|_{C^1(K)} \int_K |u(y)|_y \: d^4y \:. \]
Again applying~\eqref{intes}, we conclude that
\[ \Big| \int_\scrM \overline{\eta(x)} \big( {\mathfrak{R}}_\varepsilon(u) - u \big)(x)\: d^4x \Big|
\leq \delta\, |\eta|_{C^1(K)}\: \sqrt{\mu(K)} \: \sqrt{C\, (t_1-t_0)}\; \|u\|_\H \:, \]
proving~\eqref{reges2}.
}} \QEDrem
\end{Example}

Given a family of regularization operators, we can construct
causal fermion systems as follows. We fix~$\varepsilon \in (0, \varepsilon_{\max})$.
For any~$x \in \scrM$, we consider the bilinear form
\beq \label{bxdef}
b_x \::\: \H \times \H \rightarrow \C\:,\quad
b_x(u, v) = - \overline{({\mathfrak{R}}_\varepsilon \,u)(x)} ({\mathfrak{R}}_\varepsilon \,v)(x) \:.
\eeq
This bilinear form is well-defined and bounded because~${\mathfrak{R}}_\varepsilon$ 
is defined pointwise and because evaluation at~$x$ gives a linear operator of finite rank.
Thus for any~$v \in \H$, the anti-linear form~$b_x(.,v) : \H \rightarrow \C$
is continuous. By the Fr{\'e}chet-Riesz theorem (see for example~\cite[Section~6.3]{lax}),
there is a unique vector~$w \in \H$
such that~$b_x(u,v) = \la u | w \ra_\H$ for all~$u \in \H$.
The mapping~$v \mapsto w$ is linear and bounded. We thus obtain a bounded linear
operator~$F^\varepsilon(x)$ on~$\H$ such that
\[ b_x(u, v) = \la u \,|\, F^\varepsilon(x)\, v \ra_\H \qquad
\text{for all~$u,v \in \H$}\:. \]
Taking into account that the inner product on the Dirac spinors at~$x$ has signature~$(2,2)$,
the local correlation operator~$F^\varepsilon(x)$ is a symmetric operator on~$\H$
of rank at most four, which has at most two positive and at most two negative eigenvalues.
Finally, we introduce the {\em{universal measure}}~$\rho^\varepsilon=
F^\varepsilon_* \mu$ as the push-forward
of the volume measure on~$\scrM$ under the mapping~$F^\varepsilon$.
In this way, for every~$\varepsilon \in (0, \varepsilon_0)$ we obtain 
a causal fermion system~$(\H, \F, \rho^\varepsilon)$ of spin dimension~$n=2$.

\subsection{Correspondence of Space-Time} \label{seccorst}
We now explain the connection between points of Minkowski space and points
of space-time~$M^\varepsilon := \supp \rho^\varepsilon$ of the corresponding
causal fermion system~$(\H, \F, \rho^\varepsilon)$. We begin with a general
characterization of~$M^\varepsilon$.

\begin{Prp} \label{prpscrMM}
For any~$\varepsilon \in (0, \varepsilon_{\max})$, there is a subset~$E \subset \scrM$
of $\mu$-measure zero such that the mapping~$F^\varepsilon|_{\scrM \setminus E} \::\: \scrM \setminus E \rightarrow
 \F$ is continuous. Moreover, the support of the universal
measure~$M^\varepsilon:= \supp \rho^\varepsilon$ is given by
\beq \label{suppprop}
M^\varepsilon = \overline{F^\varepsilon(\scrM \setminus E)}^{\Lin(\H)} \:.
\eeq
\end{Prp}
\Proof To show continuity, we need to to estimate the sup-norm~$\|F^\varepsilon(x)-F^\varepsilon(y)\|$.
We first write the expectation value of the corresponding operator by
\begin{align*}
\la u &\,|\, \big(F^\varepsilon(x)-F^\varepsilon(y) \big) \,v \ra_\H
= - \overline{({\mathfrak{R}}_\varepsilon \,u)(x)} ({\mathfrak{R}}_\varepsilon \,v)(x)
+ \overline{({\mathfrak{R}}_\varepsilon \,u)(y)} ({\mathfrak{R}}_\varepsilon \,v)(y) \\
&= - \overline{({\mathfrak{R}}_\varepsilon \,u)(x)} \big( ({\mathfrak{R}}_\varepsilon \,v)(x) - ({\mathfrak{R}}_\varepsilon \,v)(y) \big) 
- \overline{ \big( ({\mathfrak{R}}_\varepsilon \,u)(x) - ({\mathfrak{R}}_\varepsilon \,u)(y) \big)} 
({\mathfrak{R}}_\varepsilon \,v)(y) \:,
\end{align*}
giving rise to the estimate
\begin{align*}
\big| \la u &\,|\, \big(F^\varepsilon(x)-F^\varepsilon(y) \big) v \ra_\H \big| \\
&\leq | ({\mathfrak{R}}_\varepsilon \,u)(x)|\: \big| ({\mathfrak{R}}_\varepsilon \,v)(x)
- {\mathfrak{R}}_\varepsilon \,v)(y) \big|
+ \big| ({\mathfrak{R}}_\varepsilon \,u)(x) - ({\mathfrak{R}}_\varepsilon \,u)(y) \big|\:
|({\mathfrak{R}}_\varepsilon \,v)(y)|\:.
\end{align*}

We now estimate the resulting spinor norms with the help of properties~(i) and~(ii)
of Definition~\ref{defreg}. First, we denote the exceptional set of $\mu$-measure zero where~\eqref{reges1}
does not hold by~$E \subset \scrM$. Combining~\eqref{reges0} and~\eqref{reges1},
one immediately sees that every point~$x \in \scrM \setminus E$ has a neighborhood~$U$
such that the boundedness property~\eqref{reges0} holds uniformly
on~$U$ (i.e.\ $|({\mathfrak{R}}_\varepsilon u)(y)| \leq c \,\|u\|_\H$ for all~$y \in U$).
We thus obtain the estimate
\begin{align*}
\big| \la u &\,|\, \big(F^\varepsilon(x)-F^\varepsilon(y) \big) v \ra_\H \big| \leq
2 c\, \delta \: \|u\|_\H\: \|v\|_\H \:,
\end{align*}
valid for all~$y \in U$ and~$u,v \in \H$.
Hence the sup-norm is bounded by~$\|F^\varepsilon(x)-F^\varepsilon(y)\| \leq 2 c \delta$,
showing that~$F^\varepsilon$ is continuous on~$\scrM \setminus E$.

It remains to prove~\eqref{suppprop}.
Since~$\mu(E)=0$, the set~$E$ can be disregarded when forming the push-forward measure.
Therefore, taking into account that the support of a measure is by definition a closed set,
it suffices to show that for every~$x \in \scrM \setminus E$,
the operator~$p:=F^\varepsilon(x)$ lies in the support of~$\rho^\varepsilon$.
Let~$U \subset \F$ be an open neighborhood of~$p$.
Then the continuity of~$F^\varepsilon$ at~$x$ implies that the
preimage~$(F^\varepsilon)^{-1}(U)$ is an open
subset of~$\scrM$. Hence the Lebesgue measure of this
subset is non-zero, $\mu((F^\varepsilon)^{-1}(U))>0$.
By definition of the push-forward measure, it follows that~$\rho^\varepsilon(U)>0$. 
Hence every neighborhood of~$p$ has a non-zero measure, implying that~$p \in \supp \rho^\varepsilon$.
This concludes the proof.
\QED

In order to have a convenient notation, in what follows we always identify a point
in Minkowski space with the corresponding operator of the causal fermion system,
\beq \label{Midentify}
\text{identify} \quad x \in \scrM \qquad \text{with} \qquad F^\varepsilon(x) \in \F \:.
\eeq
In general, this identification is not one-to-one, because the mapping~$F^\varepsilon$
need not be injective. In the latter case, there are two points~$x,y \in \scrM$
such that the bilinear forms~$b_x$ and~$b_y$ coincide (see~\eqref{bxdef}).
In other words, all correlations between regularized wave functions coincide
at the points~$x$ and~$y$.
Using a more physical language, this means that the points~$x, y$ of Minkowski space
are not distinguishable by any experiments performed on the fermionic wave functions.
We take the point of view that in such situations, the points~$x$ and~$y$ should not
be distinguished physically, and that it is reasonable and desirable that the two points
are identified in the causal fermion system with the same space-time point~$F^\varepsilon(x)
= F^\varepsilon(y) \in M^\varepsilon := \supp \rho^\varepsilon$.
In philosophical terms, our construction realizes the principle of the 
identity of indiscernibles.

We also remark that, due to the closure in~\eqref{suppprop}, it may happen that
the space-time~$M^\varepsilon$ contains a point~$z$ which does {\em{not}} lie in the image of~$F^\varepsilon$,
but is merely a limit point in~$F^\varepsilon(\scrM)$. In this case, the corresponding bilinear
form~$b(u,v) := \la u | z v \ra_\H$ can be approximated with
an arbitrarily small error by bilinear forms~$b_x$ with~$x \in \scrM$.
Since experiments always involve small imprecisions, we take the point of view that
it is again reasonable and desirable mathematically to include~$z$ into the space-time points.

Generally speaking, the just-discussed cases that~$F^\varepsilon$ is not injective or its
image is not closed seem mostly of academic interest. 
In most applications,
the mapping~$F^\varepsilon$ will be injective and closed. In all these situations,
Proposition~\ref{prpscrMM} will give us a one-to-one correspondence between
points~$x \in \scrM$ and points~$F^\varepsilon(x) \in M^\varepsilon$.

We finally note that, working with the push-forward measure~\eqref{pushforward},
the volume measure on space-time~$M^\varepsilon$ as defined by the universal measure~$d\rho^\varepsilon$
always agrees under the identification~\eqref{Midentify} with the Lebesgue measure~$d\mu$ on~$\scrM$.

\subsection{Correspondence of Spinors and Wave Functions} \label{seccorsw}
We proceed by explaining the connection between the spinor space~$S_x\scrM$
at a point~$x \in \scrM$ of Minkowski space
and the corresponding spin space~$S_xM \subset \H$ of the causal fermion system
(where we use the identification~\eqref{Midentify}).
This will also make it possible to get a connection between Dirac wave functions
in Minkowski space and wave functions as defined in~\S\ref{secwave}.
In preparation, we derive useful explicit formulas for the local correlation
operators. To this end, for any~$x \in \scrM$ we define the {\em{evaluation map}}~$e_x^\varepsilon$ by
\beq \label{evalmap}
e^\varepsilon_x \::\: \H \rightarrow S_x\scrM \:,\qquad
e^\varepsilon_x \,\psi = ({\mathfrak{R}}_\varepsilon \psi)(x)\:.
\eeq
Its adjoint is defined as usual, taking into account the corresponding inner products on the
domain and the target space, i.e.
\[ \la (e^\varepsilon_x)^* \chi \,|\, \psi \ra_\H = \overline{\chi} \,\big( e^\varepsilon_x \,\psi) \:. \]
We denote this adjoint by~$\iota^\varepsilon_x$,
\[ \iota^\varepsilon_x := (e^\varepsilon_x)^* \::\: S_x \scrM \rightarrow \H\:. \]
Multiplying~$e^\varepsilon_x$ by~$\iota^\varepsilon_x$ gives us
back the local correlation operator~$F^\varepsilon(x)$. Namely,
\begin{align*}
\la \psi \,|\, F^\varepsilon(x)\, \phi \ra_\H = 
- \overline{({\mathfrak{R}}_\varepsilon \,\psi)(x)} ({\mathfrak{R}}_\varepsilon \,\phi)(x)
= -\overline{\big( e^\varepsilon_x \psi \big)} \big(e^\varepsilon_x \phi \big)
= - \la \psi \,|\, \iota^\varepsilon_x e^\varepsilon_x \,\phi \ra_\H
\end{align*}
and thus
\beq \label{Fepsdef}
F^\varepsilon(x) = -\iota^\varepsilon_x \,e^\varepsilon_x
= -\iota^\varepsilon_x \,\big(\iota^\varepsilon_x)^* \::\: \H \rightarrow \H \:.
\eeq

The next proposition gives the desired connection between the spinor space~$S_x\scrM$
and the corresponding spin space~$S_xM$. We first state and prove the proposition
and explain it afterwards.
\begin{Prp} \label{prpisometry}
The mapping
\[ e^\varepsilon_x|_{S_x} \::\: S_xM \rightarrow S_x \scrM \quad
\text{is an isometric embedding}\:. \]
Moreover, under this embedding, the physical wave function of a vector~$u$
at~$x$ is mapped to the regularized Dirac wave function at~$x$,
\beq \label{waveagree}
e^\varepsilon_x|_{S_x}\, \psi^u(x) = \big({\mathfrak{R}}_\varepsilon u \big)(x) \:.
\eeq
If the point~$x$ is regular (see Definition~\ref{defregular}), the inverse is given by
\beq \label{invform}
\big(e^\varepsilon_x|_{S_x}\big)^{-1} =  -\big( x|_{S_x} \big)^{-1}
\iota^\varepsilon_x\::\: S_x\scrM \rightarrow S_x M \:.
\eeq
\end{Prp}
\Proof Let~$\psi, \phi \in S_xM$. Then
\begin{align*}
\overline{\big( e^\varepsilon_x \psi \big)} \big(e^\varepsilon_x \phi \big)
&= \la \psi \:|\: (e^\varepsilon_x)^* \,e^\varepsilon_x \,\phi \ra_\H
= \la \psi \:|\: \iota^\varepsilon_x \,e^\varepsilon_x \,\phi \ra_\H
\overset{\eqref{Fepsdef}}{=} -\la \psi \:|\: x \,\phi \ra_\H = \Sl \psi | \phi \Sr \:.
\end{align*}
Moreover, since the image of~$\iota^\varepsilon_x$ coincides with~$S_xM$,
we know that~$e^\varepsilon_x$ vanishes on the orthogonal complement~$S_x^\perp \subset \H$.
Therefore,
\[ e^\varepsilon_x|_{S_x}\, \psi^u(x) = e^\varepsilon_x|_{S_x}\, \pi_x\, u = e^\varepsilon_x\, u = 
\big({\mathfrak{R}}_\varepsilon u \big)(x) \:. \]
Finally, if~$x$ is regular,
\[ -\big( x|_{S_x} \big)^{-1} \iota^\varepsilon_x \:e^\varepsilon_x|_{S_xM}
\overset{\eqref{Fepsdef}}{=} \big( x|_{S_x} \big)^{-1} \:x|_{S_x} = \1_{S_x} \:, \]
proving that the inverse of~$e^\varepsilon_x|_{S_x}$ is indeed given by the expression in~\eqref{invform}.
\QED

This proposition makes it possible to identify the spin space~$S_xM \subset \H$ 
endowed with the inner product~$\Sl .|. \Sr_x$
with a subspace of the spinor space~$S_x\scrM$ with the inner product~$\overline{\psi} \phi$.
If the point~$x$ is singular, this is all we can expect, because in this case the spaces~$S_xM$
and~$S_x\scrM$ have different dimensions and are clearly not isomorphic.
As already mentioned after Definition~\ref{defregular}, in most situations of physical interest
the point~$x$ will be regular. In this case, we even obtain an isomorphism of~$S_xM$ and~$S_x\scrM$
which preserves the inner products on these spaces.
The identity~\eqref{waveagree} shows that, under the above 
identifications, the physical wave function~$\psi^u$
(as defined by~\eqref{psiudef}) goes over to the regularized Dirac wave
function~$({\mathfrak{R}}_\varepsilon u)(x)$. This shows again that the causal fermion system
involves the {\em{regularized}} objects. Moreover, one sees that the
abstract formalism introduced in Section~\ref{secframe} indeed gives agreement with
the usual objects in Minkowski space. We remark that the above isomorphism
of~$S_xM$ and~$S_x\scrM$ also makes it possible to use unambiguously the same notation for
the corresponding inner product. Indeed, it is convenient denote the inner product on the Dirac spinors
at a time point~$x \in \scrM$ by
\[ \Sl .|. \Sr_x \::\: S_x\scrM \times S_x\scrM \rightarrow \C \:,\qquad
\Sl \psi | \phi \Sr_x = \overline{\psi} \phi \:. \]
In order to avoid confusion, we avoided this notation so far. But from now on we will sometimes use it.

In the next proposition we compute the kernel of the fermionic projector~$P^\varepsilon(x,y)$
(as defined by~\eqref{Pxydef}, where the subscript~$\varepsilon$ clarifies the dependence on the
UV regularization) in Minkowski space. Moreover, we prove that
the limit~$\varepsilon \searrow 0$ exists in the distributional sense.
\begin{Prp} \label{lemma54} Assume that the points~$x$ and~$y$ are regular. Then, under
the above identification of~$S_xM$ with~$S_x\scrM$, 
the kernel of the fermionic projector has the representation
\[ P^\varepsilon(x,y) = -e^\varepsilon_x \,\iota^\varepsilon_y \::\: S_y\scrM \rightarrow S_x\scrM \:. \]
Moreover, choosing an orthonormal basis~$(u_\ell)$ of~$\H$,
the kernel of the fermionic projector can be written as
\beq \label{Pepsbase}
P^\varepsilon(x,y) = -\sum_\ell \big({\mathfrak{R}}_\varepsilon u_\ell \big)(x)\:
\overline{\big({\mathfrak{R}}_\varepsilon u_\ell \big)(y)} \:.
\eeq

In the limit~$\varepsilon \searrow 0$, the kernel of the fermionic projector~$P^\varepsilon(x,y)$
converges as a bi-distribution to the unregularized kernel defined by
\beq \label{Pxykernel}
P(x,y) := -\sum_\ell u_\ell(x)\: \overline{u_\ell(y)} \:.
\eeq
More precisely, for every compact subset~$K \subset \scrM$ and every~$\delta>0$,
there is a constant~$\varepsilon_0>0$ such that for all~$\varepsilon \in (0, \varepsilon_0)$
and for all test wave functions~$\eta, \tilde{\eta} \in C^\infty_0(K, S\scrM)$,
\beq
\bigg| \iint_{\scrM \times \scrM} \overline{\eta(x)}\, \big( P^\varepsilon(x,y)
- P(x,y) \big)\, \tilde{\eta}(y)\: d^4x\: d^4y \,\bigg| \leq \delta\: |\eta|_{C^1(K)}\, |\tilde{\eta}|_{C^1(K)} \:.
\label{Pest}
\eeq
\end{Prp} \noindent
We remark that, since~$\H$ is separable, we can always choose an at most countable orthonormal
basis~$(u_\ell)$ of~$\H$.

\Proof[Proof of Proposition~\ref{lemma54}] We first note that
\[ P^\varepsilon(x,y) = e^\varepsilon_x \, \pi_x\, y\, \big(e^\varepsilon_y|_{S_y} \big)^{-1}
= -e^\varepsilon_x \, \pi_x\, y\, 
\big( y|_{S_y} \big)^{-1}\, \iota^\varepsilon_y = - e^\varepsilon_x \:\pi_x\: \iota^\varepsilon_y = 
- e^\varepsilon_x \:\iota^\varepsilon_y \:. \]
In an orthonormal basis~$(u)_\ell$, the completeness relation yields for any spinor~$\chi \in S_y\scrM$
\begin{align*}
P^\varepsilon(x,y)\,\chi &= -e^\varepsilon_x \,\iota^\varepsilon_y\,\chi
= - \sum_\ell \big(e^\varepsilon_x \,u_\ell \big) \la u_\ell \,|\, \iota^\varepsilon_y\, \chi \ra_\H
= - \sum_\ell \big(e^\varepsilon_x \,u_\ell \big) \:\big( \overline{e^\varepsilon_x \,u_\ell} \:\chi \big)\:,
\end{align*}
and using~\eqref{evalmap} gives~\eqref{Pepsbase}.

In order to prove~\eqref{Pest}, we introduce the functionals
\begin{align*}
\Phi^\varepsilon_\eta \:&:\: \H \rightarrow \C \:, \hspace*{-1cm} &
\Phi^\varepsilon_\eta u &= \int_\scrM \overline{\eta(x)} \big( {\mathfrak{R}}_\varepsilon u)(x)\: d^4x \\
\intertext{and similarly without UV regularization,}
\Phi_\eta \:&:\: \H \rightarrow \C \:, \hspace*{-1cm} &
\Phi_\eta u &= \int_\scrM \overline{\eta(x)} \,u(x)\: d^4x \:.
\end{align*}
Then the left side of~\eqref{Pest} can be written in the compact form
\[ \big| \Phi^\varepsilon_\eta \:\big( \Phi^\varepsilon_{\tilde{\eta}} \big)^*
- \Phi_\eta \:\big( \Phi_{\tilde{\eta}} \big)^* \big| \:, \]
which can be estimated with the triangle inequality by
\beq \label{zwischen}
\big| \Phi^\varepsilon_\eta \:\big( \Phi^\varepsilon_{\tilde{\eta}} \big)^*
- \Phi_\eta \:\big( \Phi_{\tilde{\eta}} \big)^* \big| \leq 
\|\Phi^\varepsilon_\eta \| \:\big\| \Phi^\varepsilon_{\tilde{\eta}} - \Phi_{\tilde{\eta}} \big\|
+ \big\| \Phi^\varepsilon_\eta - \Phi_\eta \big\| \: \|\Phi_{\tilde{\eta}}\| \:.
\eeq

It remains to estimate the operator norms in~\eqref{zwischen}. To this end, we
use property~(iii) of Definition~\ref{defreg} in the following way:
First, the norm of~$\Phi_\eta$ can be estimated by
\[ \big| \Phi_\eta u \big| = \int_\scrM \overline{\eta(x)} \,u(x)\: d^4x
\leq |\eta|_{C^0(K)} \sqrt{\mu(K)} \:\Big( \int_K |u(x)|\: d^4x \Big)^\frac{1}{2} \:, \]
and again by applying~\eqref{intes}. This gives
\[ \|\Phi_\eta\| \leq c\: |\eta|_{C^0(K)} \:. \]
Next, we use the triangle inequality together with~\eqref{reges2} to obtain the inequality
\[ \big\| \Phi^\varepsilon_\eta \big\| \leq \big\| \Phi^\varepsilon_\eta - \Phi_\eta \big\|
\leq \delta\,|\eta|_{C^1(K)} + c\, |\eta|_{C^0(K)} \leq  2c\, |\eta|_{C^1(K)} \:, \]
valid uniformly for all~$\varepsilon \in (0, \varepsilon_0)$
(note that property~(i) cannot be used to obtain such a uniform estimate
because we have no control on how the constant~$c$ in~\eqref{reges0} depends on~$\varepsilon$).
Finally, again applying~\eqref{reges2}, we also know that
\[ \big\| \Phi^\varepsilon_\eta - \Phi_\eta \big\| \leq \delta\,|\eta|_{C^1(K)} \:. \]
Using these inequalities in~\eqref{zwischen} gives the result.
\QED

\subsection{Correspondence of the Causal Structure} \label{seccorcs}
We now explain how the causal structure of Minkowski space is related to
corresponding notions of a causal fermion system (see Definition~\ref{def2}
and the time direction~\eqref{tdir}). To this end, we need to specify~$\H$ as a closed
subspace of the solution space of the vacuum Dirac equation~\eqref{Dirfree}.
Clearly, this Dirac equation can be solved by the plane-wave ansatz
\[ \psi(x) = e^{-i k x}\: \chi_k \]
with a constant spinor~$\chi_k$. Evaluating the resulting algebraic equation for~$\chi$
shows that the momentum~$k$ must lie on the mass shell~$k^2=m^2$.
The solutions on the upper and lower mass shell are the solutions of positive respectively
negative energy. In order to avoid potential confusion with other notions of energy
(like energy densities or energy expectation values), we here prefer the notion of
solutions of positive and negative {\em{frequency}}.
 Taking Dirac's original concept literally, we here describe
the vacuum in Minkowski space by the completely filled Dirac sea.
Thus we choose~$\H$ as the subspace of the solution space spanned
by all plane-wave solutions of negative frequency.
We refer to this choice as a {\em{Dirac sea configuration}}.

\begin{Lemma} \label{lemmaDiracsea}
If~$\H$ is the subspace of the solution space of the Dirac equation~\eqref{Dirfree}
spanned by all negative-frequency solutions, then the unregularized kernel of the fermio\-nic projector
as defined by~\eqref{Pxykernel} is the tempered bi-distribution
\beq \label{Pxyvac}
P(x,y) = \int \frac{d^4k}{(2 \pi)^4}\:(\slashed{k}+m)\: \delta(k^2-m^2)\: \Theta(-k_0)\: e^{-ik(x-y)} \:,
\eeq
where~$\Theta$ is the Heaviside function, and $k (x-y)$ is a short notation for the Minkowski
inner product~$k_j\,(x-y)^j$.
\end{Lemma}
\Proof The integrand in~\eqref{Pxyvac} clearly is a tempered distribution.
Hence its Fourier transform~$P(x,y)$ is also a tempered distribution (in the vector~$y-x$
and also in both vectors~$x$ and~$y$). In addition, one verifies by direct computation
that~$P(x,y)$ is a distributional solution of the Dirac equation,
\begin{align*}
(i \Pdd_x - m)\, P(x,y)
&= \int \frac{d^4k}{(2 \pi)^4}\:(\slashed{k}-m) (\slashed{k}+m)\: \delta(k^2-m^2)\: \Theta(-k_0)\: e^{-ik(x-y)} \\
&= \int \frac{d^4k}{(2 \pi)^4}\:\big(k^2 - m^2 \big)\: \delta(k^2-m^2)\: \Theta(-k_0)\: e^{-ik(x-y)} = 0 \:.
\end{align*}
Due to the factor~$\Theta(-k_0)$, the distribution~$P(x,y)$
is composed of solutions of negative frequency. Moreover, since the matrix~$(\slashed{k}+m)$
has rank two, one sees that~$P(x,y)$ is indeed composed of {\em{all}} negative-frequency solutions.
It remains to show that the normalization of~$P(x,y)$ is compatible with~\eqref{Pxykernel}, meaning that
\[ -2 \pi \int_{\R^3} P\big( x,(t,\vec{y}) \big) \,\gamma^0\, P \big( (t,\vec{y}), z \big)\: d^3y = P(x,z) \:. \]
This identity follows by a straightforward computation: First,
\beq \label{Pnorm} \begin{split}
\int_{\R^3} & P \big( x, (t, \vec{y}) \big) \:\gamma^0\: P \big( (t, \vec{y}), z \big)\: d^3y \\
&= \int_{\R^3} d^3y \int \frac{d^4k}{(2 \pi)^4}\: e^{-i k(x-y)} \int \frac{d^4q}{(2 \pi)^4}\: e^{-i q(y-z)}
\: P_m(k)\:\gamma^0\: P_m(q) \\
&= \int \frac{d^4k}{(2 \pi)^4} \int_\R \frac{d \lambda}{2 \pi}\; e^{-i k x + i q z}
\: P_m(k)\:\gamma^0\: P_m(q) \Big|_{q = (\lambda, \vec{k})} \:.
\end{split} \eeq
Setting~$k=(\omega, \vec{k})$, we evaluate the $\delta$-distributions inside the factors~$P_m$,
\begin{align*}
\delta(k^2-m^2)& \, \delta(q^2-m^2) \big|_{q = (\lambda, \vec{k})}
= \delta \big( \omega^2 - |\vec{k}|^2 -m^2 \big) \:
\delta \big( \lambda^2 - |\vec{k}|^2 -m^2 \big) \\
&= \delta(\lambda^2 - \omega^2)\: \delta \big( \omega^2 - |\vec{k}|^2 -m^2 \big) \:.
\end{align*}
This shows that we only get a contribution if~$\lambda=\pm \omega$.
Using this fact together with the mass shell property~$\omega^2-|\vec{k}|^2=m^2$, 
we can simplify the Dirac matrices according to
\begin{align*}
(\slashed{k}+m) &\:\gamma^0\: (\slashed{q} + m)
= (\omega \gamma^0 + \vec{k} \vec{\gamma} + m) \,\gamma^0\,
(\pm \omega \gamma^0 + \vec{k} \vec{\gamma} + m) \\
&= (\omega \gamma^0 + \vec{k} \vec{\gamma} + m) \,
(\pm \omega \gamma^0 - \vec{k} \vec{\gamma} + m)\,\gamma^0 \\
&= \Big( (\pm \omega^2 + |\vec{k}|^2 + m^2) \,\gamma^0
+(1 \pm 1) \,\omega\,(\vec{k} \vec{\gamma}) + (1 \pm 1)\, m \omega \Big) \\
&= \left\{ \begin{array}{cl}
2 \omega\, (\slashed{k}+m) & \text{in case~$+$} \\
0 & \text{in case~$-\:.$} \end{array} \right.
\end{align*}
Hence we only get a contribution if~$\lambda=\omega$, giving rise to the identity
\[  \delta(\lambda^2 - \omega^2) = \frac{1}{2 |\omega|}\: \delta(\lambda-\omega)\:. \]
Putting these formulas together, we obtain
\begin{align*}
\int_{\R^3} & P \big( x, (t, \vec{y}) \big) \:\gamma^0\: P \big( (t, \vec{y}), z \big)\: d^3y \\
&= \int \frac{d^4k}{(2 \pi)^4} \int_\R \frac{d \lambda}{2 \pi}\; e^{-i k (x-z)}
\:\delta(\lambda - \omega)\: \delta( k^2 -m^2)\:
\frac{2 \omega}{2 |\omega|}\, (\slashed{k}+m)\: \Theta(-k_0) \\
&= -\frac{1}{2 \pi} \int \frac{d^4k}{(2 \pi)^4} \; e^{-i k (x-z)}
\: \delta( k^2 -m^2)\: (\slashed{k}+m)\: \Theta(-k_0) \:.
\end{align*}
This gives the result.
\QED

The Fourier integral~\eqref{Pxyvac} can be computed in closed form, giving
an expression involving Bessel functions.
In preparation, it is useful to pull the Dirac matrices out of the Fourier integral.
To this end, one rewrites the factor~$(\slashed{k}+m)$ in~\eqref{Pxyvac}
in terms of a differential operator in position space,
\beq \label{Pdiff}
P(x,y) = (i \Pdd_x + m) \,T_{m^2}(x,y) \:,
\eeq
where~$T_{m^2}$ is the scalar bi-distribution
\[ T_{m^2}(x,y) := \int \frac{d^4k}{(2 \pi)^4}\: \delta(k^2-m^2)\: \Theta(-k_0)\: e^{-ik(x-y)} \:. \]
In the next lemma, we determine the singular structure of this distribution.
The method is to subtract an explicit singular distribution and to show that the difference
is a {\em{regular distribution}} (i.e.\ a locally integrable function, denoted by~$L^1_\text{\rm{loc}}$).
The distribution~$\text{PP}/\xi^2$, denoted by {\em{principal value}}, is defined by
evaluating weakly with a test function $\eta \in C^\infty_0(\scrM)$
and by removing the positive and negative parts of the pole in a symmetric way.
There are different equivalent ways of writing the principal part, each of which could
serve as a possible definition:
\begin{align*}
\int &\frac{\text{PP}}{\xi^2} \: \eta(\xi)\: d^4\xi
= \lim_{\varepsilon \searrow 0} \int \Theta\big( |\xi^2| - \varepsilon \big)\;
\frac{1}{\xi^2} \: \eta(\xi)\: d^4\xi \\
&= \lim_{\varepsilon \searrow 0} \frac{1}{2} \sum_{\pm}
\int \frac{1}{\xi^2 \pm i \varepsilon} \: \eta(\xi)\: d^4\xi
= \lim_{\varepsilon \searrow 0} \frac{1}{2} \sum_{\pm}
\int \frac{1}{\xi^2 \pm i \varepsilon \xi^0} \: \eta(\xi)\: d^4\xi \:.
\end{align*}

\begin{Lemma} \label{lemmaTintro}
On the light cone, the bi-distribution~$T_{m^2}$ has the following singularity
structure,
\beq \label{Tsingular}
T_{m^2}(x,y) + \frac{1}{8 \pi^3}
\left( \frac{\text{\rm{PP}}}{\xi^2} +i \pi\, \delta(\xi^2)\,\epsilon(\xi^0) \right) \in
L^1_\text{\rm{loc}}(\scrM \times \scrM) \:,
\eeq
where we set~$\xi := y-x$.
Away from the light cone (i.e.\ for~$\xi^2 \neq 0$), $T_{m^2}(x,y)$ is a smooth function given by
\begin{align} \label{Taway}
T_{m^2}(x,y) = \left\{ \begin{array}{cl} 
\displaystyle \frac{m}{16 \pi^2} \:\frac{Y_1\big(m\sqrt{\xi^2} \,\big)}{\sqrt{\xi^2}}
+\frac{i m}{16 \pi^2}\: \frac{J_1 \big(m\sqrt{\xi^2} \,\big)}{\sqrt{\xi^2}}\: \epsilon(\xi^0)
& \text{if~$\xi$ is timelike} \\[1em]
\displaystyle \frac{m}{8 \pi^3} \frac{K_1 \big(m\sqrt{-\xi^2} \,\big)}{\sqrt{-\xi^2}} & \text{if~$\xi$ is spacelike}\:,
\end{array} \right. \hspace*{-0.3em}
\end{align}
where~$J_1$, $Y_1$ and~$K_1$ are Bessel functions.
\end{Lemma}
\Proof The Fourier integral is computed most conveniently by inserting
a convergence-generating factor. Thus for any~$\varepsilon>0$ we consider the Fourier integral
\[ T^\varepsilon_{m^2}(x,y) := \int \frac{d^4k}{(2 \pi)^4}\: \delta(k^2-m^2)\: \Theta(-k_0)\: e^{-ik(x-y)}\:
e^{-\varepsilon \,|k_0|} \:. \]
This Fourier integral can be computed pointwise, showing that~$T^\varepsilon(x,y)$ is a regular distribution.
Taking the limit~$\varepsilon \searrow 0$ in the distributional sense, we will then obtain~$T_{m^2}(x,y)$.

Setting~$\xi=y-x$ and~$t=\xi^0$, we first carry out
the integral over~$k_0$ to obtain
\begin{align*}
T^\varepsilon_{m^2}(x,y)&= \int \frac{d^4k}{(2\pi)^4}\:
	\delta(k^2-m^2)\:\Theta(-k_0) \:e^{ik \xi} \:e^{-\varepsilon \,|k_0|} \nonumber \\
	&= \int_{\R^3} \frac{d^3k}{(2\pi)^4}\:\frac{1}{2\sqrt{\vec{k}^2+m^2}}
	\: e^{-i\sqrt{\vec{k}^2+m^2}\, t-i \vec{k}\vec{\xi}}\:e^{-\varepsilon \sqrt{\vec{k}^2+m^2}} \:.
\end{align*}
Next, for the spatial momentum~$\vec{k}$ we introduce polar coordinates~$(p=|\vec{k}|,
\vartheta, \varphi)$, where~$\vartheta$ is the angle between~$\vec{k}$ and~$\vec{\xi}$,
and~$\varphi$ is the azimuthal angle. Also setting~$r=|\vec{\xi}|$, we get
\begin{align}
T^\varepsilon_{m^2}(x,y)&= \int_0^\infty \frac{dp}{2(2\pi)^3} \int_{-1}^1d\cos\theta
	\: \frac{p^2}{\sqrt{p^2+m^2}} \:e^{-(\varepsilon+it) \sqrt{p^2+m^2}}\: e^{-ipr\cos\theta} \nonumber \\
	&= \frac{1}{r}\int_0^\infty \frac{dp}{(2\pi)^3}\:
	\frac{p}{\sqrt{p^2+m^2}}\: e^{-(\varepsilon+it) \sqrt{p^2+m^2}}\: \sin(pr) \nonumber \\
	&= \frac{m^2}{(2\pi)^3} \frac{K_1 \big(m\sqrt{r^2+(\varepsilon+it)^2}
		\,\big)}{m\sqrt{r^2+(\varepsilon+it)^2}}, \label{Tbessel}
\end{align}
where the last integral was carried out using~\cite[formula (3.961.1)]{gradstein}.
Here the square root and the Bessel function~$K_1$ is defined as usual using a branch cut along
the negative real axis.

When taking the limit~$\varepsilon \searrow 0$, one must be careful for two reasons.
First, a pole forms on the light cone~$t=\pm r$. Second, the Bessel function~$K_1$
involves logarithms, which must be evaluated in the complex plane using the
branch cut along the negative real axis. For clarity, we treat these two issues after each other.
The asymptotic expansion of the Bessel function (see~\cite[(10.31.1)]{DLMF})
\[ K_1(z) = \frac{1}{z} + \O\big( z \log z \big) \]
yields that the pole on the light cone is of the form
\[ T^\varepsilon_{m^2}(x,y) = \frac{1}{(2\pi)^3} \:\frac{1}{r^2+(\varepsilon+it)^2} 
+ \O\big( \log|\xi^2|) \:, \]
uniformly in~$\varepsilon$. Therefore, after subtracting the pole, we can take the
limit~$\varepsilon \searrow 0$ as a locally integrable function, i.e.
\[ \lim_{\varepsilon \searrow 0}
\bigg( T^\varepsilon_{m^2}(x,y) - \frac{1}{(2\pi)^3} \:\frac{1}{r^2+(\varepsilon+it)^2} \bigg)
\in L^1_\text{\rm{loc}}(\scrM \times \scrM)\:. \]
For the subtracted pole, the limit~$\varepsilon \searrow 0$ can be computed in the
distributional sense by
\begin{align*}
\lim_{\varepsilon \searrow 0} \frac{1}{r^2+(\varepsilon+it)^2}
= \lim_{\varepsilon \searrow 0} \frac{1}{r^2 - t^2 + i \varepsilon t }
= -\frac{\text{PP}}{\xi^2} -i \pi\, \delta(\xi^2)\,\epsilon(\xi^0)\:,
\end{align*}
where we used the distributional equation
\[ \lim_{\varepsilon \searrow 0} \left( \frac{1}{x - i \varepsilon} - \frac{1}{x + i \varepsilon} \right)
= 2 \pi i \: \delta(x) \:. \]
Here ``PP'' again denotes the principal value, and~$\epsilon$ is the step function $\epsilon(x)=1$
for $x \geq 0$ and $\epsilon(x)=-1$ otherwise. This gives~\eqref{Tsingular}.

In order to compute the regular part of the distribution~$T_{m^2}$,
we may disregard the singularity on the light cone and may consider the case
that~$\xi$ is either spacelike or timelike. In the first case,
the argument~$m\sqrt{r^2+(\varepsilon+it)^2}$ of the Bessel function converges to
the positive real axis, where the Bessel function is analytic.
This gives the lower equation in~\eqref{Taway}.
In the remaining case that~$\xi$ is timelike, the
argument~$m\sqrt{r^2+(\varepsilon+it)^2}$ converges to the imaginary axis
(more precisely, to the upper imaginary axis if~$t>0$ and to the lower imaginary axis if~$t<0$).
Using the relations~\cite[(10.27.9) and~(10.27.10)]{DLMF}
\[ i \pi J_1(z) = -i K_1(-iz) -i K_1(iz) \qquad \text{and} \qquad 
-\pi Y_1(z) = -i K_1(-iz) + i K_1(iz) \]
(valid if~$|\arg z|<\frac{\pi}{2}$), one can express~$K_1$ near the upper and lower imaginary axis by
\[ K_1(\pm iz) = -\frac{\pi}{2} \big( J_1(z) \mp i Y_1(z) \big) \:. \]
Using these identities in~\eqref{Tbessel} and using that the Bessel functions~$J_1$ and~$K_1$
are analytic in a neighborhood of the positive real axis, one can take the limit~$\varepsilon \searrow 0$
to obtain the upper equation in~\eqref{Taway}.
\QED
We point out that the Bessel functions in~\eqref{Taway} are all real-valued.
In particular, one sees that~$T(x,y)$ is real-valued if the vector~$\xi$ is spacelike.

Using the result of Lemma~\ref{lemmaTintro} in~\eqref{Pdiff}, one
can derive corresponding formulas for~$P(x,y)$.
In particular, differentiating~\eqref{Tsingular}, one sees that~$P(x,y)$
has an even stronger singularity on the light cone which involves
terms of the form~$1/\xi^4$ and~$\delta'(\xi^2)$.
Differentiating~\eqref{Taway}, carrying out the derivatives
with the chain rule and using formulas for the
derivatives of Bessel functions (see~\cite[(10.6.6) and~(10.29.4)]{DLMF}),
one can also express the fermionic projector~$P(x,y)$ in terms of Bessel functions.
We do not give the resulting formulas, because we do not need the detailed form later on.
Instead, we here prefer to argue with general properties of the distribution~$P(x,y)$.
This makes it possible to infer qualitative properties of the eigenvalues of~$A_{xy}$,
even without referring to the detailed form of the formulas in Lemma~\ref{lemmaTintro}.
From Lorentz symmetry, we know that for all~$x$ and~$y$ with spacelike or timelike separation,
$P(x,y)$ can be written as
\beq \label{Pxyrep}
P(x,y) = \alpha\, \xi_j \gamma^j + \beta\:\1
\eeq
with two complex-valued functions~$\alpha$ and~$\beta$ (where again~$\xi =y-x$).
Taking the conjugate with respect to the spin scalar product, we see that
\beq \label{Pyxrep}
P(y,x) = \overline{\alpha}\, \xi_j \gamma^j + \overline{\beta}\:\1 \:.
\eeq
As a consequence,
\beq \label{1}
A_{xy} = P(x,y)\, P(y,x) = a\, \xi_j \gamma^j + b\, \1
\eeq
with two real parameters~$a$ and $b$ given by
\beq \label{ab}
a = \alpha \overline{\beta} + \beta \overline{\alpha} \:,\qquad
b = |\alpha|^2 \,\xi^2 + |\beta|^2 \:.
\eeq
Applying the formula~$(A_{xy} - b \1)^2 = a^2\:\xi^2\,\1$,
the roots of the characteristic polynomial of~$A_{xy}$ are computed by
\beq \label{root}
b \pm \sqrt{a^2\: \xi^2} \:.
\eeq
Therefore, the eigenvalues of the closed chain are either real, or else they
form a complex conjugate pair.
Which of the two cases appears is determined by the sign of the factor~$\xi^2$.
This gives the agreement of the different notions of causality in the following sense:
\begin{Prp} \label{prpcorrespond}
Assume that~$P(x,y)$ is the unregularized kernel of the fermionic projector of
the vacuum~\eqref{Pxyvac}, and that the eigenvalues~$\lambda^{xy}_1,
\ldots, \lambda^{xy}_4$ are computed as the eigenvalues of the closed chain~\eqref{Axydef}.
Then the following statements hold:
If the points~$x, y \in \scrM$ have spacelike separation in Minkowski space,
then they are also spacelike separated in the sense of Definition~\ref{def2}.
If, on the other hand, the points~$x, y \in \scrM$ have timelike separation in Minkowski space,
then they are also timelike separated in the sense of Definition~\ref{def2}.
Even more, they are properly timelike separated (see Definition~\ref{defproptl})
in the sense that the closed chain~$A_{xy}$ has strictly positive eigenvalues and
definite eigenspaces.
Finally, if the points~$x, y \in \scrM$ have lightlike separation in Minkowski space,
then the causal structure of Definition~\ref{def2} is ill-defined.
\end{Prp} \noindent
The fact that the causal structure is ill-defined for lightlike separation
again explains why an UV regularization must be introduced.

\Proof[Proof of Proposition~\ref{prpcorrespond}]
If the vector~$\xi=y-x$ is spacelike, then the term~$\xi^2$ is negative. Thus
the eigenvalues in~\eqref{root} form a complex conjugate
pair, implying that they all have the same absolute value.
Thus the points are spacelike separated in the sense of Definition~\ref{def2}.

If the vector~$\xi$ is timelike, the term~$\xi^2$ in~\eqref{root} is positive, so that
the~$\lambda_j$ are all real.
In order to show that they do not have the same absolute value,
we need to verify that the parameters~$a$ and~$b$ are both non-zero.
This makes it necessary to refer to the explicit formula involving
Bessel functions~\eqref{Taway}: The Bessel functions~$Y_1$ and~$J_1$
do not have joint zeros on the positive real axis.
As a consequence, the parameter~$\beta$ in~\eqref{Pxyrep} is non-zero.
Likewise, the derivatives~$Y_1'$ and~$J_1'$ do not have joint zeros
(as can again be verified from the fact that the Bessel functions form a fundamental system).
This implies that the parameter~$\alpha$ in~\eqref{Pxyrep} is non-zero.
We conclude that the parameter~$b$ in~\eqref{ab} is non-zero.
The combination of~$\alpha$ and~$\beta$ in the formula for~$a$ in~\eqref{ab}
can be rewritten in terms of a Wronskian of the Bessel function.
This Wronskian can be computed explicitly using~\cite[(10.5.2)]{DLMF},
implying that~$a$ is non-zero.
We conclude that the points~$x$ and~$y$ are timelike separated in the sense of Definition~\ref{def2}.

In order to get the connection to proper timelike separation,
recall that if~$\xi$ is a timelike vector of Minkowski space, then
the closed chain has the form~\eqref{ab} with~$a,b \neq 0$.
A direct computation shows that this matrix is diagonalizable and that
the eigenspaces are definite with respect to the spin scalar product.
Moreover, applying the Schwarz inequality to the explicit formulas~\eqref{ab},
one obtains
\beq \label{starin}
|a| \, \sqrt{\xi^2} = 2 \re \Big( \alpha \, \sqrt{\xi^2} \: \overline{\beta} \Big)
\overset{(\star)}{\leq} |\alpha|^2 \xi^2 + |\beta|^2 = b \:,
\eeq
proving that the eigenvalues in~\eqref{root} are non-negative. 
It remains to show that none of these eigenvalues vanishes.
To this end, it suffices to show that the inequality~($\star$) in~\eqref{starin}
is strict, which in turn is equivalent to proving that
\[ \im \big( \alpha \overline{\beta} \big) \neq 0 \:. \]
This inequality follows by a detailed analysis of the Bessel functions
(see~\cite[proof of Lemma~4.3]{lqg}).
We conclude that~$x$ and~$y$ are indeed properly timelike separated.

If the vector~$\xi$ is lightlike, then~$P(x,y)$ is not defined pointwise.
As a consequence, the closed chain is ill-defined.
\QED

This proposition cannot be applied directly to causal fermion systems because,
as explained in~\S\ref{secuvintro} and~\S\ref{secuvreg},
constructing a causal fermion system makes it necessary to introduce an
UV regularization. Nevertheless, the above proposition also gives
correspondence of the different notions of causality for
causal fermion systems describing the Minkowski vacuum, as we now explain.
Thus let us consider the causal fermion system corresponding to the regularized
fermionic projector of the vacuum~$P^\varepsilon(x,y)$.
In the limit~$\varepsilon \searrow 0$, the kernel of the fermionic
projector~$P^\varepsilon(x,y)$ converges to the unregularized kernel~$P(x,y)$
(see~\eqref{Pest} in Proposition~\ref{prpisometry}).
If this convergence is pointwise, i.e.\ if for given space-time points~$x,y \in \scrM$,
\beq \label{pointwise}
\lim_{\varepsilon \searrow 0} P^\varepsilon(x,y) = P(x,y) \:,
\eeq
then the results of Proposition~\ref{prpcorrespond} also apply to the causal
fermion system, up to error terms which tend to zero as~$\varepsilon \searrow 0$.
Thinking of~$\varepsilon$ as the Planck scale, this means physically that
the notion of causality of Definition~\ref{def2} agrees with the usual notion of
causality in Minkowski space, up to corrections which are so small that they
cannot be observed.
The subtle point of this argument is that it requires pointwise convergence~\eqref{pointwise}.
Clearly, such a pointwise convergence cannot hold if~$x$ and~$y$ are lightlike separated,
because the right side of~\eqref{pointwise} is ill-defined pointwise.
Expressed for a causal fermion system for fixed~$\varepsilon$ on the Planck scale,
this means that the notion of causality of Definition~\ref{def2} does {\em{not}} agree
with the usual notion of
causality if the vector~$\xi$ is almost lightlike in the sense that~$\big| |\xi^0| - |\vec{\xi}| \big|
\lesssim \varepsilon$. This is not surprising because we cannot expect
that the notion of causality in Minkowski space holds with a higher resolution than
the regularization scale~$\varepsilon$.
The remaining question is whether we have pointwise convergence~\eqref{pointwise}
if the points~$x$ and~$y$ have timelike or spacelike separation.
The answer is yes for a large class of regularizations (like for example the
regularization by mollification in Example~\ref{exmollify}).
However, the general notion of Definition~\ref{defreg}
only gives weak convergence of the kernels~\eqref{Pest}.
This shortcoming could be removed by adding a condition to Definition~\ref{defreg}
which ensures pointwise convergence away from the light cone.
On the other hand, such an additional condition
seems unnecessary,
and therefore it seems preferable not to impose it.
Nevertheless, the physical picture is that the regularized kernel should 
converge pointwise, at least for generic points~$x$ and~$y$ which lie
sufficiently far away from the light cone.
With this in mind, Proposition~\ref{prpcorrespond} indeed
shows that the notion of causality of Definition~\ref{def2} corresponds to the usual notion of
causality in Minkowski space, up to corrections which
are so small that they are irrelevant in most situations of interest.

We conclude this section by explaining why the functional~${\mathscr{C}}$
introduced in~\eqref{Cform} gives information on the time direction.
Our first task is to rewrite this functional in terms of the regularized kernel of the fermionic
projector~$P^\varepsilon(x,y)$.

\begin{Lemma} Assume that the operator~$P^\varepsilon(x,x) : S_x\scrM \rightarrow S_x\scrM$
is invertible. Then, setting
\beq \label{nuxdef}
\nu(x)= P^\varepsilon(x,x)^{-1} \::\: S_x\scrM \rightarrow S_x\scrM \:,
\eeq
the functional~${\mathscr{C}}$, \eqref{Cform}, can be written as
\beq \label{Ccomm}
{\mathscr{C}}(x, y) = i \Tr_{S_x} \Big(  P^\varepsilon(x,y) \:\nu(y)\:P^\varepsilon(y,x)\:
\big[ \nu(x), A_{xy} \big] \Big)\:.
\eeq
\end{Lemma}
\Proof Since~$P(x,x) = \pi_x x|_{S_x} = x|_{S_x}$, we know that~$\nu(x) = ( x|_{S_x} )^{-1}$. Thus
\begin{align*}
\pi_x\, y\,x \,\pi_y\, \pi_x|_{S_x} &=
\pi_x y\: \pi_y x\: \pi_x y\: \nu(y)\: \pi_y x\: \nu(x) |_{S_x} \\
&= P^\varepsilon(x,y)\: P^\varepsilon(y,x)\: P^\varepsilon(x,y) \: \nu(y)\: P^\varepsilon(y,x)\: \nu(x) |_{S_x} \:.
\end{align*}
Using this formula in~\eqref{Cform}, we obtain
\begin{align*}
{\mathscr{C}}(x, y) &= i \Tr_{S_x} \big( y\,x \,\pi_y\, \pi_x|_{S_x} - y\,\pi_x \,\pi_y\,x|_{S_x}
\big) \\
&= i \Tr_{S_x} \Big(  P^\varepsilon(x,y)\, P^\varepsilon(y,x) \: P^\varepsilon(x,y) \:\nu(y)\:P^\varepsilon(y,x)\: \nu(x) \\
&\qquad \quad\;\, - P^\varepsilon(x,y)\, P^\varepsilon(y,x) \:\nu(x)\:P^\varepsilon(x,y) \:\nu(y)\:P^\varepsilon(y,x) \Big) \\
&= i \Tr_{S_x} \Big(  P^\varepsilon(x,y) \:\nu(y)\:P^\varepsilon(y,x)\: \nu(x)\: P^\varepsilon(x,y)\:P^\varepsilon(y,x) \\
&\qquad \quad\;\, - P^\varepsilon(x,y)\: \nu(y)\:P^\varepsilon(y,x) \:P^\varepsilon(x,y) \:P^\varepsilon(y,x)\: \nu(x)\:  \Big) \:.
\end{align*}
This gives the result.
\QED

We point out that the operator~$\nu(x)$ in~\eqref{nuxdef} is ill-defined without
UV regularization because evaluating the distribution~$P(x,y)$ on the diagonal~$x=y$
has no mathematical meaning. As a consequence, the functional~${\mathscr{C}}$
is ill-defined without UV regularization, even if~$x$ and~$y$ have timelike separation.
This makes the following computation somewhat delicate.
In order to keep the analysis reasonably simple, we assume that the
regularized kernel of the fermionic projector has {\em{vector-scalar structure}},
meaning that it is of the general form
\beq \label{vectorscalar}
P^\varepsilon(x,y) = v^\varepsilon_j(x,y) \:\gamma^j \:+\: \beta^\varepsilon(x,y) \:\1
\eeq
with a vectorial and a scalar component. Here~$v^\varepsilon(x,y)$ is a complex vector field
(i.e.\ it can be written as~$v^\varepsilon = u^\varepsilon + i w^\varepsilon$
with Minkowski vectors~$u^\varepsilon$ and~$w^\varepsilon$ which need not be
collinear).
Then, evaluating~\eqref{vectorscalar} for~$x=y$, one sees that~$P^\varepsilon(x,x)$
can be written as
\[ P^\varepsilon(x,x) = v^\varepsilon_j(x) \:\gamma^j \:+\: \beta^\varepsilon(x) \:\1 \]
(where we set~$v^\varepsilon(x)=v^\varepsilon(x,x)$ and~$\beta^\varepsilon(x)=\beta^\varepsilon(x,x)$).
Since~$P^\varepsilon(x,x)$ is a symmetric operator on~$S_x\scrM$, it follows that~$v^\varepsilon$
is a real vector field, and~$\beta$ a real-valued function.
For a large class of regularizations, the matrix~$P^\varepsilon(x,x)$ is invertible because the
vectorial component dominates the scalar component.
With this in mind, we
here assume that~$\nu(x)$ exists. Then it is given by
\beq \label{nuform}
\nu(x) = \frac{1}{\rho(x)} \:\Big( v^\varepsilon_j(x) \:\gamma^j - \beta^\varepsilon(x) \:\1 \Big) \:,
\eeq
where~$\rho :=  v^\varepsilon_j  (v^\varepsilon)^j - (\beta^\varepsilon)^2$.
Now we can compute the composite expression in~\eqref{Ccomm}, working
for all other terms with the unregularized formulas
(which is again justified if we have pointwise convergence~\eqref{pointwise}).
This gives the following result.

\begin{Prp} Using~\eqref{nuform} and replacing~$P^\varepsilon(x,y)$, $P^\varepsilon(y,x)$
and~$A_{xy}$ by the unregularized expressions~\eqref{Pxyrep}, \eqref{Pyxrep} and~\eqref{1},
the functional~${\mathscr{C}}$ is given by
\beq \label{Cfinal}
{\mathscr{C}}(x, y) = \frac{16 a}{\rho(x)\, \rho(y)}\: \im \big(\alpha \overline{\beta}\big)\;
\Big(v^\varepsilon(x)^j\, \xi_j\: v^\varepsilon(y)^k\, \xi_k - \xi^2\:
v^\varepsilon(x)^j \,v^\varepsilon(y)_j \Big) \:.
\eeq
\end{Prp}
\Proof Using~\eqref{nuform} and~\eqref{1} in~\eqref{Ccomm} gives
\begin{align*}
{\mathscr{C}}(x, y) &= i \Tr_{S_x} \Big(  P(x,y) \:\nu(y)\:P(y,x)\:
\big[ \nu(x), A_{xy} \big] \Big) \\
&= \frac{ia}{\rho(x)} \Tr_{S_x} \Big(  P(x,y) \:\nu(y)\:P(y,x)\:
\big[  \slashed{v}^\varepsilon(x), \slashed{\xi} \big] \Big) \:,
\end{align*}
where in the last step we used that the scalar components of~$A_{xy}$ and~$\nu(x)$ drop
out of the commutator.
Taking the scalar component of~$\nu(y)$, the two factors~$P(x,y)$ and~$P(y,x)$ combine
to the closed chain, which according to~\eqref{1} has no bilinear component, so that
the trace vanishes. Therefore, we only need to take into account the vectorial component of~$\nu(y)$.
Using~\eqref{Pxyrep} and~\eqref{Pyxrep}, we obtain
\begin{align*}
{\mathscr{C}}(x, y)
&= \frac{ia}{\rho(x)\, \rho(y)} \Tr_{S_x} \Big(  \big(\alpha \slashed{\xi}+\beta\,\1\big) \: \slashed{v}^\varepsilon(y) \:\big(\overline{\alpha} \slashed{\xi}+\overline{\beta}\,\1\big)\:
\big[  \slashed{v}^\varepsilon(x), \slashed{\xi} \big] \Big) \\
&= -\frac{a}{\rho(x)\, \rho(y)}\: \im \big(\alpha \overline{\beta}\big)\:
 \Tr_{S_x} \Big(  \big[ \slashed{\xi}, \slashed{v}^\varepsilon(y) \big]\:
\big[  \slashed{v}^\varepsilon(x), \slashed{\xi} \big] \Big) \:.
\end{align*}
Computing the trace of the product of Dirac matrices gives the result.
\QED

For the interpretation of the formula~\eqref{Cfinal}, we first consider
the case that~$y$ and~$x$ have space-like separation.
In this case, it turns out that the prefactor~$\im (\alpha \overline{\beta})$ vanishes,
so that~\eqref{Cfinal} gives no information on a time direction. This is consistent with the fact that
for points in Minkowski space with space-like separation, the notions of future- and past-directed
depend on the observer and cannot be defined in a covariant manner.
However, if~$y$ and~$x$ have timelike separation, then the factors~$a$ and~$\im (\alpha \overline{\beta})$
are indeed both non-zero (see the proof of Proposition~\ref{prpcorrespond}).
Therefore, the functional~${\mathscr{C}}$ is non-zero, provided that the
vector~$\xi$ is {\em{non-degenerate}} in the sense that it is linearly independent
of both~$v^\varepsilon(x)$ and~$v^\varepsilon(y)$.
Since the set of directions~$\xi$ for which these vectors are linearly
dependent has measure zero, we may always restrict attention to
non-degenerate directions.
Moreover, the formula~\eqref{Cfinal} shows that the functional~${\mathscr{C}}$ does not change sign
for~$\xi$ inside the upper or lower light cone.
On the other hand, ${\mathscr{C}}$
is antisymmetric under sign flips of~$\xi$ because interchanging~$x$ and~$y$
in~\eqref{Cform} obviously gives a minus sign.

We conclude that for the regularized Dirac sea vacuum,
the sign of the functional~${\mathscr{C}}$ distinguishes a time direction.
Asymptotically as~$\varepsilon \searrow 0$, this time direction agrees with the distinction of the causal
past and causal future in Minkowski space.

To summarize, in this section we saw how the intrinsic structures of a causal fermion system
correspond to the usual structures in Minkowski space.
To this end, we constructed causal fermion systems from a regularized
Dirac sea configuration and analyzed the asymptotics as the UV regularization
is removed. For brevity, we only considered the topological and causal structure of space-time
as well as spinors and wave functions. The reader interested in geometric structures
like connection and curvature is referred to the detailed exposition in~\cite{lqg}.
Moreover, in Section~\ref{seclimit} below we shall explain how the methods and results
introduced in this section can be generalized to interacting systems.

\section{Underlying Physical Principles} \label{secprinciples}
In order to clarify the physical concepts, we now briefly discuss the
underlying physical principles.
Causal fermion systems evolved from an attempt to combine several physical principles
in a coherent mathematical framework. As a result, these principles appear in the
framework in a specific way:
\begin{itemize}[leftmargin=1.3em, itemsep=0.2em]
\itemD The {\bf{principle of causality}} is built into a causal fermion system in a specific way,
as was explained in~\S\ref{seccausal} above.
\itemD The {\bf{Pauli exclusion principle}} is incorporated in a causal fermion system,
as can be seen in various ways. 
One formulation of the Pauli exclusion principle states that every fermionic one-particle state
can be occupied by at most one particle. In this formulation, the Pauli exclusion principle
is respected because every wave function can either be represented in the form~$\psi^u$
(the state is occupied) with~$u \in \H$ or it cannot be represented as a physical wave function
(the state is not occupied). 
Via these two conditions, the fermionic projector encodes for every state
the occupation numbers~$1$ and~$0$, respectively, but it is
impossible to describe higher occupation numbers.
More technically, one may obtain the connection to the fermionic Fock space formalism
by choosing an orthonormal basis~$u_1, \ldots, u_f$ of~$\H$ and forming the $f$-particle Hartree-Fock state
\[ \Psi := \psi^{u_1} \wedge \cdots \wedge \psi^{u_f} \:. \]
Clearly, the choice of the orthonormal basis is unique only up to the unitary transformations
\[ u_i \rightarrow \tilde{u}_i = \sum_{j=1}^f U_{ij} \,u_j \quad \text{with} \quad U \in \U(f)\:. \]
Due to the anti-symmetrization, this transformation changes the corresponding Hartree-Fock state
only by an irrelevant phase factor,
\[ \psi^{\tilde{u}_1} \wedge \cdots \wedge \psi^{\tilde{u}_f} = \det U \;
\psi^{u_1} \wedge \cdots \wedge \psi^{u_f} \:. \]
Thus the configuration of the physical wave functions can be described by a
fermionic multi-particle wave function.
The Pauli exclusion principle becomes apparent
in the total anti-symmetrization of this wave function.
\itemD A {\bf{local gauge principle}} becomes apparent once we choose
basis representations of the spin spaces and write the wave functions in components.
Denoting the signature of~$(S_x, \Sl .|. \Sr_x)$ by~$(p(x),q(x))$, we choose
a pseudo-orthonormal basis~$(\mathfrak{e}_\alpha(x))_{\alpha=1,\ldots, p+q}$ of~$S_x$.
Then a wave function~$\psi$ can be represented as
\[ \psi(x) = \sum_{\alpha=1}^{p+q} \psi^\alpha(x)\: \mathfrak{e}_\alpha(x) \]
with component functions~$\psi^1, \ldots, \psi^{p+q}$.
The freedom in choosing the basis~$(\mathfrak{e}_\alpha)$ is described by the
group~$\U(p,q)$ of unitary transformations with respect to an inner product of signature~$(p,q)$.
This gives rise to the transformations
\[ \mathfrak{e}_\alpha(x) \rightarrow \sum_{\beta=1}^{p+q} U^{-1}(x)^\beta_\alpha\;
\mathfrak{e}_\beta(x) \qquad \text{and} \qquad
\psi^\alpha(x) \rightarrow  \sum_{\beta=1}^{p+q} U(x)^\alpha_\beta\: \psi^\beta(x) \]
with $U \in \U(p,q)$.
As the basis~$(\mathfrak{e}_\alpha)$ can be chosen independently at each space-time point,
one obtains {\em{local gauge transformations}} of the wave functions,
where the gauge group is determined to be the isometry group of the spin scalar product.
The causal action is
{\em{gauge invariant}} in the sense that it does not depend on the choice of spinor bases.
\itemD The {\bf{equivalence principle}} is incorporated in the following general way.
Space-time $M:= \supp \rho$ together with the universal measure~$\rho$ form a topological
measure space, being a more general structure than a Lorentzian manifold.
Therefore, when describing~$M$ by local coordinates, the freedom in choosing such
coordinates generalizes the freedom in choosing general reference frames in a space-time manifold.
Therefore, the equivalence principle of general relativity is respected. The causal action is {\em{generally
covariant}} in the sense that it does not depend on the choice of coordinates.
\end{itemize}

\section{The Dynamics of Causal Fermion Systems}
Similar to the Einstein-Hilbert action in general relativity, in the causal action principle
one varies space-time as well as all structures therein globally.
This global viewpoint implies that it is not obvious what the causal action principle tells us about the
dynamics of the system. The first step for clarifying the situation is to derive the
Euler-Lagrange (EL) equations corresponding to the causal action principle (\S\ref{secvary}).
Similar to the Einstein or Maxwell equations, these EL equations should
describe the dynamics. Additional insight is gained by
studying Noether-like theorems which specify the quantities which are conserved in the dynamics
(\S\ref{secnoether}). Finally, we review results on the initial value problem (\S\ref{secinitial}).
We remark that more explicit information on the dynamics is obtained by
considering limiting cases in which the EL equations corresponding to the
causal action reduce to equations of a structure familiar from classical field theory
and quantum field theory (see Section~\ref{seclimit}).

\subsection{The Euler-Lagrange Equations} \label{secvary}
We now return to the abstract setting of Section~\ref{secframe}.
Our goal is to derive the EL equations corresponding to the causal action principle
in the form most useful for our purposes.
The method is to consider so-called {\em{variations of the physical wave functions}} which we
now introduce (for more general variations see Remark~\ref{remgenEL} below).
Let~$(\H, \F, \rho)$ be a causal fermion system.
We assume that~$\rho$ is a minimizer of the causal action principle.
However, we do not want to assume that the total volume~$\rho(\F)$ be finite.
Instead, we merely assume that~$\rho$ is {\em{locally finite}} in the sense that~$\rho(K)< \infty$
for every compact subset~$K \subset \F$.
Our starting point is the wave evaluation operator~$\Psi$ introduced in~\eqref{weo},
\[ \Psi \::\: \H \rightarrow C^0(M, SM)\:, \qquad u \mapsto \psi^u \:. \]
We now vary the wave evaluation operator.
Thus for any~$\tau \in (-\delta, \delta)$ we consider a mapping~$\Psi_\tau : \H \rightarrow C^0(M)$.
For~$\tau=0$, this mapping should coincide with the wave evaluation operator~$\Psi$.
The family~$(\Psi_\tau)_{\tau \in (-\delta, \delta)}$ can be regarded as a simultaneous variation
of all physical wave functions of the system.
In fact, for any~$u \in \H$, the variation of the corresponding physical wave function is given by
\[ \psi^u_\tau := \Psi_\tau(u) \in C^0(M, SM) \:. \]
Next, we introduce the corresponding local correlation operators~$F_\tau$ by
\[ F_\tau(x) := - \Psi_\tau(x)^* \Psi_\tau(x) \qquad \text{so that} \qquad
F_\tau \::\: M \rightarrow \F \:. \]
In view of~\eqref{Fid}, we know that~$F_0(x)=x$. Therefore, the family~$(F_\tau)_{\tau \in (-\delta, \delta)}$
is a variation of the local correlation operators. Taking the push-forward measure gives rise to a family
of universal measures,
\beq \label{pushtau}
\rho_\tau := (F_\tau)_* \rho \:.
\eeq
Since~$F_0$ is the identity, 
we know that~$\rho_0 = \rho$. Therefore, the family~$(\rho_\tau)_{\tau \in (-\delta, \delta)}$
is indeed a variation of the universal measure.

We now work out the EL equations for the resulting class of variations of the universal
measure. In order for the constructions to be mathematically well-defined, we need
a few technical assumptions which are summarized in the following definition.
\begin{Def} \label{defvarc}
The {\bf{variation of the physical wave functions}}
is {\bf{smooth}} and {\bf{compact}} if the family of operators~$(\Psi_\tau)_{\tau \in (-\delta, \delta)}$
has the following properties:
\begin{itemize}
\item[(a)] The variation is trivial on the orthogonal complement of a finite-dimensional
subspace~$I \subset \H$, i.e.
\[ \Psi_\tau |_{I^\perp} = \Psi \qquad \text{for all~$\tau \in (-\delta, \delta)$} \:. \]
\item[(b)] There is a compact subset~$K \subset M$ outside which the variation is trivial, i.e.
\[ \big( \Psi_\tau(u) \big) \big|_{M \setminus K} = \big( \Psi(u) \big) \big|_{M \setminus K}
\qquad \text{for all~$\tau \in (-\delta, \delta)$ and~$u \in \H$} \:. \]
\item[(c)] The Lagrangian is continuously differentiable in the sense that the derivative
\beq \label{ccond}
\frac{d}{d\tau} \L\big( x, F_\tau(y) \big) \big|_{\tau=0}
\eeq
exists and is continuous on~$M \times M$.
\end{itemize}
\end{Def} \noindent
With the conditions~(a) and~(b) we restrict attention to variations
which are sufficiently well-behaved (similar as
in the classical calculus of variations, where one restricts attention to
smooth and compactly supported variations).
It is a delicate point to satisfy the condition~(c), because (due to the absolute values of the eigenvalues in~\eqref{Lagrange}) the Lagrangian is only Lipschitz continuous on~$\F \times \F$.
Therefore, the derivative in~\eqref{ccond} does not need to exist, even if~$F_\tau(y)$ is smooth.
This means that in the applications, one must verify that the condition~(c) holds
(for details see the computations in~\cite{cfs}).
Here we simply assume that the variation of the wave functions is smooth and compact.

By definition of the push-forward measure~\eqref{pushtau}, we know that for any integrable
function~$f$ on~$\F$,
\beq \label{pushid}
\int_\F f(x)\: d\rho_\tau = \int_\F f(F_\tau(x) \big)\: d\rho \:.
\eeq
In this way, the variation of the measure can be rewritten as a variation of the
arguments of the integrand. In particular, the variation of the action
can be written as
\[ \iint_{M \times M} \L\big(F_\tau(x), F_\tau(y) \big) \: d\rho(x)\, d\rho(y) \]
(and similarly for the other integrals).
Another benefit of working with the push-forward measure~\eqref{pushtau}
is that the total volume is preserved. Namely, combining the identity~\eqref{pushid}
with the assumption in Definition~\ref{defvarc}~(b), one readily verifies that
the volume constraint~\eqref{volconstraint} is satisfied in the sense that~$\rho_\tau$
satisfies the conditions~\eqref{totvol}.

We consider first variations, treating the constraints with Lagrange multipliers
(this procedure is justified in~\cite{lagrange}).
Since the volume constraint is already respected, it remains to
consider the trace constraint~\eqref{trconstraint} and the
boundedness constraint~\eqref{Tdef}. We conclude that first variations of the functional
\beq \label{Skaplam}
\Sact_{\kappa,\lambda}
:= \Sact + \kappa \,\big( \T - C \big) - \lambda \left( \int_\F \tr (x) \: d\rho - c \right)
\eeq
vanish for suitable values of the Lagrange parameters~$\kappa, \lambda \in \R$,
where the constants~$C$ and~$c$ are the prescribed values of the constraints.
For clarity, we point out that the boundedness constraint merely is an inequality.
The method for handling this inequality constraint is to choose~$\kappa=0$ if~$\T(\rho)<C$, whereas
in the case~$\T(\rho)=C$ the Lagrange multiplier~$\kappa$ is in general non-zero
(for details see again~\cite{lagrange}). Introducing the short notation
\[ \L_\kappa(x,y) := \L(x,y) + \kappa \, |xy|^2 \:, \]
we can write the effective action as
\[ \Sact_{\kappa,\lambda}(\rho_\tau)
= \iint_{M \times M} \L_\kappa(x,y) \: d\rho(x)\, d\rho(y) - \lambda
\int_M \tr \big(F_\tau(x) \big)\: d\rho(x) - \kappa C + \lambda c \:. \]

Now we can compute the first variation by differentiating with respect to~$\tau$.
It is most convenient to express the causal action and the
constraints in terms of the kernel of the fermionic projector
(just as explained at the beginning of~\S\ref{secker}).
Moreover, it is preferable to consider the Lagrangian~$\L_\kappa(x,y)$ as a function only
of~$P_\tau(x,y)$ by writing the closed chain as
\beq \label{Atauex}
A_{xy} = P_\tau(x,y)\, P_\tau(x,y)^*
\eeq
(where~$P_\tau(x,y)^*$ denotes similar to~\eqref{Pxysymm} the adjoint with respect to the spin scalar
product). We use the notation
\[ \delta P(x,y) = \frac{d}{d\tau}\, P_\tau(x,y) \Big|_{\tau=0} \:, \]
and similarly for other functions.
When computing the variation of the Lagrangian, one must keep in mind that~$\L_\kappa(x,y)$
depends both on~$P_\tau(x,y)$ and on its adjoint~$P_\tau(x,y)^*$ (cf.~\eqref{Atauex}).
Therefore, when applying the chain rule, we obtain contributions which are complex linear
and complex anti-linear in~$\delta P_\tau(x,y)$. We write the first variation with traces as
\[ \delta \L_\kappa(x,y) = \Tr_{S_y} \big( B\, \delta P(x,y) \big) + \Tr_{S_x} \big( C\, \delta P(x,y)^* \big) \]
with linear operators~$B : S_x \rightarrow S_y$ and~$C : S_y \rightarrow S_x$.
Since~$\delta P(x,y)$ can be chosen arbitrarily, this equation uniquely defines
both~$B$ and~$C$.
Since the variation of the Lagrangian is always real-valued, it follows that~$C=B^*$.
Using furthermore the
symmetry of the Lagrangian in the arguments~$x$ and~$y$, we conclude that
the first variation of the Lagrangian can be written as (see also~\cite[Section~5.2]{PFP})
\beq \label{delLdef}
\delta \L_\kappa(x,y) = 
\Tr_{S_y} \big( Q(y,x)\, \delta P(x,y) \big) + \Tr_{S_x} \big( Q(x,y)\, \delta P(x,y)^* \big)
\eeq
with a kernel~$Q(x,y) : S_y \rightarrow S_x$ which is symmetric in the sense that
\beq \label{Qsymm}
Q(x,y)^* = Q(y,x)\:.
\eeq

The EL equations are expressed in terms of the kernel~$Q(x,y)$ as follows.
\begin{Prp} {\bf{(Euler-Lagrange equations)}} \label{prpEL}
Let~$\rho$ be a minimizer of the causal action principle. Then for a suitable choice of the
Lagrange parameters~$\lambda$ and~$\kappa$, the integral operator~$Q$ with kernel defined by~\eqref{delLdef}
satisfies the equations
\beq \label{Qrel}
\int_M Q(x,y)\, \psi^u(y)\: d\rho(y) = \frac{\lambda}{2}\: \psi^u(x) \qquad \text{for all~$u \in \H$ and~$x \in M$}\:.
\eeq
\end{Prp} \noindent
We note for clarity that by writing the equation~\eqref{Qrel} we imply that
the integral must exist and be finite.
\Proof[Proof of Proposition~\ref{prpEL}] Using~\eqref{delLdef}, the first variation of~$\Sact_{\kappa, \lambda}$
is computed by
\begin{align*}
\delta \Sact_{\kappa,\lambda}
=\;& \iint_{M\times M} \Big(
\Tr_{S_y} \big( Q(y,x)\, \delta P(x,y) \big) + \Tr_{S_x} \big( Q(x,y)\, \delta P(x,y)^* \big) \Big) \:d\rho(x) \: d\rho(y) \\
&- \lambda \int_M \Tr \big(\delta P(x,x) \big)\: d\rho(x) \:.
\end{align*}
Noting that~$\delta P(x,y) = \delta P(y,x)$, after renaming the integration variables in the first summand
of the double integral, we obtain
\beq \begin{split}
\delta \Sact_{\kappa,\lambda}
= 2 \iint_{M\times M} \Tr_{S_x} \big( Q(x,y)\, \delta P(y,x) \big) 
- \lambda \int_M \Tr_{S_x} \big(\delta P(x,x) \big)\: d\rho(x) \:.
\end{split} \label{delS}
\eeq

Next, we express~$\delta P$ in terms of the variation of the physical wave
functions. By Lemma~\ref{lemmaPxyrep}, we know that
\[ P_\tau(x,y) = -\Psi_\tau(x) \Psi_\tau(y)^* \:. \]
Differentiating this relation gives
\[ \delta P(x,y) = -(\delta \Psi)(x) \:\Psi(y)^* -\Psi(x) \:(\delta \Psi)(y)^* \:. \]
We now specialize to the case that the variation is trivial on the
orthogonal complement of a one-dimensional subspace~$I=\text{span}(u) \subset \H$.
Then for any~$\phi \in S_y$,
\[ \delta P(x,y) \, \phi = -\delta \psi^u(x) \;\Sl \:\psi^u(y) \,|\, \phi \Sr_y
-\psi^u(x) \;\Sl \delta \psi^u(y) \,|\, \phi \Sr_y \:. \]
By inserting a phase factor according to
\[ \delta \psi^u \rightarrow e^{i \varphi}\: \delta \psi^u \:, \]
one sees that~$\delta \psi^u$ can be varied independently inside and outside the spin scalar product.
Therefore, it suffices to consider variations inside the spin scalar product.
Thus the vanishing of the first variation~\eqref{delS} yields the condition
\begin{align*}
0 = 2 \iint_{M\times M} \Sl \delta \psi^u(x) \,|\, Q(x,y) \,\psi^u(y) \Sr_x
- \lambda \int_M \Sl \delta \psi^u(x) \,|\, \psi^u(x) \Sr_x \:.
\end{align*}
Since the variation~$\delta \psi^u$ is
arbitrary (within the class of smooth and compactly supported variations), the result follows.
\QED

We remark that the kernel~$Q(x,y)$ also gives rise to an operator on the
one-particle Krein space~$(\K, \bra .|. \ket)$ as introduced in~\S\ref{secKrein}.
Thus, in analogy to~\eqref{Pdef}, one sets
\[ Q \::\: \D(Q) \subset \K \rightarrow \K \:,\qquad (Q \psi)(x) =
\int_M Q(x,y)\, \psi(y)\, d\rho(y)\:, \]
where the domain~$\D(Q)$ can be chosen for example as the continuous
wave functions with compact support.
The symmetry property of the kernel~\eqref{Qsymm} implies that the operator~$Q$
is symmetric on the Krein space~$(\K, \bra .|. \ket)$.
The equation~\eqref{Qrel} can be written in a compact form as the operator equation
\beq \label{Qrel2}
\big( 2 Q -\lambda \1 \big)\, \Psi = 0
\eeq
(where~$\Psi$ is again the wave evaluation operator~\eqref{weo}).
In words, this equation means that the operator~$(2 Q - \lambda \1)$ vanishes
on the physical wave functions.
However, the operator equation~\eqref{Qrel2} is not satisfying mathematically
because the physical wave functions in the image of~$\Psi$ are in general not vectors
of the Krein space~$(\K, \bra .|. \ket)$ (see~\S\ref{secKrein}).
Nevertheless, \eqref{Qrel2} is useful as a short notation for the EL equations~\eqref{Qrel}.

\begin{Remark} {\bf{(more general variations)}} \label{remgenEL} {\em{
Clearly, only a special class of variations of the universal measure can be
described by variations of the physical wave functions.
As a consequence, the resulting EL equations~\eqref{Qrel} are only {\em{necessary}} conditions
for~$\rho$ to be a critical point of the action~\eqref{Skaplam}.
We now explain how these necessary conditions are related to the
stronger EL equations as derived in~\cite{lagrange}.

As an example of variations which are not covered by the ansatz~\eqref{pushtau},
one can multiply the universal measure by weight functions
\beq \label{weight}
d\rho_\tau = f_\tau\: d\rho \:,
\eeq
where~$(f_\tau)_{\tau \in (-\delta, \delta)}$ is a family of non-negative
functions which are integrable and have mean zero, i.e.
\[ f_\tau \geq 0 \qquad \text{and} \qquad \int_M f_\tau\, d\rho = 0 \:. \]
Computing first variations of the action~\eqref{Skaplam} gives rise to the equation
\beq \label{ELextra}
2 \int_M \L_\kappa(x,y)\: d\rho(y) + \lambda\, \tr(x) = \text{const} \qquad \text{on~$M$}\:.
\eeq
This is an additional EL equation which minimizers of the causal action principle must satisfy.
It turns out that in the limiting case of an interacting system in Minkowski space
(to be discussed in~\S\ref{seccl} and~\S\ref{secQFT} below), this
equation can be satisfied simply by a rescaling of the local correlation operators.

Variations of the physical wave functions as well as variations of the form~\eqref{weight}
have the property that the support of the universal measure changes continuously
(in the sense that for every compact set~$K \subset \F$ and every open neighborhood~$U$
of~$K \cap \supp \rho$ there is~$\varepsilon>0$ such that~$\supp \rho_\tau \cap K \subset U$
for all~$\tau$ with~$|\tau|<\varepsilon$).
Such variations can be regarded as the analogs of variations of the potentials, the metric or the wave functions
in classical field theory or quantum mechanics.
However, in the setting of causal fermion systems there are also more general smooth variations
for which the support of the measure~$\rho_\tau$
changes discontinuously. A typical example is to let~$\rho$ be a bounded measure and to set
\beq \label{rhosc}
\rho_\tau = (1-\tau^2) \:\rho + \tau^2 \,\rho(\F)\: \delta_x \:,
\eeq
where~$\delta_x$ is the Dirac measure supported at~$x \not \in \supp \rho$.
The EL equations corresponding to such variations have a different mathematical structure,
which we cannot explain in detail here.
Generally speaking, for interacting systems in Minkowski space,
the EL equations of Proposition~\ref{prpEL} give rise to an effective interaction
via {\em{classical}} gauge fields (this so-called {\em{continuum limit}} will be discussed in~\S\ref{seccl}).
The EL equations corresponding to more general variations
like~\eqref{rhosc}, however, give rise to an effective interaction via bosonic {\em{quantum}} fields.
We will come back to this point in~\S\ref{secQFT}.
}} \QEDrem
\end{Remark} 

\subsection{Symmetries and Conserved Surface Layer Integrals} \label{secnoether}
In~\cite{noether} it is shown that symmetries of the Lagrangian give rise to conservation
laws. These results can be understood as adaptations of Noether's theorem
to the causal action principle. Since the mathematical structure of the causal action principle
is quite different from that of the Lagrangian formulation of classical field theory,
these adaptations are not straightforward.
We now explain a few concepts and results from~\cite{noether}
which are important for understanding the general physical picture.

\begin{figure}
\psscalebox{1.0 1.0} 
{
\begin{pspicture}(0,-1.511712)(10.629875,1.511712)
\definecolor{colour0}{rgb}{0.8,0.8,0.8}
\definecolor{colour1}{rgb}{0.6,0.6,0.6}
\pspolygon[linecolor=black, linewidth=0.002, fillstyle=solid,fillcolor=colour0](6.4146066,0.82162136)(6.739051,0.7238436)(6.98794,0.68384355)(7.312384,0.66162133)(7.54794,0.67939913)(7.912384,0.7593991)(8.299051,0.8705102)(8.676828,0.94162136)(9.010162,0.9549547)(9.312385,0.9371769)(9.690162,0.8571769)(10.036829,0.7371769)(10.365718,0.608288)(10.614607,0.42162135)(10.614607,-0.37837866)(6.4146066,-0.37837866)
\pspolygon[linecolor=black, linewidth=0.002, fillstyle=solid,fillcolor=colour1](6.4146066,1.2216214)(6.579051,1.1616213)(6.770162,1.1127324)(6.921273,1.0905102)(7.103495,1.0816213)(7.339051,1.0549546)(7.530162,1.0638436)(7.721273,1.0993991)(7.8857174,1.1393992)(8.10794,1.2060658)(8.299051,1.2549547)(8.512384,1.3038436)(8.694607,1.3260658)(8.890162,1.3305103)(9.081273,1.3393991)(9.379051,1.3216213)(9.659051,1.2593992)(9.9746065,1.1705103)(10.26794,1.0460658)(10.459051,0.94384354)(10.614607,0.82162136)(10.610162,0.028288014)(10.414606,0.1660658)(10.22794,0.26828802)(10.010162,0.37051025)(9.663495,0.47273245)(9.356829,0.53051025)(9.054606,0.548288)(8.814607,0.54384357)(8.58794,0.5171769)(8.387939,0.48162135)(8.22794,0.44162133)(7.90794,0.34828803)(7.6946063,0.29939914)(7.485718,0.26828802)(7.272384,0.26828802)(7.02794,0.28162134)(6.82794,0.3171769)(6.676829,0.35273245)(6.543495,0.38828802)(6.4146066,0.42162135)
\pspolygon[linecolor=black, linewidth=0.002, fillstyle=solid,fillcolor=colour0](0.014606438,0.82162136)(0.3390509,0.7238436)(0.5879398,0.68384355)(0.9123842,0.66162133)(1.1479398,0.67939913)(1.5123842,0.7593991)(1.8990508,0.8705102)(2.2768288,0.94162136)(2.610162,0.9549547)(2.9123843,0.9371769)(3.290162,0.8571769)(3.6368287,0.7371769)(3.9657176,0.608288)(4.2146063,0.42162135)(4.2146063,-0.37837866)(0.014606438,-0.37837866)
\psbezier[linecolor=black, linewidth=0.04](6.4057174,0.8260658)(7.6346064,0.45939913)(7.8634953,0.8349547)(8.636828,0.92828804)(9.410162,1.0216213)(10.165717,0.7927325)(10.614607,0.42162135)
\psbezier[linecolor=black, linewidth=0.04](0.005717549,0.8260658)(1.2346064,0.45939913)(1.4634954,0.8349547)(2.2368286,0.92828804)(3.0101619,1.0216213)(3.7657175,0.7927325)(4.2146063,0.42162135)
\rput[bl](2.0101619,0.050510235){$\Omega$}
\rput[bl](8.759051,0.0016213481){\normalsize{$\Omega$}}
\psline[linecolor=black, linewidth=0.04, arrowsize=0.09300000000000001cm 1.0,arrowlength=1.7,arrowinset=0.3]{->}(1.9434953,0.85495466)(1.8057176,1.6193991)
\rput[bl](2.0946064,1.1705103){$\nu$}
\psbezier[linecolor=black, linewidth=0.02](6.4146066,0.42384356)(7.6434956,0.057176903)(7.872384,0.43273246)(8.645718,0.52606577)(9.419051,0.61939913)(10.174606,0.39051023)(10.623495,0.019399125)
\psbezier[linecolor=black, linewidth=0.02](6.410162,1.2193991)(7.639051,0.8527325)(7.86794,1.228288)(8.6412735,1.3216213)(9.414606,1.4149547)(10.170162,1.1860658)(10.619051,0.8149547)
\rput[bl](8.499051,0.9993991){\normalsize{$y$}}
\rput[bl](7.8657174,0.49273247){\normalsize{$x$}}
\psdots[linecolor=black, dotsize=0.06](8.170162,0.65273243)
\psdots[linecolor=black, dotsize=0.06](8.796828,1.1327325)
\psline[linecolor=black, linewidth=0.02](6.1146064,1.2216214)(6.103495,0.82162136)
\rput[bl](5.736829,0.8993991){\normalsize{$\delta$}}
\rput[bl](3.6146064,0.888288){$\scrN$}
\rput[bl](1.1146064,-1.4117119){$\displaystyle \int_\scrN \cdots\, d\mu_\scrN$}
\rput[bl](5.7146063,-1.511712){$\displaystyle \int_\Omega d\rho(x) \int_{M \setminus \Omega} d\rho(y)\: \cdots\:\L(x,y)$}
\psline[linecolor=black, linewidth=0.02](6.0146065,1.2216214)(6.2146063,1.2216214)
\psline[linecolor=black, linewidth=0.02](6.0146065,0.82162136)(6.2146063,0.82162136)
\end{pspicture}
}
\caption{A surface integral and a corresponding surface layer integral.}
\label{fignoether1}
\end{figure}
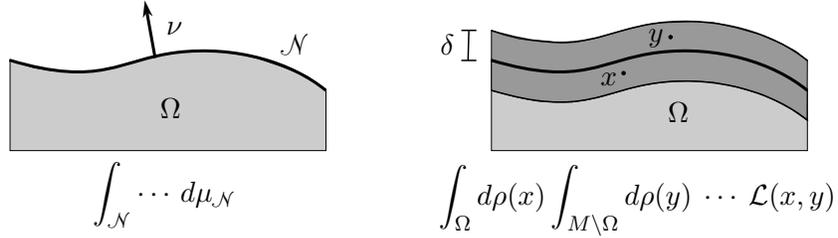

We first recall that the conservation laws obtained from the classical Noether theorem
state that the integral of a certain density over a Cauchy surface~$\scrN$ does not
depend on the choice of~$\scrN$. For example, charge conservation states that
the spatial integral of the charge density gives a constant. As another example, energy conservation
states that in a static space-time background, the integral of the energy density is a constant.
In general terms, the conserved quantities are spatial integrals over a Cauchy surface~$\scrN$
(see the left of Figure~\ref{fignoether1}).
In the setting of causal fermion systems, it is unclear how such surface integrals should be
defined, in particular because we do not have a measure on hypersurfaces and because it
is not clear what the normal~$\nu$ on the hypersurface should be.
This is the reason why in the Noether-like theorems in~\cite{noether} one works instead 
of surface integrals with so-called {\em{surface layer integrals}} where one integrates over a boundary layer
of a set~$\Omega \subset M$ (see the right of Figure~\ref{fignoether1}).
The width~$\delta$ of this layer is the length scale on which~$\L(x,y)$
decays. For a system composed of Dirac particles (similar as explained in Section~\ref{secmink}
for the Minkowski vacuum and in~\S\ref{seccl} for interacting systems),
this length scale can be identified with the {\em{Compton scale}}~$\sim m^{-1}$
of the Dirac particles. Thus the width of the surface layer is a non-zero macroscopic
length scale. In particular, the surface layer integrals cannot be identified with
or considered as a generalization of the surface integrals of the classical Noether theorem.
However, in most situations of interest, when the surface~$N$ is almost flat on the
Compton scale (like for a spatial hyperplane in Minkowski space), the surface layer
integral can be well-approximated by a corresponding surface integral.
It turns out that in this limiting case, the conservation laws obtained from the
Noether-like theorems in~\cite{noether} go over to corresponding classical
conservation laws.

From the conceptual point of view, the most interesting conservation law
is {\em{charge conservation}}. In order to construct the underlying symmetry, we let~$\scrA$
be a bounded symmetric operator on~$\H$ and let
\[ \scrU_\tau := \exp(i \tau \scrA) \]
be the corresponding one-parameter family of unitary transformations.
We introduce the family of transformations
\[ \Phi_\tau \,:\, \F \rightarrow \F\:,\qquad \Phi_\tau(x) = \scrU_\tau \,x\, \scrU_\tau^{-1} \:. \]
Since the Lagrangian is defined via the spectrum of operators on~$\H$, it clearly remains unchanged
if all operators are unitarily transformed, i.e.
\beq \label{Lsymm}
\L\big( \Phi_\tau(x), \Phi_\tau(y) \big) = \L(x,y) \:.
\eeq
In other words, the transformations~$\Phi_\tau$ describe a {\em{symmetry of the Lagrangian}}.
Next, one constructs a corresponding one-family of universal measures by taking the
push-forward,
\[ \rho_\tau := (\Phi_\tau)_* \rho \:. \]
As a consequence of the symmetry~\eqref{Lsymm}, this variation of the universal measure
leaves the action invariant. Under suitable differentiability assumptions, this
symmetry gives rise to the identity
\beq \label{conserve}
\frac{d}{d\tau} \int_\Omega d\rho(x) \int_{M \setminus \Omega} d\rho(y) \:\Big( \L\big( \Phi_\tau(x), y\big) -
\L\big(\Phi_{-\tau}(x), y \big) \Big) \Big|_{\tau=0} = 0 \:,
\eeq
valid for any compact subset~$\Omega \subset M$.

We now explain how the identity~\eqref{conserve} is related to a conservation law.
To this end, for simplicity we consider a system in Minkowski space
(similar as explained for the vacuum in Section~\ref{secmink}) and choose a sequence
of compact sets~$\Omega_n$ which exhaust the region between two Cauchy surfaces at times~$t=t_0$
and~$t=t_1$. Then the surface layer integral~\eqref{conserve} reduces to the difference
of integrals over surface layers at times~$t \approx t_0$ and~$t \approx t_1$.
Next, we choose~$\scrA =\pi_{\la u \ra}$ as the projection operator on the one-dimensional subspace
generated by a vector~$u \in \H$. Then in the limit~$\varepsilon \searrow 0$ in which the
UV regularization is removed, the resulting surface layer integral at time~$t \approx t_0$
reduces to the integral
\[ \int_{\R^3} \Sl u(t_0,\vec{x}) \,|\, \gamma^0 u(t_0, \vec{x}) \Sr_{(t_0, \vec{x})}\: d^3x\:, \]
thereby reproducing the probability integral in Dirac theory.
As a consequence, the representation of the scalar product~$\la .|. \ra_\H$ as an integral over a
Cauchy surface~\eqref{sprodMin} has a natural generalization to the setting of causal fermion systems,
if the surface integral is replaced by a corresponding surface layer integral.
This result also shows that the spatial normalization of the fermionic projector
(where one works with spatial integrals of the form~\eqref{Pnorm}; for details see~\cite{norm})
really is the correct normalization method which reflects the intrinsic conservation laws of the
causal fermion system.

The conservation laws in~\cite{noether} also give rise to the
{\em{conservation of energy and momentum}}, as we now outline.
In the classical Noether theorem, these conservation
laws are a consequence of space-time symmetries as described most conveniently using
the notion of Killing fields. Therefore, one must extend this notion to the setting of causal
fermion systems. Before explaining how this can be accomplished, we recall the
procedure in the classical Noether theorem:
In the notion of a Killing field, one distinguishes the background geometry from the
additional particles and fields. The background geometry must have a symmetry
as described by the Killing equation. The additional particles and fields, however, do not
need to have any symmetries. Nevertheless, one can construct a symmetry of the whole system
by actively transporting the particles and fields along the flow lines of the Killing field.
The conservation law corresponding to this symmetry transformation gives rise to
the conservation of energy and momentum.

In a causal fermion system, there is no clear-cut distinction between the background geometry
and the particles and fields of the system, because all of these structures are encoded in the
underlying causal fermion system and mutually depend on each other.
Therefore, instead of working with a symmetry of the background geometry,
we work with the notion of an approximate symmetry. By actively transforming
those physical wave functions which do not respect the symmetry,
such an approximate symmetry again gives rise to an exact
symmetry transformation, to which the Noether-like theorems in~\cite{noether} can be applied.
More precisely, one begins with a $C^1$-family of transformations~$(f_\tau)_{\tau \in (-\delta, \delta)}$ of space-time,
\beq \label{ftdef}
f_\tau \::\: M \rightarrow M \qquad \text{with} \qquad f_0 = \1 \:,
\eeq
which preserve the universal measure in the sense that~$(f_\tau)_* \rho = \rho$.
The family~$(f_\tau)$ can be regarded as the analog of a flow in space-time along a classical Killing field.
Moreover, one considers a family of unitary transformations~$(\scrU_\tau)_{\tau \in (-\delta, \delta)}$
on~$\H$ with the property that
\[ \scrU_{-\tau} \,\scrU_\tau = \1 \qquad \text{for all~$\tau \in (-\delta, \delta)$}\:. \]
Combining these transformations should give rise to an
{\em{approximate symmetry}} of the wave evaluation operator~\eqref{weo}
in the sense that if we compare the transformation of the space-time point with the unitary transformation
by setting
\beq \label{Edef}
E_\tau(u,x) := (\Psi u)\big(f_\tau(x) \big) - (\Psi \scrU^{-1}_\tau u)(x) \qquad (x \in M, u \in \H) \:,
\eeq
then the operator~$E_\tau : \H \rightarrow C^0(M, SM)$ should be sufficiently small. Here ``small'' means for example
that~$E$ vanishes on the orthogonal complement of a finite-dimensional subspace of~$\H$; for details see~\cite[Section~6]{noether}. Introducing the variation~$\Phi_\tau$ by
\[ \Phi_\tau \::\: M \rightarrow \F \:,\qquad \Phi_\tau(x) = \scrU_\tau \,x\, \scrU^{-1}_\tau \:, \]
we again obtain a symmetry of the Lagrangian~\eqref{Lsymm}.
This gives rise to conserved surface layer integrals of the form~\eqref{conserve}.
In order to bring these surface layer integrals into a computable form, one decomposes the first variation
of~$\Phi_\tau$ as
\beq \label{decompose}
\delta \Phi(x) := \partial_\tau \Phi_\tau(x) \big|_{\tau=0} = \delta f(x) + v(x) \:,
\eeq
where~$\delta f$ is the first variation of~$f_\tau$, \eqref{ftdef}, and~$v(x)$ is a vector field on~$\F$ 
along~$M$ which is transversal to~$M \subset \F$. Expressing~$v$ in terms of the operator~$E$
in~\eqref{Edef} shows that~$v$ is again small, making it possible to compute
the corresponding variation of the Lagrangian in~\eqref{conserve}.
We remark that in the decomposition~\eqref{decompose}, the vector field~$\delta f$ describes
a transformation of the space-time points. The vector field~$v$, however,
can be understood as an active transformation of all the objects in space-time which
do {\em{not}} have the space-time symmetry (similar as described above for
the parallel transport of the particles and fields along the flow
lines of the Killing field in the classical Noether theorem).

In order to get the connection to classical conservation laws, one again
studies a system in Minkowski space and considers the limiting case where a sequence~$\Omega_n$
exhausts the region between two Cauchy surfaces
at times~$t=t_0$ and~$t=t_1$. In this limiting case, the conserved surface layer integral
reduces to the surface integral
\[ \int_{\R^3} T_{i0} \,K^i\: d^3x\:, \]
where~$T_{ij}$ is the energy-momentum tensor of the Dirac particles and~$K= \delta f$ is a Killing field.
This shows that the conservation of energy and momentum is a 
special case of more general conservation laws which are intrinsic to causal fermion systems.

\subsection{The Initial Value Problem and Time Evolution} \label{secinitial}
In order to get a better understanding of the dynamics described by the causal
action principle, it is an important task to analyze the initial value problem.
The obvious questions are: What is the initial data? Is it clear that a solution exists? Is the solution unique?
How do solutions look like? Giving general answers to these questions is a difficult
mathematical problem. In order to evaluate the difficulties, one should recall that~$\rho$ describes
space-time as well as all structures therein. Therefore, similar as in the Cauchy problem for the Einstein equations, solving the initial value problem involves finding the geometry of space-time
together with the dynamics of all particles and fields.
In view of the complexity of this problem, the only results known at present
are contained in the paper~\cite{cauchy},
where an initial value problem is formulated and some existence and
uniqueness theorems are proven. We now review a few methods and results
of this paper. Moreover, at the end of this section we mention an approach
proposed in~\cite{jet} for obtaining more explicit information on the dynamics by
analyzing perturbations of a given minimizing measure.

Since the analysis of the causal action principle is technically demanding,
in~\cite{cauchy} one considers instead so-called {\em{causal variational principles in the
compact setting}}.
In order to get into this simplified setting, one replaces~$\F$ by a compact metric space
(or a smooth manifold).
The Lagrangian is replaced by a non-negative continuous function~$\L \in C^{0,1}(\F \times \F, \R^+_0)$
which is symmetric in its two arguments. Similar to~\eqref{Sdef} one minimizes the action
\[ \Sact(\rho) = \iint_{\F \times \F} \L(x,y)\: d\rho(x)\: d\rho(y) \]
in the class of all normalized regular Borel measures on~$\F$, but now
leaving out the constraints~\eqref{trconstraint} and~\eqref{Tdef}.
Space-time is again defined by~$M:= \supp \rho$.
The resulting causal structure is defined by saying that
two space-time points~$x, y \in M$ are called
{\em{timelike}} separated if $\L(x,y)>0$, and {\em{spacelike}} separated if~$\L(x,y)=0$.
Clearly, in this setting there are no wave functions. Nevertheless,
causal variational principles in the compact setting incorporate
basic features of the causal action principle and are therefore a
good starting point for the analysis
(for a more detailed introduction and structural results on the minimizing measures see~\cite{support}).

When solving the classical Cauchy problem,
instead of searching for a global solution, it is often easier to look for a local solution
around a given initial value surface. This concept of a local solution also reflects the
common physical situation where the physical system under consideration is only a small
subsystem of the whole universe. With this in mind, we would like to
``localize'' the variational principle to a subset~$\mathfrak{I} \subset \F$, referred to
as the {\em{inner region}}.
There is the complication that the Lagrangian~$\L(x,y)$ is nonlocal in the sense that
it may be non-zero for points~$x \in \mathfrak{I}$ and~$y \in \F \setminus \mathfrak{I}$.
In order to take this effect into account, one describes the influence
of the ``outer region'' $\F \setminus \mathfrak{I}$ by a so-called {\em{external potential}}
$\phi : \F \rightarrow \R^+_0$. In the limiting case when the
outer region becomes large, this gives rise to the so-called {\em{inner variational principle}},
where the action defined by
\beq \label{SIdef}
\Sact_{\mathfrak{I}}[\rho, \phi] = \iint_{\mathfrak{I} \times \mathfrak{I}} \L(x,y)\: d\rho(x)\: d\rho(y)
+ 2 \int_{\mathfrak{I}} \big( \phi(x) - \mathfrak{s} \big) \: d\rho(x)
\eeq
is minimized under variations of~$\rho$ in the class of regular Borel measures on~$\mathfrak{I}$
(not necessarily normalized because the volume constraint is now taken care of 
by the corresponding Lagrange parameter~$\mathfrak{s}>0$).

The {\em{initial values}} are described by a regular Borel measure~$\rho_0$
(which is to be thought of as the universal measure restricted to a time slice around
the initial value surface in space-time). The {\em{initial conditions}} are implemented
by demanding that
\beq \label{initial}
\rho \geq \rho_0 \:.
\eeq
The naive method of minimizing~\eqref{SIdef} under the constraint~\eqref{initial}
is not a sensible concept because the constraint~\eqref{initial} would give rise to undesirable
Lagrange multiplier terms in the EL equations.
Instead, one minimizes~\eqref{SIdef} without constraints, but chooses the external potential~$\phi$ in such
a way that the minimizing measure satisfies the initial values~\eqref{initial}.
It turns out that this procedure does not determine the external potential uniquely.
Therefore, the method proposed in~\cite{cauchy} is to {\em{optimize the external potential}}
by making it in a suitable sense ``small.''
As is made precise in~\cite{cauchy} in various situations, the resulting
interplay between minimizing the action and optimizing the external potential
gives rise to unique solutions of the initial-value problem with an optimal external potential.

We point out that, due to the mathematical simplifications made, the
results in~\cite{cauchy} do not apply to physically interesting situations
like the initial value problem for interacting Dirac sea configurations.
Moreover, the methods in~\cite{cauchy} do not seem to give
explicit information on the dynamics of causal fermion systems.
Therefore, it is a promising complementary approach to consider perturbations of a given minimizing measure
(which should describe the ``vacuum configuration'') and to analyze
the dynamics of the perturbations by studying the resulting EL equations.
This approach is pursued in~\cite{jet} in the following way.
In order to describe the perturbations of the minimizing measure~$\rho$,
one considers smooth variations for which the support of~$\rho$ changes
continuously. Combining~\eqref{pushtau} and~\eqref{weight}, 
these variations can be written as
\[ \tilde{\rho}_\tau = (F_\tau)_* \big( f_\tau \, \rho \big) \]
with a family of mappings~$F_\tau : M \rightarrow \F$ and
a family of non-negative functions~$f_\tau$.
Expanding in powers of~$\tau$, these variations can be described conveniently
in terms of sections of {\em{jet bundles}} over~$M$. The EL equations yield conditions
on the jets, which can be rewritten as dynamical equations in space-time.

\section{Limiting Cases} \label{seclimit}
We now discuss different limiting cases of causal fermion systems.
\subsection{The Quasi-Free Dirac Field and Hadamard States} \label{secquasifree}
We now turn attention to interacting systems. The simplest interaction is
obtained by inserting an {\em{external potential}} into the Dirac equation~\eqref{Dirfree},
\beq \label{Cont:DiracEq}
\big( i \gamma^j \partial_j + \B - m \big) \,\psi(x) = 0 \:.
\eeq
Another situation of physical interest is to consider the Dirac equation in an external
classical gravitational field as described mathematically by a globally hyperbolic
Lorentzian manifold~$(\scrM, g)$.
In this section, we explain how the methods and results of Section~\ref{secmink}
generalize to the situation when an external field is present.
This will also give a connection to quasi-free Dirac fields and Hadamard states.
In order to keep the explanations as simple as possible, we here restrict attention to
an external potential~$\B$ in Minkowski space, but remark that many methods and results could or have
been worked out also in the presence of a gravitational field.

The obvious conceptual difficulty when extending the constructions of Section~\ref{secmink}
is that one no longer has the notion of ``negative-frequency solutions''
which were essential for introducing Dirac sea configurations (see Lemma~\ref{lemmaDiracsea}).
In order to overcome this difficulty, 
one needs to decompose the solution space of the Dirac equation~\eqref{Cont:DiracEq}
into two subspace, in such a way that without external potential the two subspaces
reduce to the subspaces of positive and negative frequency.
This {\em{external field problem}} was solved perturbatively in~\cite{sea, grotz}
and non-perturbatively in~\cite{finite, infinite, hadamard} (for a more detailed
exposition see~\cite[\S2.1]{PFP}).

We now briefly outline the non-perturbative treatment, which relies on the
construction on the so-called {\em{fermionic signature operator}}.
Choosing again the scalar product~\eqref{sprodMin}, the solution space of the
Dirac equation~\eqref{Cont:DiracEq} forms a Hilbert space denoted by~$(\H_m, (.|.)_m)$.
Moreover, on the Dirac wave functions (not necessarily solutions of the Dirac equations)
one may introduce a dual pairing by integrating the spin scalar product over all of space-time,
\beq
\bra .|. \ket \::\: C^\infty(\scrM, S\scrM) \times C^\infty_0(\scrM, S\scrM) \rightarrow \C \:, \quad
\bra \psi|\phi \ket = \int_\scrM \Sl \psi | \phi \Sr_x \: d^4x\:. \label{stip} 
\eeq
The basic idea is to extend this dual pairing to a bilinear form on the Hilbert space~$\H_m$
and to represent this bilinear form in terms of the Hilbert space scalar product
\[ \bra \phi_m | \psi_m \ket = ( \phi_m \,|\, \Sig\, \psi_m)_m \:. \]
If~$\scrM$ is a space-time of {\em{finite lifetime}}, this construction can indeed be carried out
and defines the {\em{fermionic signature operator}}~$\Sig$
being a bounded symmetric operator on~$\H_m$ (see~\cite{finite}).
The positive and negative spectral subspaces of~$\Sig$ give the desired
decomposition of~$\H_m$ into two subspaces.
We remark that the fermionic signature operator makes it possible to study
{\em{spectral geometry}} for Lorentzian signature
(see~\cite{drum} and~\cite{index} for the connection to index theory).

In space-times of infinite lifetime like Minkowski space, the above method does not work
because~\eqref{stip} does not extend to a continuous bilinear form on~$\H_m \times \H_m$.
The underlying problem is that the time integral in~\eqref{stip} in general diverges for
solutions of the Dirac equation. In order to circumvent this problem, one considers
families of Dirac solutions~$(\psi_m)_{m \in I}$ (for an open interval~$I=(m_a, m_b)
\subset (0, \infty)$) and makes use of the fact that integrating over the mass parameter
generates decay of the wave functions for large times
(for details see~\cite{infinite}). As a result, one can make sense of the equation
\[ \bra \int_I \psi_m \,dm \,|\, \int_I \psi_{m'} \,dm' \ket = \int_I (\psi_m \,|\, \Sig_m \,\phi_m)_m\: dm \:, \]
which uniquely defines a family of bounded symmetric operators~$(\Sig_m)_{m \in I}$.
Now the positive and negative spectral subspaces of the operator~$\Sig_m$ again give the desired
decomposition of~$\H_m$ into two subspaces.

Having decomposed the solution space, one may choose the Hilbert space~$\H$
of the causal fermion system as one of the two subspaces of the solution space.
Choosing an orthonormal basis~$(u_\ell)$ of~$\H$ and introducing
the unregularized kernel of the fermionic projector again by~\eqref{Pxykernel},
one obtains a two-point distribution~$P(x,y)$. Using that this two-point distribution
comes from a projection operator in the Hilbert space~$\H_m$, 
there is a canonical construction which gives a quasi-free
Dirac field together with a Fock representation such that the two-point distribution
coincides with~$P(x,y)$. In the language of algebraic quantum field theory,
this result is stated as follows (see~\cite[Theorem~1.4]{hadamard}):
\begin{Thm} \label{thmstate}
There is an algebra of smeared fields generated by~$\Psi(g)$, $\Psi^*(f)$
together with a quasi-free state~$\omega$ with the following properties: \\[0.3em]
(a) The canonical anti-commutation relations hold:
\[ \{\Psi(g),\Psi^*(f)\} = \bra g^* \,|\, \tilde{k}_m\, f \ket \:,\qquad
\{\Psi(g),\Psi(g')\} = 0 = \{\Psi^*(f),\Psi^*(f')\} \:. \]
(b) The two-point function of the state is given by
\[ \omega \big( \Psi(g) \,\Psi^*(f) \big) = -\iint_{\scrM \times \scrM} g(x) P(x,y) f(y) \: d^4x\, d^4y\:. \]
\end{Thm}
This theorem means that before introducing an UV regularization, the description
of the Dirac system using the fermionic projector is equivalent to the usual description
of a quasi-free Dirac field in quantum field theory.

Moreover, it is shown in~\cite{hadamard} that the two-point distribution~$P(x,y)$ is of {\em{Hadamard form}},
provided that~$\B$ is smooth, not too large and decays faster than quadratically for large times
(for details see~\cite[Theorem~1.3]{hadamard} and the references in this paper).
This result implies that the representation of the quasi-free Dirac field as obtained
from the fermionic projector is a suitable starting point for a perturbative treatment
of the resulting interacting theory (see for example~\cite{brunetti+dutsch+fredenhagen}).

In our context, the fact that~$P(x,y)$ is of Hadamard form implies that that the results
in~\S\ref{seccausal} also apply in the presence of an external potential,
as we now explain. The Hadamard property means in words that the
bi-distribution~$P(x,y)$ in the presence of the external potential has the same singularity structure
as in the Minkowski vacuum. As a consequence, the arguments in~\S\ref{seccausal} remain true
if the points~$x$ and~$y$ are sufficiently close to each other. More precisely, the relevant length
scale is given by the inverse of the amplitude~$|\B(x)|^{-1}$ of the external
potential. On the other hand, the separation of the points~$x$ and~$y$ must be larger than
the scale~$\varepsilon$ on which regularization effects come into play.
Therefore, the causal structure of a causal fermion system agrees with that of Minkowski
space on the scale~$\varepsilon \ll \big|x^0-y^0\big| +\big|\vec{x}-\vec{y} \big| \ll |\B|^{-1}$
(where~$|\B|$ is any matrix norm).
Thinking of~$\varepsilon$ as being at least as small as the Planck length, in most situations
of interest the lower bound is no restriction. The upper bound is also unproblematic because
the causal structure on the macroscopic scale can still be recovered by considering
paths in space-time and subdividing the path on a scale~$\delta \ll |\B|^{-1}$
(similar as explained in~\cite[Section~4.4]{lqg} for the spin connection).
With this in mind, we conclude that the causal structure of a causal fermion system
indeed agrees with that of Minkowski space, even in the presence of an external potential.

\subsection{Effective Interaction via Classical Gauge Fields} \label{seccl}
We now outline how to describe interacting systems in Minkowski space
by analyzing the EL equations corresponding to the causal action principle
as worked out in Proposition~\ref{prpEL}. In this so-called {\em{continuum limit}}
the interaction is described by classical gauge fields.
For brevity, we can only explain a few basic concepts
and refer the interested reader to the detailed computations in the book~\cite{cfs}.

Let us begin with the Minkowski vacuum. As shown in~\S\ref{secuvreg},
regularizing a vacuum Dirac sea configuration gives rise to a
causal fermion system~$(\H, \F, \rho^\varepsilon)$.
Moreover, we saw in the following sections~\S\ref{seccorst}--\S\ref{seccorsw} that
the inherent structures of the causal fermion system can be identified with those
of Minkowski space (in particular, see~\eqref{Midentify} as well as Propositions~\ref{prpisometry}
and~\ref{lemma54}). This makes it possible to write the EL equations~\eqref{Qrel} as
\beq \label{ELreg}
\int_\scrM Q^\varepsilon(x,y)\, \big({\mathfrak{R}}_\varepsilon u_\ell \big)(y)\: d^4y
= \frac{\lambda}{2}\: \big({\mathfrak{R}}_\varepsilon u_\ell \big)(x) \qquad \text{for all~$u \in \H$}\:,
\eeq
where the regularized kernel~$Q^\varepsilon(x,y)$ is again defined via~\eqref{delLdef}
as the derivative of the Lagrangian.
Next, one chooses the Hilbert space~$\H$ as in~\S\ref{seccorcs} as the Dirac sea configuration
formed of all negative-energy solutions of the Dirac equation.
Then~$P^\varepsilon(x,y)$ can be computed explicitly by regularizing the
distribution~$P(x,y)$ as given in momentum space by~\eqref{Pxyvac} and
in position space by~\eqref{Pdiff} and Lemma~\ref{lemmaTintro}.
Computing~$Q^\varepsilon(x,y)$, it turns out that the
EL equations are mathematically well-defined if the convolution integral in~\eqref{ELreg} is
rewritten with the help of Plancherel's theorem as a multiplication in momentum space.
The analysis of the {\em{continuum limit}} gives a procedure for studying these equations in the
asymptotics~$\varepsilon \searrow 0$ when the regularization is removed.
The effective equations obtained in this asymptotic limit are evaluated most conveniently in
a formalism in which the unknown microscopic structure of space-time (as described by the regularization)
enters only in terms of a finite (typically small) number of so-called {\em{regularization parameters}}.
According to the method of variable regularization (see Remark~\ref{remmvr}),
one needs to analyze the dependence of the regularization parameters in detail.
It turns out that the causal fermion systems obtained from the vacuum Dirac sea configuration satisfy
the EL equations in the continuum limit, for any choice of the regularization parameters.

The first step towards interacting systems is to consider systems involving {\em{particles}}
and/or {\em{anti-particles}}. To this end, one simply modifies the constructions in~\S\ref{seccorcs}
by choosing the Hilbert space~$\H$ differently. Namely, instead of choosing
all negative-energy solutions, one chooses~$\H$ as a subspace of the solution space which
differs from the space of all negative-energy solutions by a finite-dimensional subspace.
In other words, $\H$ is obtained from the space of all negative-energy solutions by taking
out a finite number~$\na$ of states and by adding a finite number of states~$\np$ of positive energy.
Thus, denoting the regularized kernel of the fermionic projector of the Minkowski vacuum for
clarity by~$P^\varepsilon_\text{sea}(x,y)$, the kernel of the fermionic projector~\eqref{Pepsbase} can be written as
\beq \label{Pepsmod}
P^\varepsilon(x,y) = P^\varepsilon_\text{sea}(x,y)
-\sum_{k=1}^{\np} \big({\mathfrak{R}}_\varepsilon \psi_k \big)(x)
\overline{\big({\mathfrak{R}}_\varepsilon \psi_k \big)(y)}
+\sum_{l=1}^{\na} \big({\mathfrak{R}}_\varepsilon \phi_l \big)(x)
\overline{\big({\mathfrak{R}}_\varepsilon \phi_l \big)(y)} \:,
\eeq
where~$\psi_k$ and~$\phi_l$ are suitably normalized bases of the particle and anti-particle states,
respectively. In this procedure, we again take Dirac's concept of a ``sea'' of particles literally
and describe particles and anti-particles by occupying positive-energy states and creating ``holes'' in the
Dirac sea, respectively. We also remark that the construction~\eqref{Pepsmod}
modifies the kernel of the fermionic projector only by smooth contributions and thus
preserves the singularity structure of~$P^\varepsilon(x,y)$ as~$\varepsilon \searrow 0$.
As a consequence, the correspondence of the inherent structures of the causal fermion systems
to the structures in Minkowski space remains unchanged (just as explained at the end of~\S\ref{secquasifree}
for an external potential).

According to~\eqref{Pepsmod}, the particle and anti-particle states modify the
kernel of the fermionic projector. It turns out that this has the effect that
the EL equations in the continuum limit no longer hold.
In order to again satisfy these equations, we need to introduce an interaction.
In mathematical terms, this means that the universal measure~$\rho$ must be modified.
The basic question is how to modify the universal measure in such a way that the
EL equations in the continuum limit again hold.
It turns out that it is a useful first step
to insert an external potential~$\B$ into the Dirac equation~\eqref{Dirfree}
by going over to the Dirac equation~\eqref{Cont:DiracEq}.
Choosing~$\H$ as a subspace of the solution space of this Dirac equation,
the constructions of Section~\ref{secmink} again apply and give rise to causal fermion
systems~$(\H, \F, \rho^\varepsilon)$. The potential~$\B$ modifies the dynamics of all physical wave functions
in a collective way. Now one can ask the question whether the resulting causal fermion systems satisfy
the EL equations in the continuum limit. It turns out that this is the case
if and only if the potential~$\B$ satisfies certain equations, which can be identified
with classical field equations for the potential~$\B$. In this way, the causal action principle
gives rise to classical field equations.
In order to make our concepts clear, we point out that the 
potential~$\B$ merely is a convenient device in order to describe the collective
behavior of all physical wave functions. It should not be considered as a fundamental
object of the theory. We also note that, in order to describe variations of the physical wave functions,
the potential in~\eqref{Cont:DiracEq} can be chosen arbitrarily (in particular, the potential
does not need to satisfy any field equations). Each choice of~$\B$
describes a different variation of the physical wave functions.
It is the EL equations in the continuum limit which single out the physically admissible potentials
as being those which satisfy the field equations.

Before going on, we briefly explain how the subspace~$\H$ is chosen.
Clearly, the Dirac equation~\eqref{Cont:DiracEq}
cannot in general be solved in closed form. Therefore, for an explicit analysis
one must use perturbative methods. When performing the perturbation expansion,
one must be careful about the proper normalization of the fermionic states
(in the sense that spatial integrals of the form~\eqref{Pnorm} should be preserved).
Moreover, one must make sure that the singular structure of~$P(x,y)$ in
position space is compatible with the causal action principle
(meaning that the light-cone expansion of~$P(x,y)$ only involves bounded
integrals of~$\B$ and its derivatives).
Satisfying these two requirements leads to the {\em{causal perturbation expansion}}
(see~\cite{norm} and the references therein).
We also mention that regularizing the perturbation expansion is a delicate issue.
This can already be understood for the simple regularization by mollification
in Example~\ref{exmollify}, in which case it is not clear whether one should first
mollify and then introduce the interaction or vice versa.
The correct method for regularizing the perturbation expansion is obtained
by demanding that the behavior under gauge transformations should be preserved
by the regularization. This leads to the {\em{regularized causal perturbation expansion}}
as developed in~\cite[Appendix~D]{PFP} and~\cite[Appendix~F]{cfs}.

We proceed with a brief overview of the results of the analysis of the continuum limit.
In~\cite{cfs} the continuum limit is worked out in several steps beginning from simple systems
and ending with a system realizing the fermion configuration of the standard model.
For each of these systems, the continuum limit gives rise to effective equations for
second-quantized fermion fields
coupled to classical bosonic gauge fields (for the connection to second-quantized bosonic
fields see~\S\ref{secQFT} below).
To explain the structure of the obtained results, it is preferable to first describe the
system modelling the leptons as analyzed in~\cite[Chapter~4]{cfs}.
The input to this model is the configuration of the leptons in the standard model
without interaction. Thus the fermionic projector of the vacuum is assumed to be
composed of three generations of Dirac particles of masses~$m_1, m_2, m_3>0$
(describing~$e$, $\mu$, $\tau$)
as well as three generations of Dirac particles of masses~$\tilde{m}_1, \tilde{m}_2,
\tilde{m}_3 \geq 0$ (describing the corresponding neutrinos).
Furthermore, we assume that the regularization of the neutrinos breaks the chiral
symmetry (implying that we only see their left-handed components).
We point out that the definition of the model does not involve any assumptions on the interaction.

The detailed analysis in~\cite[Chapter~4]{cfs} reveals that the effective interaction in the continuum limit
has the following structure.
The fermions satisfy the Dirac equation coupled to a left-handed $\SU(2)$-gauge
potential~$A_L=\big( A_L^{ij} \big)_{i,j=1,2}$,
\[ \left[ i \Pdd + \begin{pmatrix} \Aslsh_L^{11} & \Aslsh_L^{12}\, \UMNS^* \\[0.2em]
\Aslsh_L^{21}\, \UMNS & -\Aslsh_L^{11} \end{pmatrix} \chi_L
- m Y \right] \!\psi = 0 \:, \]
where we used a block matrix notation (in which the matrix entries are
$3 \times 3$-matrices). Here~$mY$ is a diagonal matrix composed of the fermion masses,
\beq \label{mY}
mY = \text{diag} (\tilde{m}_1, \tilde{m}_2, \tilde{m}_3,\: m_1, m_2, m_3)\:,
\eeq
and~$\UMNS$ is a unitary $3 \times 3$-matrix (taking the role of the
Maki-Nakagawa-Sakata matrix in the standard model).
The gauge potentials~$A_L$ satisfy a classical Yang-Mills-type equation, coupled
to the fermions. More precisely, writing the isospin dependence of the gauge potentials according
to~$A_L = \sum_{\alpha=1}^3 A_L^\alpha \sigma^\alpha$ in terms of Pauli matrices,
we obtain the field equations
\beq \label{l:YM}
\partial^k \partial_l (A^\alpha_L)^l - \Box (A^\alpha_L)^k - M_\alpha^2\, (A^\alpha_L)^k = c_\alpha\,
\overline{\psi} \big( \chi_L \gamma^k \, \sigma^\alpha \big) \psi\:,
\eeq
valid for~$\alpha=1,2,3$ (for notational simplicity, we wrote the Dirac current for one Dirac
particle; for a second-quantized Dirac field, this current is to be replaced by the expectation value
of the corresponding fermionic field operators). Here~$M_\alpha$ are the bosonic masses and~$c_\alpha$
the corresponding coupling constants.
The masses and coupling constants of the two off-diagonal components are
equal, i.e.\ $M_1=M_2$ and~$c_1 = c_2$,
but they may be different from the mass and coupling constant
of the diagonal component~$\alpha=3$. Generally speaking, the mass ratios~$M_1/m_1$, $M_3/m_1$ as
well as the coupling constants~$c_1$, $c_3$ depend on the regularization. For a given regularization,
they are computable.

Finally, our model involves a gravitational field described by the Einstein equations
\beq \label{l:Einstein}
R_{jk} - \frac{1}{2}\:R\: g_{jk} + \Lambda\, g_{jk} = \kappa\, T_{jk} \:,
\eeq
where~$R_{jk}$ denotes the Ricci tensor, $R$ is scalar curvature, and~$T_{jk}$
is the energy-momentum tensor of the Dirac field. Moreover, $\kappa$ and~$\Lambda$ denote the
gravitational and the cosmological constants, respectively.
We find that the gravitational constant scales like~$\kappa \sim \delta^{-2}$, where~$\delta \geq \varepsilon$ is
the length scale on which the chiral symmetry is broken.

In~\cite[Chapter~5]{cfs} a system is analyzed which realizes the
configuration of the leptons and quarks in the standard model.
The result is that the field equation~\eqref{l:YM} is replaced by
field equations for the electroweak and strong interactions after spontaneous
symmetry breaking (the dynamics of the corresponding Higgs field has not yet been analyzed).
Furthermore, the system again involves gravity~\eqref{l:Einstein}.

A few clarifying remarks are in order. First, the above field equations come with
corrections which for brevity we cannot discuss here (see~\cite[Sections~3.8, 4.4 and~4.6]{cfs}).
Next, it is worth noting that, although
the states of the Dirac sea are explicitly taken into account in our analysis, they do not
enter the field equations. More specifically, in a perturbative treatment,
the divergences of the Feynman diagram
describing the vacuum polarization drop out of the EL equations of the causal action.
Similarly, the naive ``infinite negative energy density'' of the
sea drops out of the Einstein equations, making it unnecessary to subtract any counter terms.
We finally remark that the only free parameters of the theory are the masses in~\eqref{mY}
as well as the parameter~$\delta$ which determines the gravitational constant.
The coupling constants, the bosonic masses and the mixing matrices
are functions of the regularization parameters
which are unknown due to our present lack of knowledge on the microscopic structure of space-time.
The regularization parameters cannot be chosen arbitrarily because they must satisfy certain
relations. But except for these constraints, the regularization parameters are currently treated
as free empirical parameters.

To summarize, the dynamics in the continuum limit is described by Dirac spinors
coupled to classical gauge fields and gravity. The effective continuum theory is manifestly covariant
under general coordinate transformations.
The only limitation of the continuum limit is that the bosonic fields are merely classical.
We shall come back to second-quantized bosonic fields in~\S\ref{secQFT} below.

\subsection{Effective Interaction via Bosonic Quantum Fields} \label{secQFT}
In~\S\ref{seccl} it was outlined that and in which sense
the regularized Dirac sea vacuum satisfies the EL equations~\eqref{Qrel}.
In simple terms, these results mean that the regularized Dirac sea vacuum is a critical
point of the causal action under variations of the physical wave functions
(see Definition~\ref{defvarc}).
We now explain why the regularized Dirac sea vacuum is  {\em{not a minimizer}} of the causal action
principle. This argument will lead us to a method for further decreasing the causal action.
It also gives some insight on the structure of the minimizing measure.
In particular, we shall see that the effective interaction in the resulting space-time
is to be described effectively by bosonic {\em{quantum}} fields.

Suppose that~$(\H, \F, \rho)$ is a causal fermion system describing a regularized
Dirac sea configuration (see~\S\ref{seccorcs}).
In order to explain the basic idea, it suffices to consider the case that~$\rho$ has
finite total volume (which can be arranged for example by considering the system
in a four-dimensional box).
For a unitary transformation~$V \in \U(\H)$, we define the measure~$V(\rho)$ by
\beq \label{Vrho}
(V \rho)(\Omega) = \rho(V \Omega V^{-1}) \:.
\eeq
We choose a finite number of unitary transformations~$V_1, \ldots, V_L$
and introduce a new measure~$\tilde{\rho}$ as the convex combination of the
unitarily transformed measures,
\[ \tilde{\rho} = \frac{1}{L} \sum_{\as=1}^L V_\as \rho \:. \]
Obviously, all linear constraints like the volume
constraint~\eqref{volconstraint} and the trace constraint~\eqref{trconstraint} are
preserved by this transformation. The action becomes
\begin{align}
\Sact(\tilde{\rho}) &= \frac{1}{L^2} \sum_{\as, \bs=1}^L \iint_{\F \times \F} \L(x,y)\: d(V_\as \rho)(x)\:
d(V_\bs \rho)(y) \notag \\
&= \frac{\Sact(\rho)}{L} + \frac{1}{L^2} \sum_{\as \neq \bs} \iint_{\F \times \F} \L(x,y)\: d(V_\as \rho)(x)\:
d(V_\bs \rho)(y) \:. \label{Smix}
\end{align}
Due to the factor~$1/L$, the first summand becomes small as~$L$ increases.
The second summand involves all the contributions for~$\as \neq \bs$.
If we can arrange that these contributions become small, then the action
of the new measure~$\tilde{\rho}$ will indeed be smaller than the action of~$\rho$.

Let us consider the contributions for~$\as \neq \bs$ in more detail. In order to simplify the explanations,
it is convenient to assume that the measures~$V_\as \rho$ have mutually disjoint supports
(this can typically be arranged by a suitable choice of the unitary transformations~$V_\as$).
Then the space-time~$\tilde{M} := \supp \tilde{\rho}$ can be decomposed into~$L$ ``sub-space-times''
$M_\as := \supp \rho_\as$,
\[ \tilde{M} = M_1 \cup \cdots \cup M_L \qquad \text{and} \qquad M_\as \cap M_\bs = \varnothing \quad
\text{if $\as \neq \bs$}\:. \]
Likewise, a physical wave function~$\psi^u$ can be decomposed into the contributions in
the individual sub-space-times,
\[ \psi^u = \sum_{\as=1}^L \psi^u_\as \qquad \text{with} \qquad
\psi^u_\as := \chi_{M_\as} \,\psi^u \]
(and~$\chi_{M_\as}$ is the characteristic function). This also gives rise to a corresponding
decomposition of the fermionic projector:
\begin{Lemma} Every sub-space-time~$M_\as$ of~$\tilde{M}$ is homeomorphic to~$M$,
with a homeomorphism given by
\[ \phi_\as \::\: M \rightarrow M_\as \:,\qquad
\phi_\as(x) := V_\as^* \,x\, V_\as \:. \]
Moreover, the mapping
\beq \label{Vasiso}
V_\as^* \big|_{S_x} \,:\, S_x \rightarrow S_{\phi_\as(x)}
\eeq
is an isomorphism of the corresponding spinor spaces.
Identifying the spinor spaces in different sub-space-times via this isomorphism,
the fermionic projector can be written as
\begin{align}
P(x,y) &= -\sum_{\as, \bs=1}^L \chi_{M_\as}(x) \: P_{\as, \bs}(x,y)\:\chi_{M_\bs}(y) \qquad \text{with} \\
P_{\as, \bs}(x,y) \:&\!:= \Psi(x) \:V_\as \,V_\bs^* \:\Psi(y)^* \:. \label{Pxymix}
\end{align}
\end{Lemma}
\Proof The definition of~$V \rho$, \eqref{Vrho}, immediately implies that the
transformation~\eqref{psiudef} maps~$M$ to~$M_\as$ and is a homeomorphism.
By definition of the physical wave function~\eqref{psiudef},
\[ \psi^u(\phi_\as(x)) = \pi_{\phi_\as(x)} = \pi_{V_\as^* x V_\as} u
= V_\as^* \,\pi_x\, V_\as u \:. \]
The identification~\eqref{Vasiso} makes it possible to leave out the factor~$V_\as^*$.
Then we can write the wave evaluation operator~\eqref{weo} as
\[ \tilde{\Psi}(x) = \sum_{\as=1}^L \chi_{M_\as}(x)\: \Psi(x)\: V_\as \:. \]
Applying~\eqref{Pid} gives the result.
\QED

This lemma makes it possible to rewrite the action~\eqref{Smix} as
\beq \label{SPmix}
\Sact(\tilde{\rho}) = \frac{\Sact(\rho)}{L} + \frac{1}{L^2} \sum_{\as \neq \bs}
\iint_{M \times M} \L\big[P_{\as, \bs}(x,y)\big] \: d\rho(x)\: d\rho(y) \:,
\eeq
where the square bracket means that the Lagrangian is computed
as a function of the kernel of the fermionic projector~$P_{\as, \bs}(x,y)$
(just as explained after~\eqref{Axydef} for the kernel~$P(x,y)$).
The identities~\eqref{Pxymix} and~\eqref{SPmix} give a good intuitive understanding of how
the action depends on the unitary operators~$V_\as$.
We first note that in the case~$\as = \bs$, the unitary operators in~\eqref{Pxymix} drop out,
so that~$P_{\as,\as}(x,y) = P(x,y)$. This also explains why the first summand in~\eqref{SPmix}
involves the original action~$\Sact(\rho)$.
In the~$\as \neq \bs$, however, the unitary operators in~\eqref{Pxymix} do not drop out.
In particular, this makes it possible to introduce phase factors into the fermionic projector.
For example, one may change the phase of each physical wave function~$\psi^u_\as$
arbitrarily while keeping the physical wave functions~$\psi^u_\bs$ for~$\bs \neq \as$ unchanged.
Choosing the resulting phases randomly, one gets destructive interference, implying that
the kernel~$P_{\as,\bs}(x,y)$ becomes small.
Making use of this {\em{dephasing effect}}, one can make the summands
in~\eqref{SPmix} for~$\as \neq \bs$ small. A detailed analysis of the involved scalings
reveals that this indeed makes it possible to decrease the causal action
(see~\cite{qft}).

In words, this result means that minimizing the causal action triggers a mechanism
which tends to decompose space-time~$M$ into many small sub-space-times
$M_1, \ldots, M_L$. The physical wave functions in the different sub-space-times
involve relative phases, with the effect that the correlations between the sub-space-times
(as described by the kernels~$P_{\as,\bs}(x,y)$) become small.
Since the dephasing takes place on a microscopic length scale, this effect is
referred to as {\em{microscopic mixing}}.

Let us discuss what microscopic mixing implies for the effective macroscopic interaction.
One must distinguish two situations. One limiting case is complete dephasing, in which
case~$P_{\as,\bs}$ is approximately zero. As a result, there are no relations
or structures between the two sub-space-times
(note that for example the causal structure is encoded in the kernel of the fermionic projector;
see~\S\ref{secker}). This entails that the two sub-space-times do not interact with each other.
The resulting picture is that space-time looks effectively like a ``superposition'' of
the different sub-space-times. This scenario is referred to as the {\em{microscopic mixing of
space-time regions}}. The dephasing can be understood similar to
decoherence effects in standard quantum field theory (see for example~\cite{joos}).

If each of the microscopically mixed sub-space-times involves a different
classical bosonic field, one obtains effectively a superposition of classical field configurations.
This makes it possible to describe second-quantized bosonic fields
(see~\cite{entangle}). However, as the different sub-space-times do not interact with each other,
each sub-space-time has it own independent dynamics. This dynamics is described by the
classical bosonic field in the corresponding sub-space-time.

In order to obtain an interaction via second-quantized bosonic fields,
one needs to consider another limiting case in which
the dephasing involves only some of the physical
wave functions. In this case, the fermionic projector~$P_{\as,\bs}$ is not necessarily small.
This also implies that relations arising as a consequence of the collective behavior
of all physical wave functions (like the causal relations or classical bosonic fields)
still exist between the sub-space-times~$M_\as$ and~$M_\bs$.
In more physical terms, the sub-space-times still interact with each other.
This scenario is studied in~\cite{qft} and is referred to as the {\em{microscopic mixing
of wave functions}}. In order to describe the effective interaction, one describes the
unitary operators~$V_\as$ by random matrices. Taking averages over the random
matrices, one finds that the effective interaction can be described perturbatively
in terms of Feynman diagrams which involve both fermionic and bosonic loops.
The appearance of bosonic loops can be understood by working with second-quantized bosonic fields.
Working out the detailed combinatorics and the implications of the resulting quantum
field theory is work in progress (for the first step in this program see~\cite{qed}).


\Thanks {{\em{Acknowledgments:}}
I would like to thank the referee for helpful comments on the manuscript.

\providecommand{\bysame}{\leavevmode\hbox to3em{\hrulefill}\thinspace}
\providecommand{\MR}{\relax\ifhmode\unskip\space\fi MR }
\providecommand{\MRhref}[2]{%
  \href{http://www.ams.org/mathscinet-getitem?mr=#1}{#2}
}
\providecommand{\href}[2]{#2}

\end{document}